\documentclass[12pt]{article}
\usepackage[a4paper, total={6in, 10in}]{geometry}
\usepackage{amssymb}
\usepackage{latexsym}
\usepackage{authblk}
\usepackage{url}
\usepackage{soul}
\usepackage{color}
\usepackage[dvipsnames]{xcolor}
\usepackage{amsmath}
\usepackage{booktabs}
\usepackage{subfigure}
\usepackage{hyperref}
\usepackage{tabularx}
\usepackage{array}

\DeclareMathOperator*{\argmin}{arg\,min}
\DeclareMathOperator{\erfc}{erfc}
\usepackage{pdflscape}
\usepackage{rotating}
\newcolumntype{Y}{>{\centering\arraybackslash}X}

\begin{document}

\title{Calibration of cross-field transport models in SOLPS-ITER on the TCV-X21 case}

\author{Stefano Carli$^{1,2}$, Reinart Coosemans$^3$, Claudia Colandrea$^3$ and Wouter Dekeyser$^{1}$}

\affil{$^1$Department of Mechanical Engineering, KU Leuven, Leuven, Belgium}

\affil{$^2$LPP-ERM/KMS, Brussels, Belgium}

\affil{$^3$Swiss Plasma Center, École Polytechnique Fédérale de Lausanne, Lausanne, Switzerland}
\date{}
\maketitle
\begin{abstract}
Cross-field turbulent transport remains one of the largest uncertainties in edge plasma simulations and is commonly approximated through empirical transport coefficients. In this work, we calibrate and assess several cross-field transport models implemented in SOLPS-ITER using measurements from the TCV-X21 reference case. The considered models range from conventional constant-diffusivity descriptions to the self-consistent $\kappa$-model, in which anomalous diffusivities evolve along with plasma conditions. Model parameters are estimated through gradient-based optimization by minimizing discrepancies between simulated and experimental upstream and divertor profiles in forward field configuration. The calibration results show that increasing the number of free parameters substantially improves agreement with the calibration dataset. However, these more flexible models exhibit poor predictive capability when applied to the reversed field configuration, indicating overfitting. In contrast, the simplest constant-diffusivity model provides the best overall predictive performance while requiring only a small number of calibrated parameters. The $\kappa$-model achieves a calibration quality comparable to the constant-diffusivity model and reproduces the experimental profiles with similar accuracy, while simultaneously providing a physics-based description of the spatial variation of anomalous transport. Predictions for a density scan reveal differences that are not apparent near the calibration point. The $\kappa$-model predicts increasing transport levels around the separatrix with increasing density, leading to broader upstream profiles and an earlier onset of divertor rollover compared to the constant-diffusivity model. The presented framework provides a systematic and efficient route for model calibration in SOLPS-ITER, and a set of calibrated $\kappa$-model parameters to be employed in future studies.

\end{abstract}

\section{Introduction}\label{sc:intro}
Reliable operation of magnetically confined nuclear fusion reactors requires a robust solution to the yet unsolved problem of plasma particle and heat exhaust \cite{roadmap}. In this context, the key component is the divertor, subject to mitigated heat loads of 10 MW/m$^2$ and beyond in steady-state operation, along with highly energetic particle fluxes. As such, significant challenges are posed to its lifetime and reliability concerning cooling capabilities, surface material sputtering, erosion, and melting. To tackle this open issue, one typically employs plasma edge codes, aiming at accurately describing the plasma behavior in the Scrape-Off Layer (SOL) and especially in the divertor. With these codes, interpretation of current experiments, predictions for future reactors, and design of the divertor and other plasma facing components can be carried out.

A key player in the SOL and divertor plasma behavior is the cross-field turbulent transport, which drives the magnitude of the heat flux decay length $\lambda_q$, and in turn the peak heat load on the divertor targets. 3D edge (fluid) turbulence codes like GRILLIX \cite{grillix}, SOLEDGE3X \cite{soledge3x}, and GBS \cite{gbs} have seen rapid developments in recent years, thanks to both novel numerical techniques and increase in the available computational power. Despite these advances, 3D edge turbulence codes still require huge computational resources to resolve the fine spatial and time scales of plasma turbulence. As such, reactor-scale simulations of e.g. ITER are only now within reach, and their day-to-day use for experiment interpretation, scenario development, or divertor design is unfeasible. Furthermore, a detailed plasma-neutrals and plasma-wall interactions model is of utmost importance to truthfully capture plasma edge behavior\cite{detachment1,detachment2,detachment3}. This is only approximately included in turbulence codes \cite{soledgeneut, grillixneut, gbsneut}, and coupling with state-of-the-art neutral particles models such as EIRENE \cite{reiter} remains computationally too expensive \cite{soledgeneut}. On top of this, plasma-wall and plasma neutral interactions strongly depend on the divertor or first wall geometry \cite{closure1,closure2,closure3}, which is not always well captured by plasma edge turbulence codes, see for example the penalization technique employed by GRILLIX. Finally, plasma impurities also play an important role in the SOL, either because they are purposely puffed to radiate the incoming power from the core before it reaches the divertor \cite{radiation1,radiation2}, or because they are unwanted He ashes and wall-sputtered atoms, which need to be pumped-away \cite{pump1,pump2}. In either case, they too require an adequate model, and first steps in this direction have only been recently taken by including multi-component closures in SOLEDGE3X and GBS \cite{soledgezhdanov, gbsimpurity}.

In view of the limitations mentioned, experiment interpretation, scenario development and divertor design is routinely carried out through mean-field plasma edge codes like SOLPS-ITER \cite{SOLPS1,SOLPS2}, SOLEDGE2D \cite{soledge}, UEDGE \cite{uedge}, and EDGE2D \cite{edge2d}. SOLPS-ITER in particular is arguably the most wide-spread plasma edge code: it has been extensively employed for the ITER divertor design and operational scenarios \cite{pitts,park}; it is routinely used for interpretation of current tokamaks like ASDEX Upgrade \cite{panasdex}, JET \cite{horsten2025}, and EAST \cite{solpseast}, and linear machines \cite{solpsgym,solpsmagnumpsi}; and it is currently employed for designing  future reactors like EU-DEMO\cite{subba}, SPARC \cite{solpssparc}, and CFEDR \cite{solpscfedr}. While limited to a 2D radial-poloidal cross section assuming toroidal axisymmetry, SOLPS-ITER provides a global consistent picture of the plasma edge, by coupling the plasma transport code B2.5 \cite{b25}, solving the Braginskii fluid equations \cite{braginskii}, with the Monte Carlo code EIRENE \cite{reiter}, solving the kinetic Boltzmann equation for neutral particles. As such, SOLPS-ITER accurately accounts for plasma-neutral and plasma-wall interactions reproducing detachment behavior \cite{reimold}, electromegnetic drift effects \cite{solps_eq} leading to divertor asymmetries \cite{paradela}, and realistic divertor geometry to study advanced divertor configurations \cite{solpsadc}.

SOLPS-ITER  has seen significant advances in recent years. From the numerical standpoint, the most important one has been the upgrade of B2.5 to a fully unstructured solver \cite{unstructured}. This has removed one of the main limitations of the code, allowing simulations up to and including realistic first wall geometries. In older structured versions of the code, implementation of a 9-point stencil scheme correctly accounted for plasma grid distortion \cite{9point}, significantly affecting simulations adopting a fluid neutral approximation. Finally, considerable code speedups have been gained tackling the neutral transport model: first, EIRENE averaging schemes have been implemented, which provide the same statistical accuracy of the solution employing fewer Monte Carlo particles \cite{eireneaveraging}; second, Advanced Fluid Neutral (AFN) \cite{afns1,afns2} and spatially hybrid fluid-kinetic neutral \cite{spatiallyhybr1,spatiallyhybr2} models allow obtaining the same accuracy of the EIRENE kinetic treatment at a considerably reduced computational effort. From the physics point of view, the multi-ion collisional Grad-Zhdanov closure now allows simulating any arbitrary plasma mixture \cite{makarov23}. However, as other mean-field plasma edge codes, SOLPS-ITER does not resolve the turbulent cross-field transport. Instead, it strongly approximates such transport adopting a diffusive or advective-diffusive model, possibly with the addition of ballooning enhancement,  prescribing ad-hoc \textit{anomalous} coefficients. These coefficients needs to be estimated from experimental data, and are in general device, scenario, and spatially dependent \cite{diffusion}, significantly hampering the predictive capabilities of the code. Following the Reynolds Average Navier Stokes (RANS) approach widely employed in hydrodynamic turbulence, Coosemans and Dekeyser developed a novel self-consistent cross-field transport model for SOLPS-ITER, called the $\kappa$-model \cite{kmodel4, kmodel}, while Baschetti et al. carried out similar work for SOLEDGE2D \cite{baschetti}. The $\kappa$-model describes the average turbulence behavior by solving an additional model equation for the turbulent kinetic energy $\kappa$, the value of which locally indicates the turbulence strength, and is thus linked to the local value of the cross-field diffusion coefficients. In this way, one obtains self-consistently a 2D distribution of such transport coefficients, thereby alleviating the limitation of imposing ad-hoc values. The new model still retains tuning parameters, albeit fewer and supposedly more universal. The first applications of this new model on COMPASS \cite{kmodelcompass} and C-Mod \cite{kmodel} proved encouraging, but the model parameters were manually adjusted until a certain degree of agreement with respect to experimental data was obtained. A more systematic approach for their calibration has not been carried out and is therefore one of the objectives of this work. We adopt as case study the TCV-X21 reference case, which has been developed for the validation of edge fluid turbulence codes  \cite{x21paper, dudsonx21}. We first perform the model calibration step on forward field data, and in a second step we assess the predictive capability of the calibrated models on reversed field data and in a density scan. The final outcome of this work is dual: first of all, a calibrated set of $\kappa$-model parameters, which can be employed in further modeling studies and as starting point for Uncertainty Quantification (UQ); second, a guideline on how systematic and efficient model calibration in plasma edge codes can be carried out using gradient-based optimization.

The remainder of this paper is organized as follows: in section \ref{sc:model_calibration} we briefly review what model calibration is, the current state-of-the-art in the plasma edge community, and the gradient-based optimization approach adopted in this work; in section \ref{sc:transport_models} we concisely describe the cross-field transport models of SOLPS-ITER; section \ref{sc:setup} then summarizes the simulation and calibration setups. Afterwards, we present and discuss the model calibration results on forward field data in section \ref{sc:calibration}. We then analyze model prediction using the calibrated parameters on reversed field data and in a density scan in section \ref{sc:prediction}. Finally, we conclude and provide suggestions for future work in section \ref{sc:conclusion}.


\section{Model calibration}\label{sc:model_calibration}

The following aims at giving a brief overview of model calibration in general, without delving too deeply in the topic itself. This introduction is based on the following references, which the interested reader may want to look into: \cite{jaynes,tarantola} give a general overview of inverse problems and probability theory; \cite{kennedyohagan} focuses on the Bayesian inference approach while \cite{dose,toussaint} further develop the Bayesian approach for fusion-specific problems.

When dealing with (numerical) models, the typical problem of interest is the so-called \textit{forward problem}, in which the model $\mathcal{F}$, which is a parametrization of a physical law, for example a numerical implementation of Braginskii equations, is evaluated to make a prediction of certain outputs or quantities of interest $y$. Concerning the model and its parametrization, one typically distinguishes between true model inputs $x$ and model \textit{parameters} $\theta$. The former are quantities which can be directly measured albeit with a certain parametric variability, for example the (discrete) space or time points at which the Braginskii equations are evaluated, or an imposed gas puff strength. The model parameters $\theta$ instead, cannot be directly measured or are the consequence of model assumptions (parametrization). The perfect example here are the particles and heat diffusion coefficients employed in the approximation of perpendicular transport in the mean-field Braginskii equations. The forward problem is then simply stated as $y=\mathcal{F}(x,\theta)$. 

Model calibration, also known as parameter estimation, inference, data assimilation or backward UQ, is instead an \textit{inverse problem}, in which the value of model parameters $\theta$ is sought so that the output $y=\mathcal{F}(x,\theta)$ reproduces as closely as possible the measurement data $\mathcal{D}$. We could thus naively look at the inverse problem as $\theta=\mathcal{F}^{-1}(x,\mathcal{D})$, (keeping in mind that this is not correct, as explained later on). One major challenge compared to the forward problem is clear, namely that the functional relationship $\mathcal{F}^{-1}$ is hardly ever available, except for very simple models. Furthermore, one should also account for the fact that uncertainties are inherently present, being in the model inputs (e.g. enforcing boundary conditions of the systems up to a certain controlling error), the measured outputs (e.g. measurement error), and the model itself (e.g. continuum vs discrete). This implies that the solution to the calibration process is not unique, but rather a spectrum of possible outcomes. Therefore, the inverse problem is more correctly approached in a probabilistic sense, where the inversion $\theta=\mathcal{F}^{-1}(x,\mathcal{D})$ is not valid.

For treating such kind of probabilistic problems, one can follow two distinct views: one is the frequentist approach, which treats the model parameters as fixed yet unknown constants; the other is the Bayesian approach, which we argue is the most adequate, and treats model parameters as random variables with a probability distribution. Following one or the other leads to quite different methods for solving the model calibration problem, as well as in the results. In the Bayesian setting, the current state of information on the parameters, expressed by the prior probability distribution, is updated through information gained from experimental data, expressed by the likelihood function, to provide the posterior probability distribution of the model parameters. The latter inherently contains uncertainty information on such parameters (e.g. their variance). Dealing with probability distributions, Bayesian model calibration typically employs sampling-based methods like Markov-Chain Monte Carlo, which are computationally very expensive in that they entail a random exploration of the whole parameter space. Approximate solutions are nevertheless available, assuming Gaussian distributions. On the other hand, frequentist methods, for example Maximum Likelihood or least squares, provide point-estimates obtained through gradient-based optimization, which is computationally cheaper. However, they are subject to local optima and only approximate uncertainty estimates can be obtained, for example Fisher information.

In this work, we focus on point estimates, obtained through a least squares estimation, and not on the full posterior distribution, as its estimation is not yet available for SOLPS-ITER and is thus left as future work. This will nonetheless provide the community with reasonable values for $\kappa$-model parameters to be used for predictions.  Incidentally, the least squares optimum is the same as the Bayesian Maximum A Posterior (MAP) if in the latter one assumes uniform priors and a Gaussian likelihood with constant standard deviation. In case of a varying (unknown) standard deviation, the least squares result is still very close to the MAP \cite{carlibayes}, and thus can be used as a hot-start for complete posterior estimation and UQ.

\subsection{Model calibration in plasma edge codes}
In the field of plasma edge modeling, model calibration has mostly been based on expensive parameter scans or manual iterative procedures \cite{xuereb}. With the high computational cost of plasma edge codes, along with the many model parameters, such procedures are excessively demanding. Moreover, visual inspection of results and modeler experience were the only \textit{subjective} stopping criteria for the procedure. In addition, uncertainty on the experimental data is not considered, and uncertainty estimates on the obtained parameters are not available.

Coster et al. and Kim et al. developed in an earlier version of SOLPS an automated nonlinear least squares regression method to estimate spatially constant anomalous diffusion coefficients and boundary conditions to match experimental data \cite{coster,kim}. They thereby introduced an \textit{objective} metric for comparing simulation results to experimental data, and made a significant step forward compared to manual tuning and visual inspection of results. Moreover, they included some aspects of measurement uncertainty. On the other hand, the use of finite differences for gradient calculation makes this approach unfeasible for large parameter sets, e.g. radial profiles of transport coefficients. In addition, no uncertainty estimates on the obtained parameters is available. In a recent study, Wang et al. employed a similar nonlinear regression procedure for estimating constant diffusion coefficients and gas puff strength in TCV \cite{wangx21}. This procedure, employing the conjugate gradient method, still relies on finite differences gradient computation.

An iterative algorithm recently proposed by Canik et al. \cite{canik} and Zito et al. \cite{zito} estimates radial profiles of transport coefficients based on physical arguments. With the advantage of using a reduced set of model evaluations, this approach lacks an objective metric quantification and quickly looses applicability for complex nonlinear models, such as the $\kappa$-model. On top of that, parameters uncertainty is not considered, and measurement uncertainty only implicitly.

Even more recently, a different approach based on Bayesian Optimization has been developed by Bowman et al. \cite{bowman2} and Hecko et al. \cite{hecko} for SOLPS-ITER, and Fu et al. \cite{fu2026} for UEDGE. In this approach, a surrogate model is constructed by running the plasma edge code at selected points in parameter space. The cheaper surrogate is then employed in a gradient-based optimization step. This approach is quite powerful as the expensive model is run for a limited number of points, and the obtained cheap surrogate can be readily exploited for UQ. On the other hand, the surrogate's accuracy strongly depends on the number of expensive model evaluations, and the approach itself is generally suited for small to medium size problems, in terms of number of calibration parameters \cite{bayesopt}. Moreover, the obtained probability distribution of the parameters primarily reflects uncertainty of the surrogate itself in predicting the actual plasma edge model.

In this work, we adopt instead an adjoint-based optimization strategy, which was first proposed by Baelmans et al. for a simplified plasma edge model \cite{baelmans2014}, and then developed for SOLPS-ITER \cite{carlibayes}, making use of Algorithmic Differentiation (AD) for gradient computation \cite{carliadjoint,carlipsi}. This adjoint-based strategy, employing the full SOLPS-ITER model throughout the optimization, is likely more computationally intensive compared to e.g. the Bayesian Optimization approach, and may converge to local optima. A second limitation of this strategy is that currently only a fluid atoms model for the neutral particles can be employed, although first steps to include the EIRENE neutral model have been taken \cite{horstenad}. The advantages of the adjoint-based optimization strategy are that the full plasma edge model details and its nonlinearities are by definition accounted for, and any problem complexity/size (in terms of number of calibration parameters) can be tackled at roughly the same computational cost thanks to adjoint gradients \cite{carliadjoint}. Incidentally, an almost identical approach using AD gradients has been deployed by Auroux et al. \cite{auroux} and Lamerand et al. \cite{lamerand} for the parameter estimation of the $\kappa$-model in Baschetti et al. \cite{baschetti}, though the procedure was restricted to the $\kappa$-model itself, decoupled from the full plasma edge model.

To conclude, we summarize in Table \ref{tb:table1} the different model calibration options for SOLPS-ITER, highlighting their features.

\begin{landscape}
	\begin{table}[p]
		\caption{Summary of available model calibration strategies for SOLPS-ITER}\label{tb:table1}
		\centering
		\begin{tabularx}{\linewidth}{m{7cm} Y Y Y Y Y Y}
			\hline
			Type \& author(s) & Problem type & Problem size & CPU effort & Objective metric & UQ capability & Distributed with SOLPS-ITER \\
			\hline
			Parameter scans, manual tuning & Any  & Small & Very high & No & No & / \\
			Nonlinear regression (Kim, Coster) & Any & Small & High & Yes & Partial & No\\
			Nonlinear regression (Wang) & Any & Small & High & Yes & No & No\\
			Physics-based (Canik, Zito) & Only radial diffusion coefficients & Small & Low & No & No & No\\
			Bayesian Optimization (Bowman, Hecko) & Any & Medium & Low & Yes & Yes & No\\
			Adjoint-based optimization (Carli) & Any & Large & Medium & Yes & Yes & Yes\\
			\hline
		\end{tabularx}
	\end{table}
\end{landscape}

\subsection{Adjoint-based optimization for model calibration}\label{sc:adjointop}

In this section, we briefly recall the adjoint-based model calibration procedure already described in \cite{carlibayes}, which entails finding the best-fitting set of model parameters $\theta$ in a least squares sense. Namely, we seek the parameters $\theta$ that minimize the scalar performance indicator $\mathcal{J}$:

\begin{equation}
	\hat{\theta} =\argmin_\theta \mathcal{J}(\theta, q \left(\theta; x\right); x, \mathcal{D}).
\end{equation}

In practice, this performance indicator $\mathcal{J}(\theta, q \left(\theta; x\right); x, \mathcal{D})$ quantifies the (dis)agreement between model outputs $y$ and experimental data $\mathcal{D}$, and is also known as the cost function. It depends on the unknown model parameters $\theta$, the model inputs $x$, and the state variables $q$:

\begin{equation} \label{eq:leastsquares}
	\mathcal{J}(\theta, q\left(\theta; x\right); x, \mathcal{D})=\sum_{l} \sum_{y} \frac{1}{\Omega_l} \frac{w_{l,y}}{\bar{\mathcal{D}}_{y,l}^2} \int_{\Omega_l} \left( \frac{\left(y(\theta, q\left(\theta; x\right); x ) - \mathcal{D}_y  \right)^2}{2} \right) \mathrm{d}\Omega.
\end{equation} 

In Eq. \ref{eq:leastsquares} summation over $l$ indicates different spatial locations such as outer midplane (OMP) or target, while summation over $y$ indicates different output quantities such as plasma density or saturation current. The symbol $\mathcal{D}_y$ indicates the experimental data related to the plasma quantity $y$.  $\Omega_l$ is the domain over which the integration is performed and $w_{l,y}$ are weighting factors. Finally, the normalizing constants $\bar{\mathcal{D}}_{y,l}=1/\Omega_l\int_{\Omega_l} \mathcal{D}_{y}\mathrm{d}\Omega$ are the average experimental value of $\mathcal{D}_{y}$ over domain $\Omega_l$. The state variables $q$ are density and parallel velocity of ionized and neutral species, ions and electrons temperature, and electrostatic potential, which are defined by the plasma transport equations and their boundary conditions, commonly indicated as the state equations $\mathcal{B}(\theta,q(\theta;x);x)=0$. The general model calibration problem now consists of finding the set of parameters $\theta$ that minimize the cost function, subject to the condition that state equations are satisfied. As such, the model calibration is cast into a PDE-constrained optimization problem of the form:
\begin{equation}
	\begin{array}{ll}
		\min\limits_{\theta,q} \mathcal{J}(\theta,q(\theta;x);x,\mathcal{D})  \\
		\mathrm{s.t.}~\mathcal{B}(\theta,q(\theta;x);x)=0. \\
	\end{array}
\end{equation}

Through the state equations, the state variables $q(\theta;x)$  are uniquely determined for a certain set of model parameters $\theta$ and model inputs $x$, namely $\mathcal{B}(\theta, q(\theta;x) ; x) \equiv 0$. Thus, dropping from now onward the dependence on model inputs $x$ and experimental data $\mathcal{D}$, we can introduce the so-called reduced cost function $\hat{\mathcal{J}}(\theta) \equiv \mathcal{J}(\theta,q(\theta))$, which implicitly contains the dependence on state variables. The optimization problem is therefore:
\begin{equation}
		\min\limits_{\theta} \hat{\mathcal{J}}(\theta).
\end{equation}
Such optimization problem can effectively be tackled with gradient-based methods, which are very efficient when dealing with expensive-to-evaluate models, but converge in general to a local optimum \cite{nocedal}. Of course, one is now in need not only of the cost function value, but also of its gradient with respect to the model parameters $\theta$, which in the optimization context are also known as \textit{design} variables.

The gradient $\nabla_\theta\hat{\mathcal{J}}(\theta)$ is easily obtained through finite differences, as discussed previously, though its cost is proportional to the number of model parameters $\theta$. A more effective approach widely employed in optimal design is the use of the adjoint method, which has been already introduced in the plasma edge community for parameter estimation and divertor design in a simplified setting \cite{baelmans2014,divshape,magnfield1}. With this method, a second set of equations, the adjoint equations, are derived from the underlying state equations, and solved to provide the desired gradient information at a cost which is roughly equal to that of the state equations themselves, independently of the number of model parameters. Since plasma edge codes like SOLPS-ITER are complex, multi-physics codes with a large developers basis, direct derivation and manual implementation of the adjoint equations would be cumbersome and error-prone. Thus, we deployed in SOLPS-ITER a semi-automatic (adjoint) derivative computation using AD \cite{griewank}. The interested reader can find the details of the implementation in \cite{carlipsi,carliadjoint}.

In addition, constraints other than the state equations can be included in the optimization problem, and are typically distinguished between state and design constraints. The former are conditions involving state variables, for example that the temperature on divertor targets must be smaller than a threshold. In gradient-based  methods, one additionally needs the full gradient of such constraints with respect to the model parameters, which is as expensive as the gradient of the cost function itself. As such, these constraints are ``difficult'' to handle, though solutions like the Augmented Lagrangian approach exist \cite{sander}. The design constraints are instead conditions directly acting on the model parameters $\theta$. This type of constraint is so to say ``easy'' to handle, as the required gradient is straightforwardly available. An example of such kind of constraint is the positiveness of diffusion coefficients, which can be expressed as a bound-constraint. In the current implementation only bound constraints on model parameters are included. The final bound-constraint optimization problem takes thus the form:
\begin{equation}\label{eq:optim}
	\begin{array}{ll}
		\min\limits_{\theta} \hat{\mathcal{J}}(\theta) \\
		\mathrm{s.t.}~\theta_{L}\le\theta\le\theta_{U},
	\end{array}
\end{equation}
where $\theta_{L}$ and $\theta_{U}$ are lower and upper bounds respectively.

\begin{figure}
	\centering
	\includegraphics[width=0.5\textwidth]{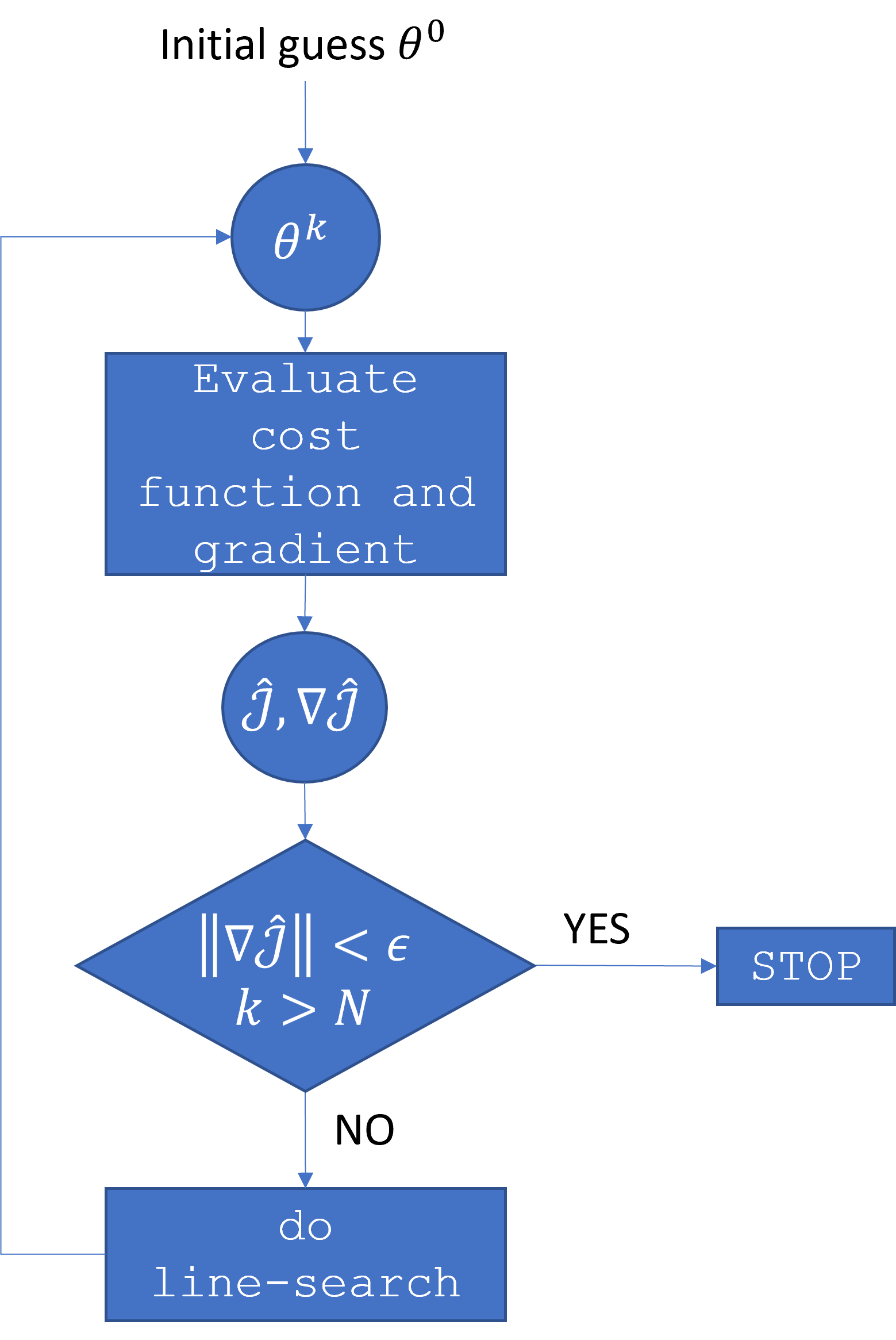}
	\caption{Schematic of the optimization algorithm in SOLPS-ITER}
	\label{fg:optiloop}
\end{figure}

The optimization problem in Eq. \ref{eq:optim} is solved in SOLPS-ITER by coupling to the PETSc/TAO library \cite{petsc1,petsc2,petsc3}, from which different optimization methods can be chosen. The general optimization algorithm follows the steps summarized in Figure \ref{fg:optiloop}: an initial guess $\theta^0$ for the model parameters is first provided, which is used to evaluate the cost function and its gradient; then, a check is done on certain convergence criteria, typically involving the gradient norm  $\vert\vert \nabla \hat{\mathcal{J}} \vert\vert$ being smaller than a threshold $\epsilon$, or the number of optimization iterations $k$ being greater than a maximum $N$. If yes, the algorithm stops, if not the line-search procedure is called to find the new guess for the model parameters based on the current cost function and gradient values: 
\begin{equation}\label{eq:linesearch}
 \theta^{k+1}=\theta^k-\alpha_k B_k^{-1}\nabla\hat{\mathcal{J}}(\theta^k),
\end{equation}
and the next iteration follows. In Eq. \ref{eq:linesearch} the matrix $B_k$ depends on the chosen method, e.g. unity for steepest descent and Hessian for Newton method, while the step length $\alpha_k$ is chosen at each $k$ in a backtracking approach to satisfy the Armijo condition and ensure convergence to an optimum \cite{nocedal}.

\section{Cross-field transport models}\label{sc:transport_models}
In this section we briefly present the cross-field transport models available in SOLPS-ITER and addressed by the model calibration procedure. First, we show what we will now call the `standard' model, i.e. the specification of ad-hoc anomalous transport coefficients, and then the $\kappa$-model.

\subsection{Standard model}\label{sc:stdmodel}
The standard cross-field transport model employed in SOLPS-ITER approximates the turbulent transport by an anomalous diffusive flux of particles $\mathbf{\Gamma}_{i,an}$, momentum $\mathbf{\Gamma}^m_{i,an}$, and heat $\mathbf{Q}_{i/e,an}$, where subscript $i$ refers to ions and $e$ to electrons, with user-defined constant diffusion coefficients $D_\perp$, $\nu_{\perp}$ and $\chi_{\perp}$ respectively. These anomalous fluxes take the following form:
\begin{gather}\label{eq:standard}
	\mathbf{\Gamma}_{i,an} = -D_{\perp} f_b \nabla n + \mathbf{v}_{\perp} f_b n,\\
	\mathbf{\Gamma}^m_{i,an} = -n m \nu_{\perp} f_b \nabla u_{\parallel},\\ 
	\mathbf{Q}_{i/e,an} = -\chi_{i/e,\perp} n f_b \nabla T_{i/e},\\
\end{gather}
where $n$ is the ion density, $m$ is the ion mass, $u_{\parallel}$ is the ion parallel velocity and $T$ is the ion or electron temperature. In addition, an anomalous convective velocity $\mathbf{v}_\perp$ can also be specified for the particle transport. The factor $f_b$ models a ballooning enhancement of all transport coefficients as
\begin{equation}\label{eq:ball}
f_b= C_a\left\vert \frac{\bar{B}}{B\left(\theta,r\right)} \right\vert^{C_b}, 
\end{equation}
with $\bar{B}$ the arithmetic average of the magnetic field in the whole domain, $\theta$ and $r$ the poloidal and radial coordinates respectively, and $C_a$ and $C_b$ model parameters. All the above mentioned transport coefficients can be defined either as constants or as radially varying. For the latter, one specifies their values at selected radial points $r-r_{sep}$ at the OMP.

\subsection{$\kappa$-model}\label{sc:kmodel}
We now synthesize the $\kappa$-model equations, the meaning behind them, and identify the model parameters where present. For the full derivation of such equations and their closures we refer to the works in \cite{kmodel1,kmodel2,kmodel3,kmodel4,kmodel5}, while the implementation in SOLPS-ITER is presented in \cite{kmodel}.

The basic idea behind the $\kappa$-model is the same as RANS models used in hydrodynamics \cite{pope}, where the fine length and time scales of turbulence are not resolved. Yet, they self-consistently capture turbulence effects in an averaged sense and at a reduced computational cost, significantly improving the predictive capability compared to ad-hoc diffusion coefficients. More specifically, the $\kappa$-model is derived by ensemble averaging the Braginskii fluid turbulent equations to find on one side the mean-field particle, current, momentum, and energy conservation equations, which eventually are the ones adopted in SOLPS-ITER \cite{solps_eq,solpsmanual}, and on the other side a new mean-field transport equation for the perpendicular turbulent kinetic energy $\kappa$, a quantity that describes the local intensity of the turbulence. This derivation ensures self-consistent coupling between transport of $\kappa$ and other plasma quantities, and gives rise to nonlinear terms which require specific closure. The most important of these terms are those related to $E \times B$ drift fluctuations, which are eventually linked to the anomalous fluxes presented above. Therefore, only such nonlinear perpendicular transport terms are retained, and closures are found by assuming an electrostatic $E \times B$ interchange turbulence regime and a diffusive transport, where the local value of the diffusion coefficients is linked to the local $\kappa$ value. In particular, the main closure term is the one relating the anomalous particle diffusion coefficient to $\kappa$:
\begin{equation}\label{eq:kdanml}
	D_\perp=D_{E\times B} + D_{min} = C_{D} \frac{\kappa/m_i}{\sqrt{\kappa/m_i}/\rho_L + C_S\left\vert S_{mean}\right\vert} + D_{min}.
\end{equation}
This closure is not an analytical expression derived from theory, but rather a model best-fitting 2D turbulence data \cite{kmodel1}. In Eq. \ref{eq:kdanml}, the turbulence-related transport is modeled via  $D_{E\times B}$, while $D_{min}$ is a lower limit mimicking classical transport. Further,  $\rho_L$ is the local gyro radius, $m_i$ is the ion mass, and $S_{mean}=\nabla_r V_{E\times B,\perp}$ is the radial shear of the diamagnetic $E\times B$ velocity. Finally, the factors $C_D$ and $C_S$ are model parameters. Based on the given model for $D_\perp$, similar closures are assumed for the ions/electron heat diffusivities and ions momentum diffusivity according to the following equations:
\begin{gather}
	\chi_{i/e,\perp} = \chi_{i/e,E\times B} + \chi_{i/e,min} = C_{\chi_{i/e}} D_{E\times B} + \chi_{i/e,min}, \label{eq:keps_heat}\\ 
	\nu_{\perp} = \nu_{E\times B} + \nu_{min} = C_{\nu} D_{E\times B} + \nu_{min}, \label{eq:keps_nu}
\end{gather}
where $C_{\chi_{i/e}}$ and $C_\nu$ are model parameters, and $\chi_{i/e,min}$ and $\nu_{min}$ are classical lower limits.

The transport equation for $\kappa$ takes the usual form of conservation equations:
\begin{equation}\label{eq:keq}
	\frac{\partial n_i\kappa }{\partial t} + \nabla \cdot \mathbf{\Gamma}_\kappa =  S_{\kappa,prod} + S_{\kappa,diss} -\mathbf{\Pi}_{RS}:\nabla \mathbf{V}_{E\times B}^T.
\end{equation}
In Eq. \ref{eq:keq} $\mathbf{\Gamma}_\kappa$ is the flux of $\kappa$, $S_{\kappa,prod}$ is the interchange source term, $S_{diss,\kappa}$ is the parallel dissipation of $\kappa$ and $\mathbf{\Pi}_{RS}$ is the Reynolds Stress. Concerning $\mathbf{\Gamma}_\kappa$, it is made up of four contributions:
\begin{equation}\label{eq:kflux}
	\mathbf{\Gamma}_\kappa = \mathbf{\Gamma}_i\kappa - n_i \chi_{\kappa,E\times B} \nabla \kappa + C_{ie} \frac{2}{3} \left(\mathbf{Q}_{i,E\times B} + \mathbf{Q}_{e,E\times B}  \right) - C_{\sigma_{\parallel},1}\sigma_{\parallel}B^2\rho_L^2\mathbf{b}\nabla_\parallel\frac{\kappa}{m_i}.
\end{equation}
The first term in Eq. \ref{eq:kflux} is the convection with the mean-field ions flux $\mathbf{\Gamma}_i$, while the second is a turbulent diffusion term linked to gradients in $\kappa$ with coefficient 
\begin{equation}\label{eq:keps_cdkt}
	\chi_{\kappa,E\times B} = C_\kappa D_{\perp},
\end{equation}
and $C_\kappa$ a model parameter. The third term in Eq. \ref{eq:kflux} is a transport contribution stemming from the turbulent heat fluxes $\mathbf{Q}_{i/e,E\times B}$, defined as
\begin{equation} \label{eq:kheatflux}
	\mathbf{Q}_{i/e,E\times B} = \frac{3}{2} D_{\perp,E\times B} T_{i/e} \nabla n_i - \chi_{i/e, E \times B} n_{i/e} \nabla T_{i/e},
\end{equation}
and $C_{ie}$ in Eq. \ref{eq:kflux} is another model parameter. The fourth and last term in the $\kappa$ flux of Eq. \ref{eq:kflux} is a parallel transport term stemming from potential and parallel current fluctuations. In this term we find the parallel conductivity $\sigma_{\parallel}$, the connection length $L_{conn}$, and another model constant $C_{\sigma_{\parallel},1}$.

Looking now at the sources and sinks on the right hand side of Eq. \ref{eq:keq}, we find first the interchange source term
\begin{equation}\label{eq:skprod}
	S_{\kappa,prod}=\left[ (T_i+T_e)D_{E\times B}\nabla n_i + \frac{2}{3}n_e\chi_{e,E\times B}\nabla T_e  + \frac{2}{3}n_i\chi_{i,E\times B}\nabla T_i  \right] \cdot \nabla(\ln(B^2)),
\end{equation}
which, for electrostatic $E \times B$ turbulence, does not need any additional closure. Note that this term naturally includes the ballooned character of interchange turbulence, namely enhanced transport on the low-field-side of the tokamak, where turbulent heat flux and magnetic field gradient are opposite, leading to a positive source term. Contrary, on the high-field-side, they both point in the same direction and this term is therefore negative, effectively suppressing transport. Contributions from ions and electrons to this term appear with opposite sign in the respective energy equations, ensuring global energy conservation.

The second term on the right hand side of Eq. \ref{eq:keq} models parallel resistive dissipation of $\kappa$, again linked to fluctuating potential and parallel current similar to the last term in Eq. \ref{eq:kflux}:
\begin{equation} \label{eq:kdiss}
	S_{\kappa,diss} = - C_{\sigma_{\parallel},2} \sigma_{\parallel}B^2\frac{\rho_L^2}{L_{conn}^2}\frac{\kappa}{m_i}.
\end{equation}
This term contains the model parameter $C_{\sigma_{\parallel},2}$ and appears with opposite sign in the electron energy equation.

The last term on the right hand side of Eq. \ref{eq:keq} corresponds to the Reynolds Stress and models $E \times B$ flow shear,  suppressing turbulence and transport:
\begin{equation} \label{eq:visrs}
	-\mathbf{\Pi}_{RS}:\nabla \mathbf{V}_{E\times B}^T \approx \frac{2}{3}n_i\kappa \mathbf{V}_{E\times B} \cdot \nabla \ln B^2 -\frac{2}{3} \eta_{E\times B} \left(\mathbf{V}_{E\times B} \cdot \nabla \ln B^2\right)^2 + \eta_{E\times B} \left( \nabla_r {V}_{E\times B,\theta} \right)^2.
\end{equation}
In Eq. \ref{eq:visrs}, $\eta_{E\times B}$ is a negative turbulent viscosity modeled as
\begin{equation} \label{eq:ceta}
	\eta_{E\times B} = -C_\eta m_i n_i D_{E\times B},
\end{equation}
with $C_{\eta}$ a model parameter. Global energy conservation is ensured in this case by an additional drift velocity $\mathbf{V}_{RS}$ in the ion perpendicular momentum equation
\begin{equation}\label{eq:rsdrift}
  \mathbf{V}_{RS}=-\frac{\nabla \cdot \mathbf{\Pi}_{RS}}{qB^2} \times \mathbf{B} \approx C_{vis,\kappa} \left( \frac{\mathbf{b}}{B}\times \nabla \left(\frac{2}{3} n_i \kappa\right) - \frac{\mathbf{e}_r}{B} \nabla \cdot \left(\eta_{E\times B} \nabla_r V _{E\times B,\perp} \mathbf{e}_r\right) \right),
\end{equation}
which tends to be very small apart from regions with strong radial $\kappa$ gradients.

To summarize, we have replaced the specification of (radial) ad-hoc diffusion coefficients, possibly with a poloidal ballooning enhancement, with the solution of the $\kappa$ transport equation \ref{eq:keq} along with the specification of ten, presumably more universal, model parameters: $C_D$, $C_S$, $C_{\chi,e}$, $C_{\chi,i}$, $C_\nu$, $C_{\kappa}$, $C_{ie}$, $C_{\sigma_\parallel,1}$, $C_{\sigma_\parallel,2}$, $C_\eta$. Closures for such parameters are needed either from theoretical derivations or comparison with turbulence simulation data. Such endeavor has already been started in \cite{kmodel2,kmodel3,dewolf,kmodel4}, where Bayesian inference on different competing closures with 2D interchange turbulence data provided values for $C_D\approx8$, $C_S\approx15$, $C_{\chi,i/e}\approx1$, $C_\kappa\approx0.8$, $C_{\sigma_{\parallel},2} \approx 37$ and $C_{ie}\approx1$. Moreover, some of these parameters have stronger impact on the solution, while other likely play a smaller role, e.g. $C_\kappa$ and $C_\nu$. The local sensitivity analysis presented later on this work already shows the parameters' impact on the obtained solution, and should be complemented in future with a global sensitivity analysis. We remark once again the self-consistent nature of the $\kappa$-model, which models the effects of electrostatic $E\times B$ interchange turbulence on cross-field transport, and automatically provides 2D distributions of diffusion coefficient.


\section{Modeling setup}\label{sc:setup}
As stated in the introduction, we target in this paper the TCV-X21 reference case described in \cite{x21paper}. This case is particularly suited for edge turbulence code validation as it features a reduced toroidal field and line integrated density, thereby increasing the turbulence length scale and limiting the effect of neutral particles in the divertor. The case is a collection of several identically programmed deuterium-fuelled L-mode discharges with ohmic heating power of 150 kW, of which approximately 120 kW cross the separatrix. Both forward (favorable) and reversed (unfavorable) toroidal field directions have been performed, and the experimental data has been collected in an open-source database \cite{x21data}.

\begin{figure}
	\centering
	\includegraphics[scale=0.6]{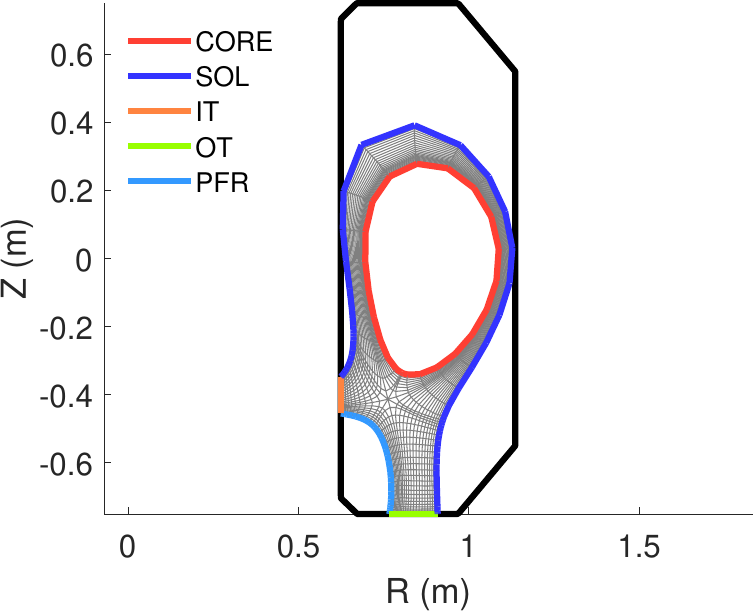}
	\caption{Geometry and computational grid for the TCV-X21 case (grey), with identification of boundaries: core (red), SOL (blue), IT (orange), OT (green), and PFR (light blue). The vessel wall is represented in black. } \label{fg:geom}
\end{figure}

To model the TCV-X21 case, we employ the SOLPS-ITER wide grid version \cite{unstructured}. Since the cost of the optimization in the model calibration step is essentially equivalent to running several tens of SOLPS-ITER simulations in serial, we make some approximations in the case setup to keep the computational cost limited. In particular, we use a relatively coarse grid made of 60 poloidal and 24 radial cells, and not extending up to the vessel wall. We assess a-posteriori the effect of considering a finer grid by doubling the number of cells in each direction, while the assessment of a fully extended grid is left for future study. The employed grid is shown on Figure \ref{fg:geom}, along with the vessel wall and the identification of boundary faces groups. We also approximate the description of neutral particles by employing the AFN model \cite{afns1,afns2}, and as such only consider D atoms, assuming Maxwellian distributions at all boundaries, a carbon wall material, and a Franck-Condon dissociation energy equal to 3 eV. Given that the neutral particles density was limited by design of the experimental scenario, we expect that the error we make with this approximation is negligible, and we verify a-posteriori the effect of including kinetic neutrals with EIRENE. Note also that the optimization procedure we employ for the calibration cannot yet deal with kinetic neutrals. Furthermore, we assume a pure deuterium plasma without carbon impurities. Given the experimental conditions, one can expect a low carbon impurity concentration, and we assess the effect of this assumption along with the kinetic neutrals one. For both deuterium species a recycling coefficient $R_c = 0.99$ is employed at each boundary, following other TCV  modeling activities \cite{wensing21, wangx21, tonello2024, colandrea, carpita2025}.

The cross-field transport model is described further on in this section, where the model calibration setup is presented. We mention here that for the standard cross-field transport model a fixed viscosity $\nu_{\perp} = 0.2$  kg/m/s is used, as we have seen a negligible effect when varying this parameter (for the $\kappa$ the viscosity is not fixed, and optimized with the parameter $C_\eta$). In addition, and regardless of the cross-field transport model, we employ an anomalous current conductivity $\sigma_{\perp} = 2.5\cdot10^{-4}en_e$ S/m (with $e$ the elementary charge and $n_e$ electron density). While this value is relatively high compared to other studies \cite{kaveeva2018,colandrea,carpita2025}, it was found that larger values would significantly hamper the simulation convergence during the optimization, especially for the $\kappa$-model, and despite using the partial flux-surface averaging proposed in \cite{kaveeva2018}. This may also be linked to the relatively high false time step (10$^{-3}$s) and relaxation factor (0.3) employed, again to spare computational time. Also in this case, we verify a-posteriori the effect of this assumption. Electrons and ions parallel heat transport coefficients are flux-limited, while for neutral atoms we impose an isotropic flux-limit to all transport coefficients. Finally, electromagnetic drifts and currents are included.

To complete the simulation case setup we list the boundary conditions (BCs):
\begin{itemize}
	\item \textit{Core}. Zero particle flux for D and fixed D$^+$ density (see calibration setup below). Constant heat flux $P_{core}=$125 kW, equally distributed between ions and electrons. Zero parallel velocity gradient for D  and zero parallel velocity for D$^+$. Imposed diamagnetic and inertia currents for the potential equation. Fixed $\kappa$ value.
	\item \textit{Inner target (IT) and outer target (OT)}. Standard Bohm-Chodura BC \cite{bohm}.
	\item \textit{SOL}. An outward particle flow is imposed for D$^+$ continuity equation, defined as a fraction of the local sound speed flux $\Gamma_{i}=\alpha c_s n_{i}$, with $\alpha=-1.5 \cdot 10^{-3}$. The same type of BC is also enforced for the energy equations: for ions and electrons we impose a flux $Q_{i/e}=\alpha c_s n_{i/e} T_{i/e}$, with factor $\alpha=-1.5 \cdot 10^{-2}$, while for $\kappa$ the imposed flux is $\Gamma_{\kappa}=\alpha c_s n_{i} \kappa$, with $\alpha=-1.5 \cdot 10^{-3}$. Zero-gradient BC is applied both to D$^+$ parallel velocity and to the potential. For D, zero particle and momentum flux is imposed. 
	\item \textit{Private flux region (PFR)}. Zero gradient is imposed to ions density and parallel velocity, ions and electrons temperatures, potential, and $\kappa$. Zero particle and momentum flux is imposed to neutral atoms.
\end{itemize}

\subsection{Model calibration setup}\label{sc:calibration_setup}
We present now the setup common to both the standard model and $\kappa$-model calibration. As stated in the BC list above, we impose a fixed D$^+$ density at the core, instead of a gas puff at the chamber wall. We also let this BC be a model parameter in the calibration, instead of imposing an experimental estimate, so that the optimizer will find a value consistent with the other model parameters and experimental data included in the cost function. Note that, in principle, we should also allow $P_{core}$ to be a model parameter, as this quantity is subject to uncertainties, e.g. in the bolometric reconstruction of the core radiated power. However, initial tests showed that the calibration was too sensitive to both these core BC, the optimization generally seeking a tradeoff between these two rather than varying other model parameters. In a pragmatic approach, we chose to only let the density vary, while future studies should, if possible, include also the heat flux, either as a state constraint or with precise bounds, to alleviate such sensitivity. In the same vain, also the recycling coefficient $R_c$ should have been treated as unknown, despite the adopted value is widely employed for D on carbon.

We include in the cost function $\hat{\mathcal{J}}$ experimental quantities from the forward field configuration at the OMP and both targets. In particular, we include OMP radial profiles of electron density $n_e$ and temperature $T_e$ from Thomson Scattering (TS)\footnote{The actual TS measurements are taken somewhat above the X-point and not at the OMP, see Figure 1 in \cite{x21paper}. However, in the current implementation it is not possible to specify cost function data at arbitrary lines of sight. As such, we assume the TS data to be taken at the OMP, given their close similarity with FHRP data.} and Fast Horizontal Reciprocating Probe (FHRP). At both targets, we include radial profiles of saturation current $j_{sat}$ and electron temperature $T_e$ from Langmuir probes (LP). Finally, we also include the parallel heat flux $q_\parallel$ radial profile at the outer target (OT) from the infrared camera (IR). While additional experimental data is in principle available \cite{x21paper,x21data}, we choose to include only \textit{directly} measured quantities, i.e. the ones not involving any model (let alone for the FHRP density). The current code capability does not allow for 2D cost function definition and we thus excluded the 2D profiles of the reciprocating divertor probe array. Thus, the cost function in Eq. \ref{eq:leastsquares} employs for $l$ and $y$ the values listed in Table \ref{tb:table2}, and equal weight $w_{l,y}=1$ for each quantity at each location.

\begin{table}
	\caption{Summary of quantities included in the cost function} \label{tb:table2}
	\centering
	\begin{tabular}{l l l l}
		\hline
		Location $l$ 		& OMP 		 & OT 		      & IT\\
		\hline
		Plasma quantity $y$ & $n_e$ (TS) & $j_{sat}$ (LP) & $j_{sat}$ (LP) \\
						    & $T_e$ (TS) & $T_{e}$ (LP) & $T_{e}$ (LP) \\
						    & $n_e$ (FHRP) & $q_\parallel$ (IR) & \\
						    & $T_e$ (FHRP) &  & \\
		\hline
	\end{tabular}
\end{table}

Finally, as optimization method we employ the Quasi-Newton BFGS algorithm \cite{nocedal}, which provides superlinear convergence when the solution is not too far from the optimum. As stopping criteria we set a tolerance $10^{-4}$ for the gradient norm ($\vert\vert \nabla \hat{\mathcal{J}} \vert\vert$), gradient relative norm (${\vert\vert \nabla \hat{\mathcal{J}} \vert\vert}/{\vert \hat{\mathcal{J}}\vert}$) and gradient reduction (${\vert\vert \nabla \hat{\mathcal{J}} \vert\vert}/{\vert\vert \nabla \hat{\mathcal{J}}_0 \vert\vert}$), the optimization being stopped when one of these has been met. Additionally, we limit the optimization iterations to a maximum of 50, after which the process is terminated. In addition, each cost function or gradient evaluation converges the SOLPS-ITER solution for 25k false time steps. While this does not ensure machine-accurate residuals, it provides sufficiently accurate cost function and gradient values at a reduced computational cost. We also ensured stable convergence of the optimization by devising safeguards as described in Appendix \ref{app:d}.

\subsubsection{Standard model calibration setup}
For the standard model presented in Section \ref{sc:stdmodel}, we consider different levels of complexity, with the aim of showing how larger parametric freedom provides better cost function performance with respect to simpler models, at the cost of poor predictive capability, the well-known overfitting problem.

The simplest model, referred in the following as Model 1, considers constant anomalous transport coefficients throughout the computational domain, with the additional ballooning enhancement of Eq. \ref{eq:ball}. While in principle Eq. \ref{eq:ball} features two coefficients, the multiplier $C_a$ and the exponent $C_b$, we only retain the latter as model parameter and set $C_a=1$. The reason is that eventually $C_a$ is simply multiplied to each transport coefficient, e.g. $D_\perp C_a$, and the two would be highly correlated, providing non-unique solutions, see for example Figure 3 of \cite{dewolf}. We leave the verification of this behavior to a future Bayesian inference and UQ effort on the same case. Further, since the $\kappa$-model does not include yet a convective component for the anomalous fluxes, we exclude $\mathbf{v}_\perp$ from the calibration procedure. Hence, for Model 1 we have five model parameters: $D_\perp$, $\chi_{e,\perp}$, $\chi_{i,\perp}$, $C_b$, and $n_{e,core}$.

The second model we consider extends the previous one by defining a radial profile for the diffusion coefficients. In doing so, we arbitrarily choose six radial locations $r-r_{sep}=[-1, -0.5, 0, 0.5, 1.5, 3]$ cm along the OMP. The optimizer is in this case free to vary the transport coefficients at these points, but not their radial location. Obviously, the optimization outcome highly depends on this choice, and future developments should include a more general parametrization, allowing also the variation of such radial locations, as already done in \cite{bowman2}. Hence, this model, hereafter referred to as Model 2, includes 20 model parameters.

The last model considered brings the parametric freedom to the extreme, extending the previous case by defining  transport coefficient in each of the 24 cells of the OMP. We do not list each of these radial locations, which can be found in the accompanying dataset \cite{repodata}. This last model thus employs 74 model parameters and will be referred to as Model 3. 

Table \ref{tb:table3} summarizes the different models and the number of parameters included, along with the arbitrarily chosen initial guess $\theta^0$, lower and upper bounds $\theta_L$, $\theta_{U}$.

\begin{table}
	\caption{Summary of the standard models and their parameters, with initial guess and bounds} \label{tb:table3}
	\centering
	\begin{tabular}{c c c c c c}
		\hline
		  Parameter $\theta$ & $D_\perp$ (m$^2$/s) & $\chi_{e,\perp}$ (m$^2$/s) & $\chi_{i,\perp}$ (m$^2$/s) & $C_b$ (-) & $n_{e,core}$ (m$^{-3}$) \\
		\hline
		Model 1 & ct. & ct. & ct.  & ct. & ct. \\ 
		Model 2 & 6 pts. & 6 pts. & 6 pts. &  ct. & ct. \\
		Model 3 & 24 pts. & 24 pts. & 24 pts. & ct. & ct. \\
		\hline
		$\theta^0$ & 0.5 & 0.5 & 0.5 & 1 & 10$^{19}$ \\
		$\theta_L$ & 10$^{-2}$ & 10$^{-2}$ & 10$^{-2}$ & 0 & 0.6$\cdot$10$^{19}$ \\
		$\theta_U$ & 10 & 25 & 5 & 10 & 1.5$\cdot$10$^{19}$ \\
		\hline
	\end{tabular}
\end{table}

\subsubsection{$\kappa$ model calibration setup}

For the $\kappa$-model we consider a total of 11 model parameters, as summarized in Table \ref{tb:table4} alongside their initial guess and lower/upper bounds. Compared to what we presented in Section \ref{sc:kmodel}, we split the multiplier to the $\kappa$ resistive dissipation in Eq. \ref{eq:kdiss} into a separate contribution inside the core and outside of it. Indeed this dissipation mechanism is strictly speaking valid only for open field lines regions, and we assume the same dissipation form is valid in the core with a separate multiplier $C_{\sigma_{\parallel},2,core}$.

In addition, the attentive reader may notice that we excluded from the calibration set two model parameters, the most important one being the multiplier $C_D$ in the closure term for $D_{E\times B}$ in Eq. \ref{eq:kdanml}. This choice stems from the experience gained in previous work \cite{carlibayes}, where we showed that $C_D$ and the $\kappa$ BC at the core are highly correlated with $C_D \propto \frac{1}{\sqrt{\kappa_{core}}}$, making the solution non-unique. In principle, one could estimate the radial profile of the turbulent kinetic energy based on the measured fluctuations in the FHRP electric field. Including this additional quantity in the cost function further constrains the optimization, and would allow for a unique solution to exist. However, it is yet to be proved whether the proxy of $\kappa$ estimated from FHRP itself is reliable. As such, we choose to leave the BC $\kappa_{core}$ free, and set $C_D=0.1$ based on our previous work \cite{kmodelcompass,carlibayes,kmodel}. A future Bayesian inference effort \cite{carlibayes,dose} will allow to correctly account for and unveil these interdependencies, which the current least-squares approach suffers from.

The second parameter excluded is the turbulent viscosity multiplier $C_\eta$ appearing in the Reynolds Stress term, see Eq. \ref{eq:visrs}-\ref{eq:ceta}. Note that while in principle the first term in Eq.\ref{eq:visrs} is not multiplied by $C_{\eta}$, the whole term is effectively zero if $C_{\eta}=0$. The reason for this choice is similar as for $C_D$: the Reynolds Stress dissipation term in the $\kappa$ equation plays a similar role as the flow shear term in Eq. \ref{eq:kdanml}, suppressing transport. As such, we expect that the model parameters connected with these two terms, $C_S$ and $C_\eta$, are correlated as well. We leave also these interdependecies and more in general the weight of this term to be assessed in a future Bayesian approach. To ensure global energy conservation, we also exclude the additional drift velocity stemming from the Reynolds Stress term, Eq. \ref{eq:rsdrift}.

At last, we assign a fixed value to the classical lower limits for the transport coefficients in Eq. \ref{eq:kdanml}, \ref{eq:keps_heat}, \ref{eq:keps_nu}. For this, we first choose $D_{min}=0.05$ to avoid instabilities in regions where $\kappa$ is suppressed. Then, based on classical arguments we assume that $\nu_{min}=D_{min}$, $\chi_{i,min}=10D_{min}$, and $\chi_{e,min}=\frac{5}{2}D_{min}$.

\begin{table}
	\caption{Summary of the $\kappa$-model parameters, with initial guess and bounds} \label{tb:table4}
	\centering
	\begin{tabular}{c c c c c c c}
		\hline
		Parameter $\theta$ & $C_{\chi_e}$ & $C_{\chi_i}$ & $C_{\sigma_{\parallel},1}$ & $C_{\sigma_{\parallel},2}$ & $C_{\sigma_{\parallel},2,core}$ &  $C_S$  \\
		\hline
		$\theta^0$ & 1 & 1 & 10$^{-1}$  & 10$^{-1}$  & 1 & 10$^{-1}$ \\
		$\theta_L$ & 10$^{-2}$ & 10$^{-2}$ & 10$^{-3}$ & 10$^{-6}$  & 10$^{-6}$  & 10$^{-4}$ \\
		$\theta_U$ & 10 & 10 & 10 & 10 & 10 & 1\\
		\hline
		Parameter $\theta$ &  $C_{\nu}$ & $C_\kappa$ & $C_{ie}$ & $\kappa_{core}$ (eV) & $n_{e,core}$ (m$^{-3}$) & \\
		\hline
		$\theta^0$ & 0.2 & 1 & 10$^{-1}$  & 1 & 10$^{19}$ & \\
		$\theta_L$ & 10$^{-4}$ & 10$^{-4}$ & 10$^{-4}$ & 10$^{-3}$  & 0.6$\cdot$10$^{19}$ & \\
		$\theta_U$ & 10$^{2}$  & 10$^{2}$ & 10$^{2}$ & 10 & 1.5$\cdot$10$^{19}$ & \\
		\hline
		& 
	\end{tabular}
\end{table}

\section{Calibration on forward field data}\label{sc:calibration}
In this section we present and discuss the results of the model calibration step, first by showing an example evolution of the optimization process in \ref{sc:optim_evol}. Then, we summarize the obtained model parameters in \ref{sc:obtained_param} and plasma profiles in \ref{sc:calibratio_prof}. Afterwards, we discuss the differences between the standard model and the $\kappa$-model in \ref{sc:model_diff}, and finally conclude this section with a sensitivity analysis in \ref{sc:sensitivity}.

\subsection{Optimization evolution} \label{sc:optim_evol}
\begin{figure}
	\centering
	\subfigure{\includegraphics[scale=0.35]{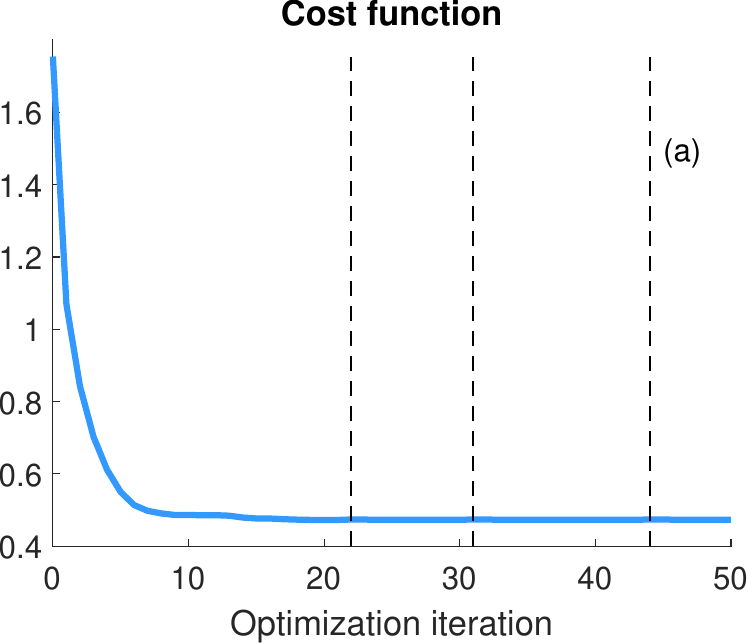}}
	\subfigure{\includegraphics[scale=0.35]{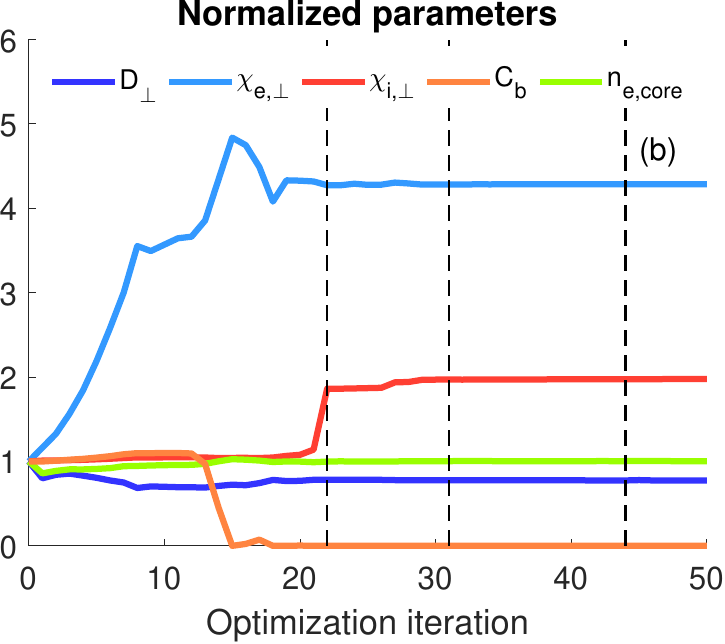}}
	\subfigure{\includegraphics[scale=0.35]{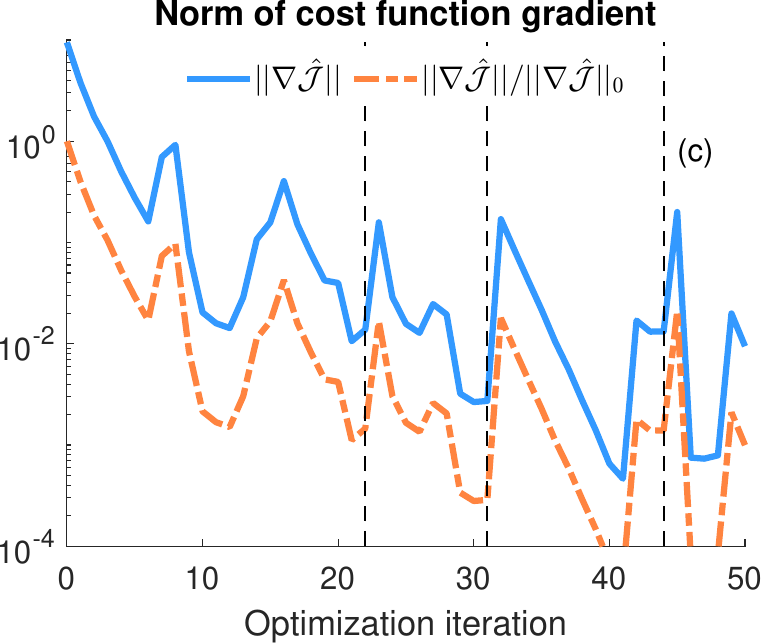}}
	\caption{Evolution of relevant quantities as a function of the optimization iteration for Model 1: (a) the cost function; (b) the calibrated model parameters, normalized with respect to their initial guess $\theta_0$; (c) the cost function gradient norm (blue solid line) and the cost function gradient norm reduction (orange dash-dotted line). The vertical black dashed lines indicate the points at which the optimization was stopped and restarted.} \label{fg:optimresult}
\end{figure}

We show an example evolution of relevant  quantities through the optimization in Figure \ref{fg:optimresult}, taken from the Model 1 case. The vertical dashed black lines in this figure indicate points at which the optimization was stopped and restarted, due to maximum wall clock time on the supercomputer. From the cost function $\hat{\mathcal{J}}$ evolution in Figure \ref{fg:optimresult}a, we notice that most of its reduction (99\% of the total) happens within the first 10 iterations, while in the remaining 40 iterations the change is negligible. We notice a similar behavior also in the evolution of model parameters, shown on Figure \ref{fg:optimresult}b, where all of them have been normalized with respect to their initial guess $\theta_0$. From a practical point of view, one could stop the optimization already at 10 iterations, having obtained most of the reasonably achievable performance improvement, despite some of the parameters display residual changes until 20 iterations. From an optimization point of view however, the optimum has not been reached, as visible from the cost function gradient norm $\vert\vert\nabla\hat{\mathcal{J}}\vert\vert$ in Figure \ref{fg:optimresult}c. Indeed, we see that the desired $10^{-4}$ threshold is not achieved. The reason is that when PETSc/TAO checks for gradient reduction $\vert\vert\nabla\hat{\mathcal{J}}\vert\vert/\vert\vert\nabla\hat{\mathcal{J}}\vert\vert_0$, after stopping and restarting the zero-th iteration becomes the initial iteration of the newly restarted batch. This is a limitation that we should alleviate in the future. If we now re-evaluate the gradient norm reduction based on the `absolute' zero-th iteration, we actually observe that around iteration 40 the desired gradient reduction is achieved.

We also list in Table \ref{tb:cfvalues1} the final values of the cost function $\hat{\mathcal{J}}$ and its difference with respect to a reference value, taken as Model 1. The striking improvement in the performance indicator can be noticed, up to 50\%, when allowing more parametric freedom, i.e. the radially varying diffusion coefficients of Model 2 and 3. In contrast, the $\kappa$-model achieves essentially the same performance as Model 1 in terms of cost function. This will be confirmed by analysis of the plasma quantities in \ref{sc:calibratio_prof}. As previously stated, allowing large parametric freedom on one hand significantly improves the match with experimental data, while on the other it likely incurs in overfitting. Eventually, this happens to be the case for both Model 2 and 3, based on their prediction on the reversed field data. We expect that a formal model comparison through Bayesian evidence will show that their improvement in cost function does not compensate the increase in model complexity \cite{dose,dewolf}. Hence, we do not discuss the outcomes for Model 2 and 3 in this and further sections, and display their results in Appendix \ref{app:a}.

\begin{table}
	\caption{Cost function value for different calibrated models and improvement compared to Model 1}\label{tb:cfvalues1}
	\centering
	\begin{tabular}{c c c c c}
		\hline
		      & Model 1 & Model 2 & Model 3 & $\kappa$-model \\
		\hline
		$\hat{\mathcal{J}}$ & 0.474 & 0.294 & 0.235 & 0.456 \\
		Difference       & 0     & -38\% & -50\% & -4\%\\
		\hline
		& 
	\end{tabular}
\end{table}

\subsection{Obtained parameters}\label{sc:obtained_param}
\begin{figure}
	\centering
	\subfigure{\includegraphics[scale=0.45]{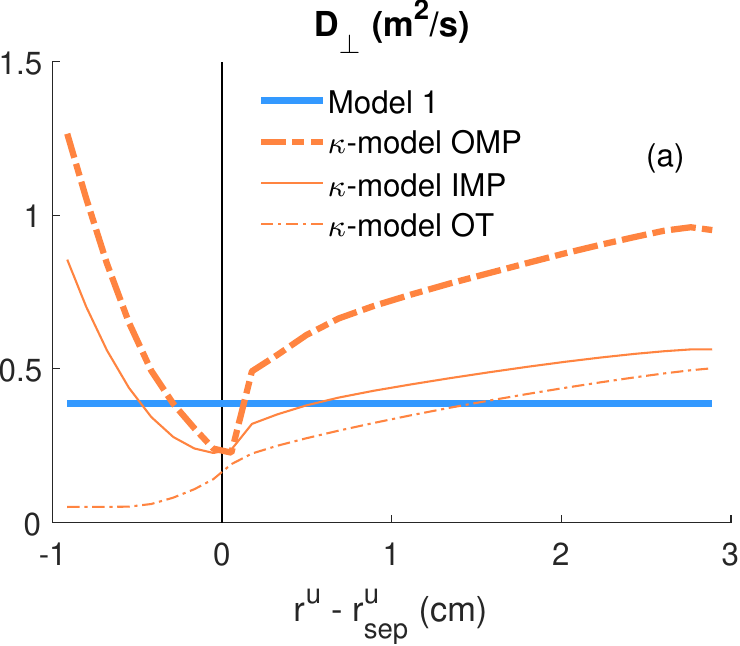}}
	\subfigure{\includegraphics[scale=0.5]{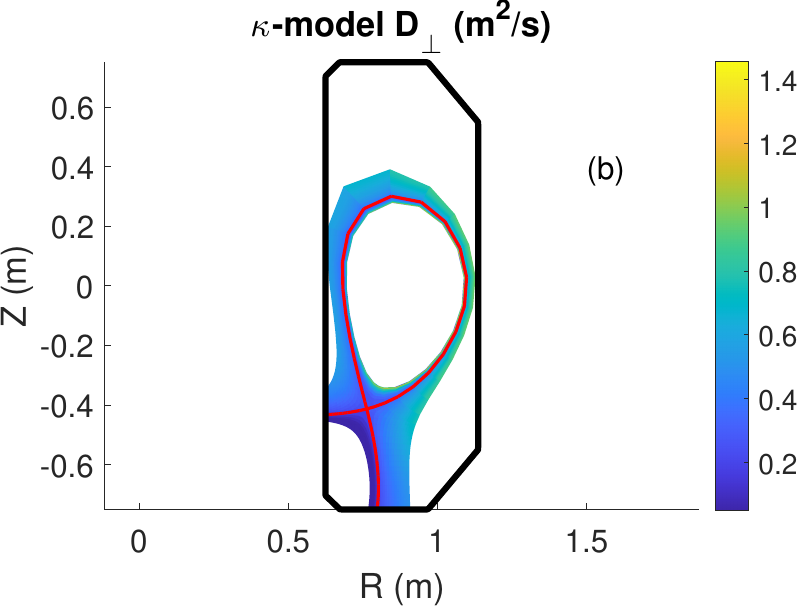}}
	\caption{(a) Profiles of the particle diffusion coefficient $D_\perp$ obtained with Model 1 (blue solid thick line) and the $\kappa$-model (orange) at the OMP (thick dash-dotted line), the IMP (thin solid line), and the OT (thin dash-dotted line). The vertical line indicates the separatrix position. (b) 2D map of the particle diffusion coefficient $D_\perp$ obtained with the $\kappa$-model. The red solid line indicates the separatrix. } \label{fg:dperp}
\end{figure}

Focusing on the results of the calibration for Model 1 and the $\kappa$ model, we list the optimal model parameters in Table \ref{tb:optimalvalues}, and show on Figure \ref{fg:dperp} the radial profiles of $D_\perp$ along with the 2D distribution of this quantity obtained by the $\kappa$-model. For comparison, we also list in Table \ref{tb:optimalvalues} the diffusion coefficient for other SOLPS-ITER modeling efforts of TCV found in literature \cite{wensing21,wangx21,tonello2024,carpita2025,colandrea}, and the $\kappa$-model parameters employed in \cite{kmodel}. Note that in \cite{kmodel} the resistive dissipation term was  multiplied by $C_{\sigma_{\parallel},1}$, and as such we report in Table \ref{tb:optimalvalues} the product $C_{\sigma_{\parallel},1}C_{\sigma_{\parallel},2}$. Concerning Model 1, we find that our optimized values are very close to those found in literature. We also notice that for Model 1 the ballooning exponent $C_b=0$, meaning no such enhancement seems necessary to match the experimental data. This is in line with the other TCV modeling efforts mentioned before, where no ballooning was employed. Interestingly, also the $\kappa$-model shows limited ballooning, as can be seen by comparing the OMP and IMP radial profiles in Figure \ref{fg:dperp}a and the 2D distribution in Figure \ref{fg:dperp}b. This is rather different compared to the more ballooned cases shown in \cite{kmodel,kmodelcompass}, and can be linked to the order of magnitude larger parallel transport constant found in this study $C_{\sigma_{\parallel},1}\sim1$, compared to the $\sim$0.1 value in \cite{kmodel}. In second instance, we notice that the profile obtained by the $\kappa$-model is quite similar to the trend in Figure 2 of \cite{kmodel}. Indeed also in this case $D_\perp$ is suppressed around the separatrix due to $E\times B$ flow shear, and tends to grow in the SOL thanks to the positive interchange source and small resistive dissipation $C_{\sigma_{\parallel},2}=10^{-6}$ (this was $\sim$0.1 in \cite{kmodel}). However, differently from what was found in \cite{kmodel}, the  core profile shows stronger $\kappa$ dissipation and hence decrease in $D_\perp$. This is linked to the order of magnitude different resistive dissipation term in the core $C_{\sigma_{\parallel},2,core}\sim 1$, which was $\sim0.1$ in \cite{kmodel}. The remaining model parameters are instead in line with \cite{kmodel}.

Interestingly, an ancillary optimization of Model 1 with the anomalous velocity $\mathbf{v}_\perp$ as additional model parameter (see Eq. \ref{eq:standard}), provided the same result as shown now for Model 1, i.e. $\mathbf{v}_\perp=0$ at the optimum.

\begin{table}
	\caption{Summary of the optimal parameters value for Model 1 and the $\kappa$-model, and values found in literature} \label{tb:optimalvalues}
	\centering
	\begin{tabular}{c c c c c c c}
		\hline
		& & & Model 1 & & & \\
		\hline
		Parameter $\theta$ & $D_\perp$ (m$^2$/s) & $\chi_{e,\perp}$ (m$^2$/s) & $\chi_{i,\perp}$ (m$^2$/s) & $C_b$ (-) & $n_{e,core}$ (m$^{-3}$) &\\
		Optimal value & 0.39 & 2.14 & 0.99  & 0  & 9.42$\cdot$10$^{18}$ & \\
		Literature & 0.2-0.4 & 0.3-1.0 & 0.4-1.0  & 0  & / & \\
		\hline
		& & & $\kappa$-model & & & \\
		\hline
		Parameter $\theta$ & $C_{\chi_e}$ & $C_{\chi_i}$ & $C_{\sigma_{\parallel},1}$ & $C_{\sigma_{\parallel},2}$ & $C_{\sigma_{\parallel},2,core}$ &  $C_S$  \\
		Optimal value & 5.22 & 2.54 &  1.47  & 10$^{-6}$  & 0.69 & 0.12 \\
		Literature & 2.0 & 2.0 &  0.1  & 0.1  & 0.1 & 10 \\
		\hline
		Parameter $\theta$ &  $C_{\nu}$ & $C_\kappa$ & $C_{ie}$ & $\kappa_{core}$ (eV) & $n_{e,core}$ (m$^{-3}$) & \\
		Optimal value & 0.18 & 1.01 & 4.2$\cdot$10$^{-4}$  & 1.27 & 9.46$\cdot$10$^{18}$ & \\
		Literature & 0.2 & / &  /  & /  & / & \\
		\hline
		& 
	\end{tabular}
\end{table}

\subsection{Plasma profiles}\label{sc:calibratio_prof}
\begin{figure}[h!]
	\centering
	\subfigure{\includegraphics[scale=0.5]{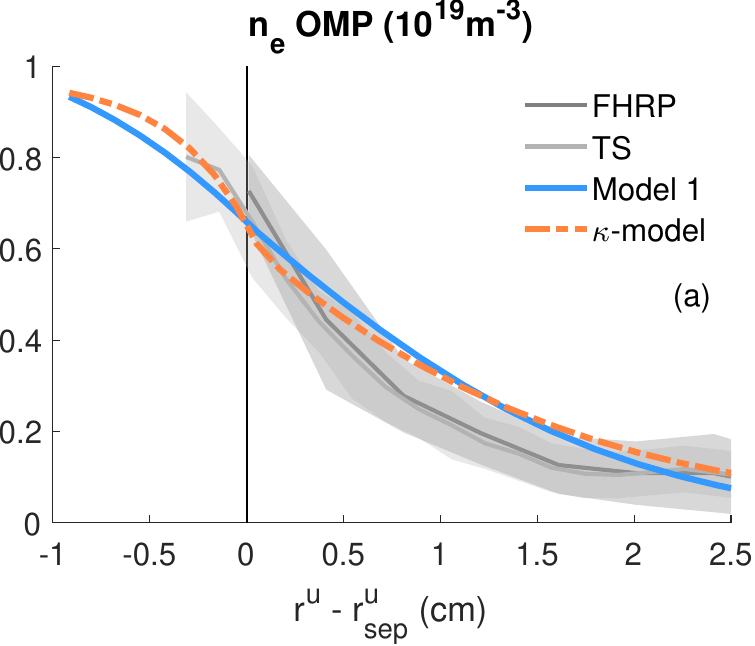}}
	\subfigure{\includegraphics[scale=0.5]{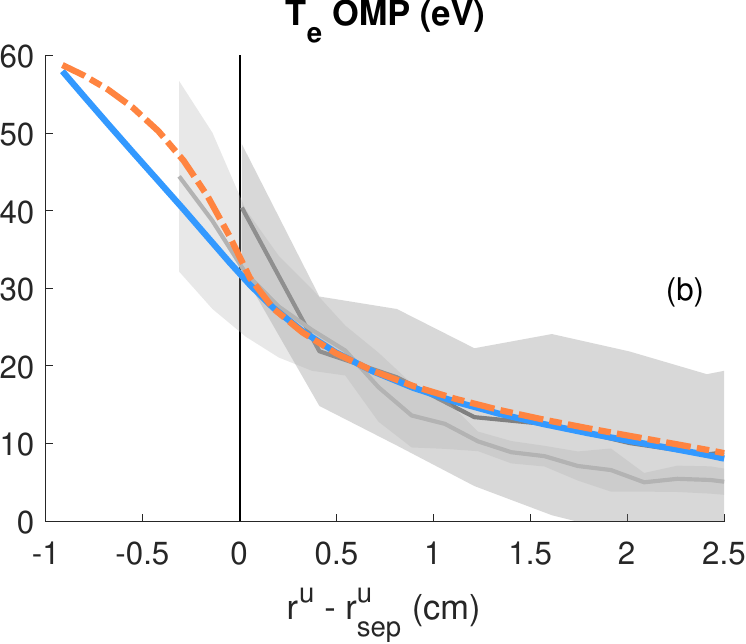}}
	\caption{Upstream profiles of electron density (a) and temperature (b) for the forward field case as a function of the OMP radial distance from the separatrix. Thin solid grey lines indicate FHRP and TS data, with shaded regions indicating experimental uncertainty. Thick solid blue lines indicate Model 1 profiles, and thick dash-dotted orange lines indicate $\kappa$-model profiles. Solid vertical black lines indicate the separatrix location.} \label{fg:ompprof1}
\end{figure}

We now move to the analysis of the obtained plasma profiles, starting with electron density and temperature at the OMP for Model 1 and the $\kappa$-model, shown on Figure \ref{fg:ompprof1}. We see small differences between the two models, given the 4\% discrepancy attained in the cost function, and such differences seem within the experimental uncertainty (shaded grey regions). The $\kappa$-model density profile is slightly steeper than Model 1 around the separatrix, and closer to the experimental one, and is flatter elsewhere. This is in accordance with the obtained $D_\perp$ profile in Figure \ref{fg:dperp}, showing smaller diffusion around the separatrix caused by the $E\times B$ flow shear. Near-SOL profiles of both models closely reproduce experiment, while far-SOL profiles are overestimated. The same differences between models is visible in the core electron temperature profile, while in the SOL they both provide the same outcome, closely following the experiment. What strikes is that, despite the largely different anomalous diffusion coefficients, both models attain the same density and temperature values at the separatrix.

In Table \ref{tb:lambdas1} we also compare the upstream density and temperature decay lengths obtained by the two models in the near-SOL $0\leq r^u-r_{sep}^u \leq 1.5$ cm, and the experimentally estimated values from \cite{x21paper}. The OMP values reflect the differences between model and experiment just discussed, while at the divertor entrance (DE) models and experiment show better agreement.

\begin{table}[h!]
	\caption{Near-SOL decay lengths at the OMP and DE for electron density $\lambda_n$ and temperature $\lambda_{T_e}$ (in cm), and OT parallel heat flux decay length ($\lambda_q$) and spreading factor ($S$) (in mm), as obtained from experimental data (Exp.) \cite{x21paper} and with the different models, for the forward field case} \label{tb:lambdas1}
	\centering
	\begin{tabular}{l c c c}
		\hline
		& Exp. & Model 1 & $\kappa$-model\\
		\hline
		$\lambda_{n,OMP}$& 0.9$\pm$0.2 & 1.4 & 1.5 \\
		$\lambda_{n,DE}$&  0.9$\pm$0.2 & 1.0 &  0.9 \\
		$\lambda_{T_e,OMP}$& 1.0$\pm$0.4 & 1.6 & 1.7 \\
		$\lambda_{T_e,DE}$&  1.0$\pm$0.1 & 1.9 & 2.0 \\
		$\lambda_{q}$&     5.5$\pm$0.2 & 5.4 & 5.1 \\
		$S$ &              1.8$\pm$0.1 & 1.1 & 0.8\\
		\hline
	\end{tabular}
\end{table}

\begin{figure}[h!]
	\centering
	\subfigure{\includegraphics[scale=0.33]{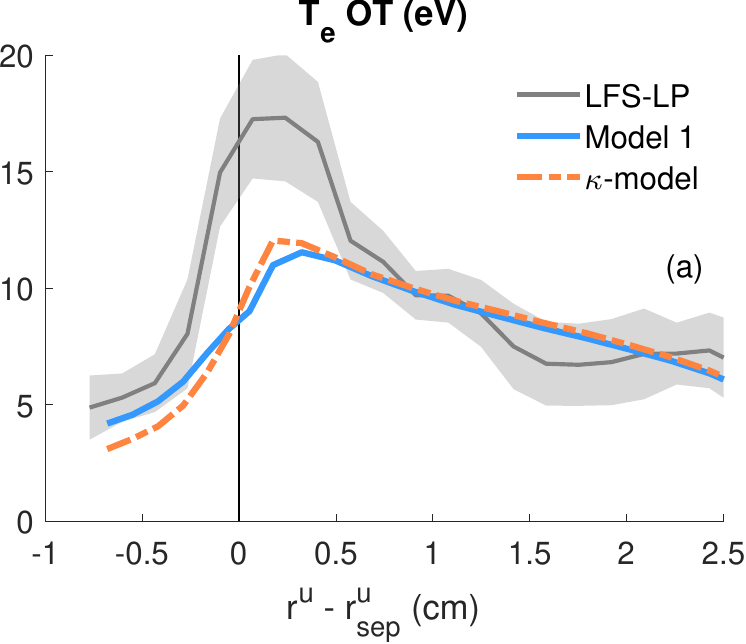}}
	\subfigure{\includegraphics[scale=0.33]{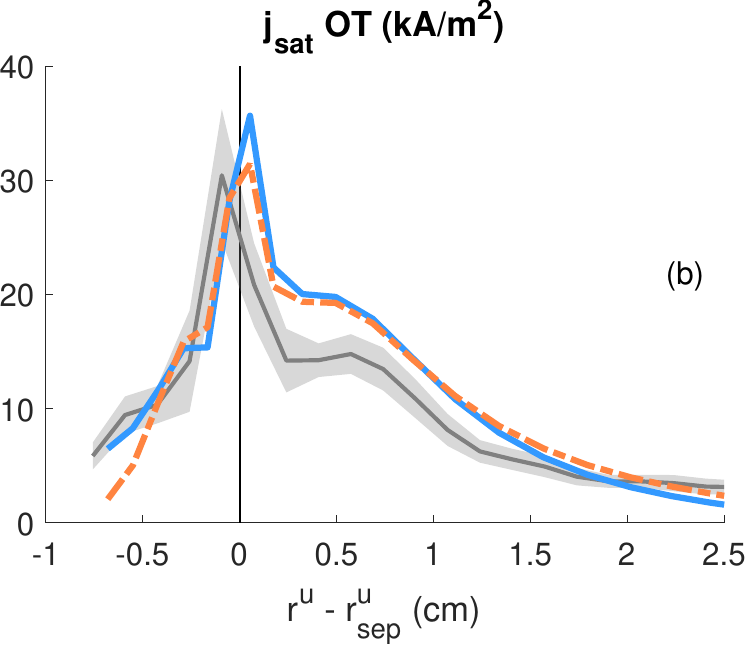}}
	\subfigure{\includegraphics[scale=0.33]{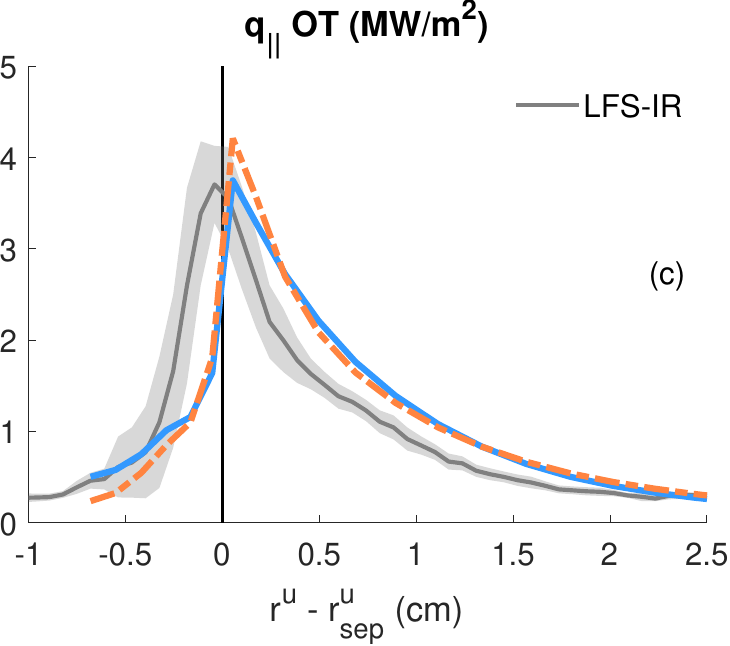}}
	\subfigure{\includegraphics[scale=0.33]{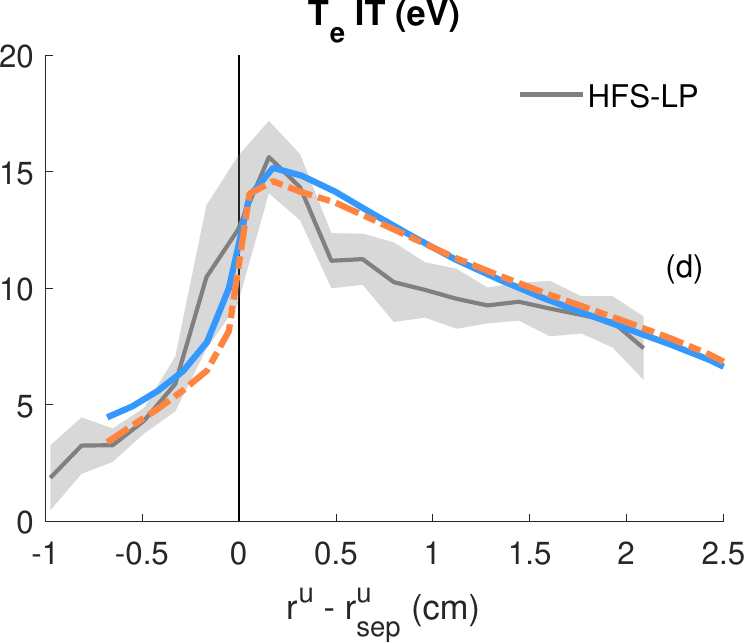}}
	\subfigure{\includegraphics[scale=0.33]{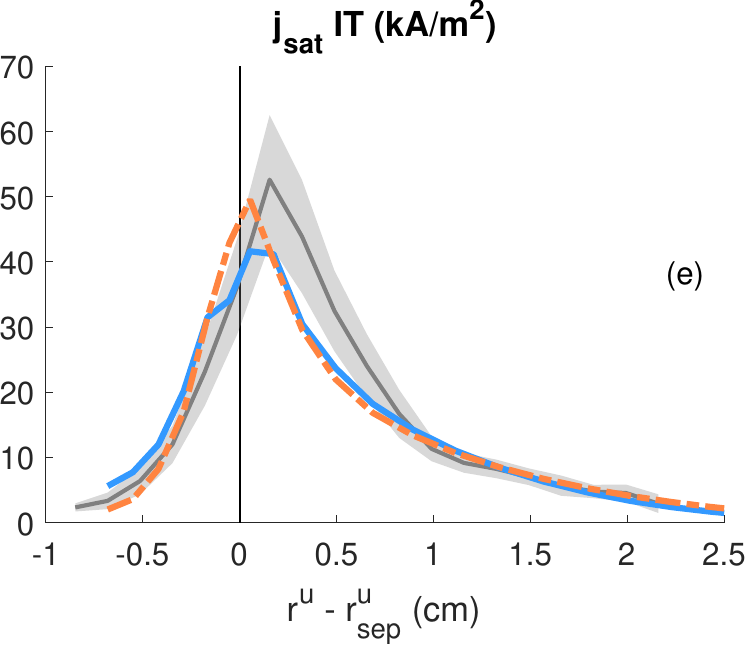}}
	\caption{Forward field case profiles of: OT electron temperature (a), saturation current (b), and parallel heat flux (c); IT electron temperature (d), and saturation current (e). All quantities are mapped to the OMP radial distance from the separatrix. Thin solid grey lines indicate LP and IR data, with shaded regions indicating experimental uncertainty. Thick solid blue lines indicate Model 1 profiles, and thick dash-dotted orange lines indicate $\kappa$-model profiles. Solid vertical black lines indicate the separatrix location} \label{fg:targets1}
\end{figure}

Profiles at the OT and inner target (IT) are displayed on Figure \ref{fg:targets1}. Some differences between Model 1 and the $\kappa$-model are visible, similar to those found for the upstream profiles. In particular, both models underestimate the OT electron temperature $T_e$ around the separatrix, with the peak $\sim $5 eV lower than the experimental data, as visible in Figure \ref{fg:targets1}a. The peak OT saturation current $j_{sat}$ is instead slightly overestimated and shifted towards the SOL in both models, compared to the Langmuir Probe data, see Figure \ref{fg:targets1}b. The profile shape is however well captured, also the bump in the SOL around $r^u-r_{sep}^u\sim$0.6 cm which is related to $E\times B$ flows. Similarly, the OT parallel heat flux $q_\parallel$ in Figure \ref{fg:targets1}c is captured in the peak magnitude by both models, the overall profiles being again slightly shifted toward the SOL compared to the experiment. When comparing the heat flux decay length $\lambda_q$ obtained by fitting using MATLAB \cite{matlab} the same Eich-like profile \cite{eich}  
\begin{equation}\label{eq:eich}
	q_\parallel(r)= \frac{q_0}{2} \exp \left[ \left(\frac{S}{2\lambda_{q}}\right)^2 -\frac{r-r_0}{\lambda_{q}}  \right] \erfc \left( \frac{S}{2\lambda_{q}} -\frac{r-r_0}{S}   \right) + q_{BG}
\end{equation}
as done in \cite{x21paper}, the models very well agree with the experiment. The spreading factor $S$ is instead much smaller for both models, due to the sharp heat flux decrease in the PFR.

At the IT the agreement between models and experiment is even better than at the OT. Both very well follow the $T_e$ peak and profile, see Figure \ref{fg:targets1}d, with $j_{sat}$ being better captured by the $\kappa$-model, despite a small shift towards the PFR, as visible in Figure \ref{fg:targets1}e.

\subsubsection{A consistent picture of upstream and downstream plasma quantities}
What strikes is that, despite the largely different anomalous diffusion coefficients, both models attain the same density and temperature values at the separatrix, see Figure \ref{fg:ompprof1}. We underline that this is a consequence of the optimization, and not of a feedback control of the separatrix density. Indeed, the optimization approach finds values providing the highest upstream-downstream consistency, as seen discussing the target profiles. This is also in accordance with a 2-point model interpretation, in which target quantities depend on power two and three of the upstream density \cite{stangeby}.

\begin{figure}[ht]
	\centering
	\subfigure{\includegraphics[scale=0.33]{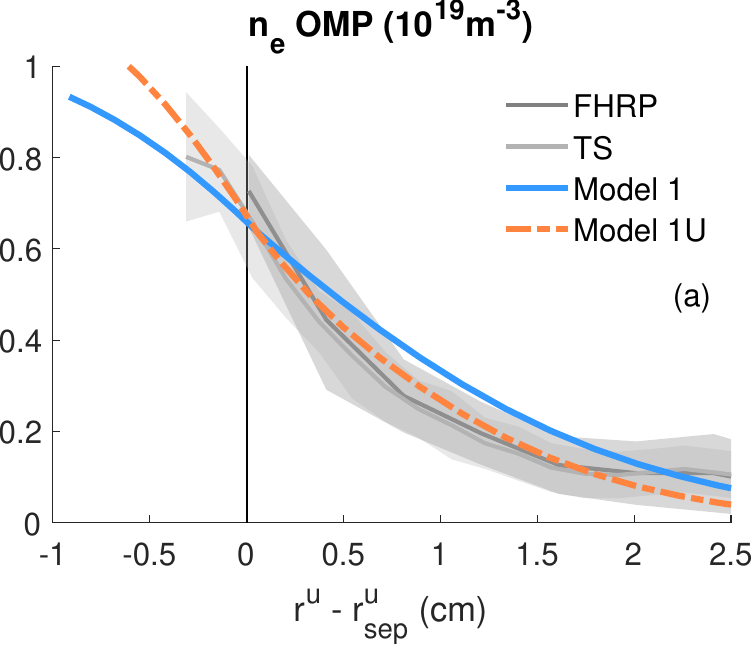}}
	\subfigure{\includegraphics[scale=0.33]{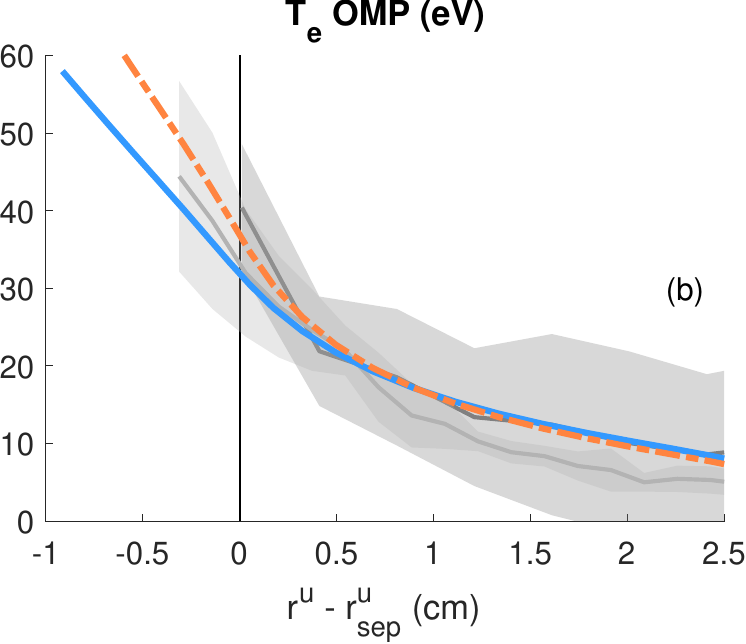}}
	\subfigure{\includegraphics[scale=0.33]{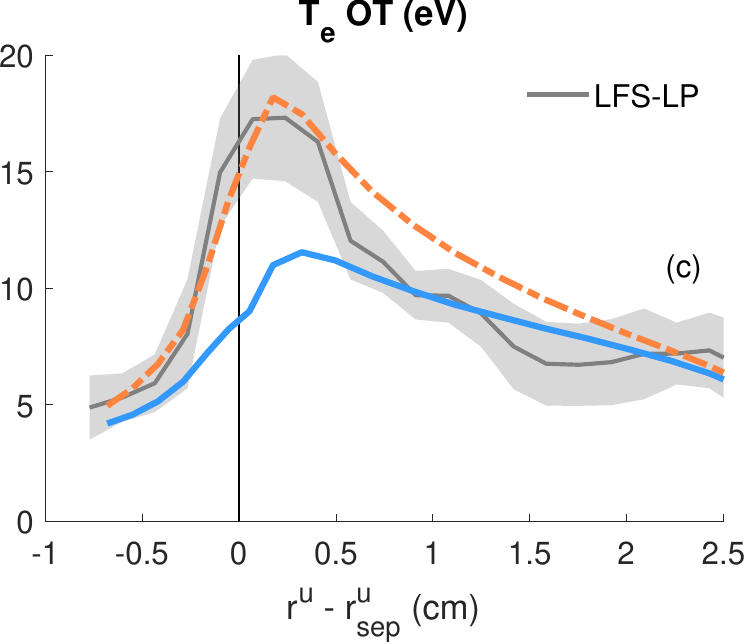}}
	\subfigure{\includegraphics[scale=0.33]{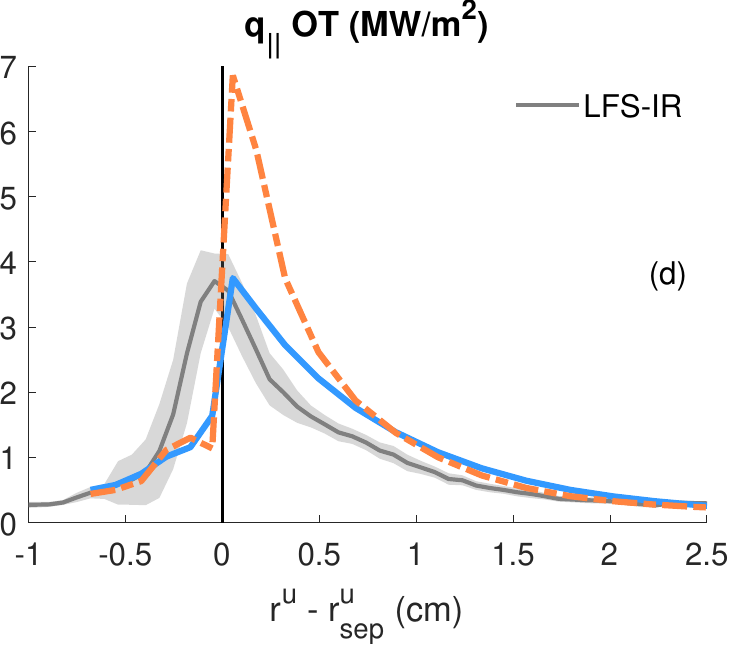}}
	\subfigure{\includegraphics[scale=0.33]{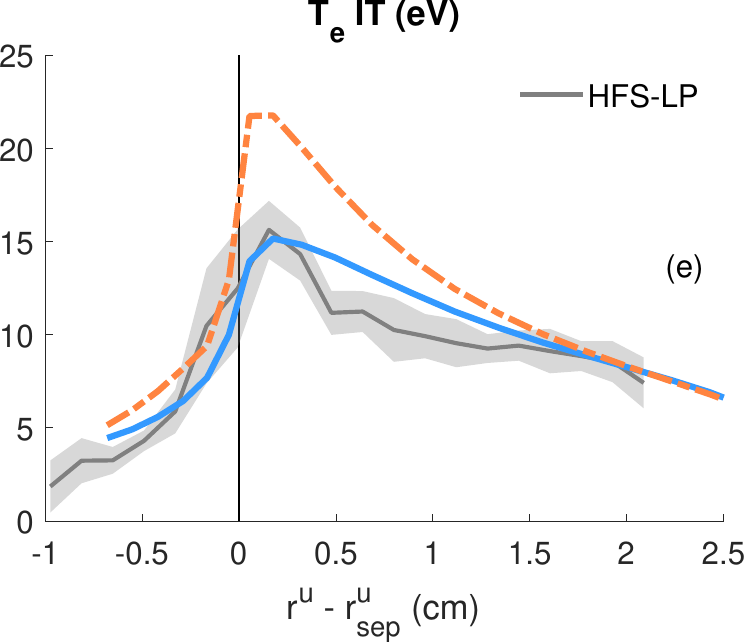}}
	\subfigure{\includegraphics[scale=0.33]{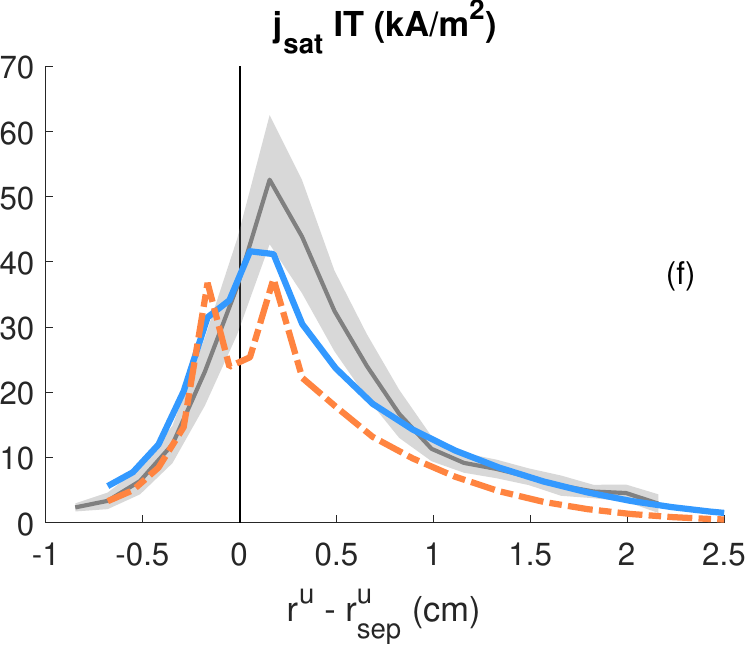}}
	\caption{Forward field case profiles as a function of the OMP radial distance from the separatrix of: uspstream electron density (a) and temperature (b); OT electron temperature (c) and parallel heat flux (d); IT electron temperature (e) and saturation current (f). Thin solid grey lines indicate experimental data, with shaded regions indicating experimental uncertainty. Thick solid blue lines indicate Model 1 and thick dash-dotted orange lines indicate Model 1U. Solid vertical black lines indicate the separatrix location.} \label{fg:OMPonly}
\end{figure}

This is in stark contrast with the current practice of manual tuning. In such case, modelers typically aim for the best match of upstream data, targeting for example the mean of the FHRP and TS data at the separatrix. Any discrepancy with target data would then be attributed to model deficiencies. While such deficiencies are certainly present and their role should be quantified in future, we show now that the best upstream match does not necessarily lead to the most consistent picture between upstream and downstream quantities. To do so, we performed the same calibration as Model 1, this time including only upstream data in the cost function, namely density and temperature from TS and FHRP. We will refer to this setup as Model 1U. If we calculate the cost function using only upstream data also for Model 1 and compare it to Model 1U, the latter is a factor $\sim$2 smaller. This means Model 1U provides a much better agreement with upstream density and temperature, as visible in Figure \ref{fg:OMPonly}a and b. Also the near-SOL decay lengths more closely resemble the experimental ones, as can be seen from Table \ref{tb:lambdasOMP}. However, the cost function calculated including also target data is $\sim$2 higher for Model 1U compared to Model 1. Indeed, target quantities predicted by Model 1U display larger discrepancies compared to experiment, see Figures \ref{fg:OMPonly}c-d: the OT $T_e$ of Model 1U actually improves compared to Model 1, and well captures the experiment, but the parallel heat flux is almost twice as large at the peak. At the IT the electron temperature is also overestimated by Model 1U, with the $j_{sat}$ profile somewhat worse compared to Model 1 and experiment. The OT $j_{sat}$ profile is instead relatively similar in both models and is not shown. The calibrated parameters in Model 1U are also a factor  $\sim$2 different compared to those of Model 1 (see Table \ref{tb:optimalvalues}): $D_\perp=0.22$, $\chi_{e,\perp}=1.3$, $\chi_{i,\perp}=10$, $C_b=3\cdot10^{-7}$ and  $n_{e,core}=1.11\cdot10^{19}$. Hence, we can conclude that when including target data in the estimation procedure we obtain the best overall fit, and a `bad fit' for one diagnostic (OT $T_e$), while not including target data provides a `bad fit' for two diagnostics (OT $q_\parallel$ and IT $T_e$). We think this an important result which should be accounted for in current and future parameter estimation frameworks.

\begin{table}[h!]
	\caption{Near-SOL decay lengths at the OMP and DE for electron density $\lambda_n$ and temperature $\lambda_{T_e}$ (in cm), as obtained from experimental data (Exp.) \cite{x21paper}, and for Model 1 and Model 1U, for the forward field case} \label{tb:lambdasOMP}
	\centering
	\begin{tabular}{l c c c}
		\hline
		& Exp. & Model 1 & Model 1U\\
		\hline
		$\lambda_{n,OMP}$  & 0.9$\pm$0.2 & 1.4 & 1.1 \\
		$\lambda_{n,DE}$   & 0.9$\pm$0.2 & 1.0 & 0.8 \\
		$\lambda_{T_e,OMP}$& 1.0$\pm$0.4 & 1.6 & 1.3 \\
		$\lambda_{T_e,DE}$ & 1.0$\pm$0.1 & 1.9 & 1.5 \\
		\hline
	\end{tabular}
\end{table}

\subsection{Discussion of model differences}\label{sc:model_diff}
\begin{figure}[h!]
	\centering
	\subfigure{\includegraphics[scale=0.33]{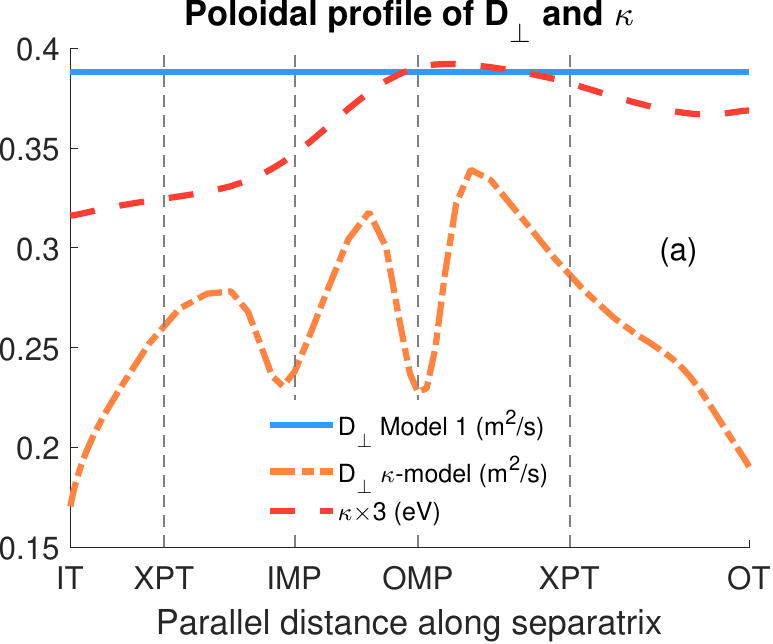}}	\subfigure{\includegraphics[scale=0.33]{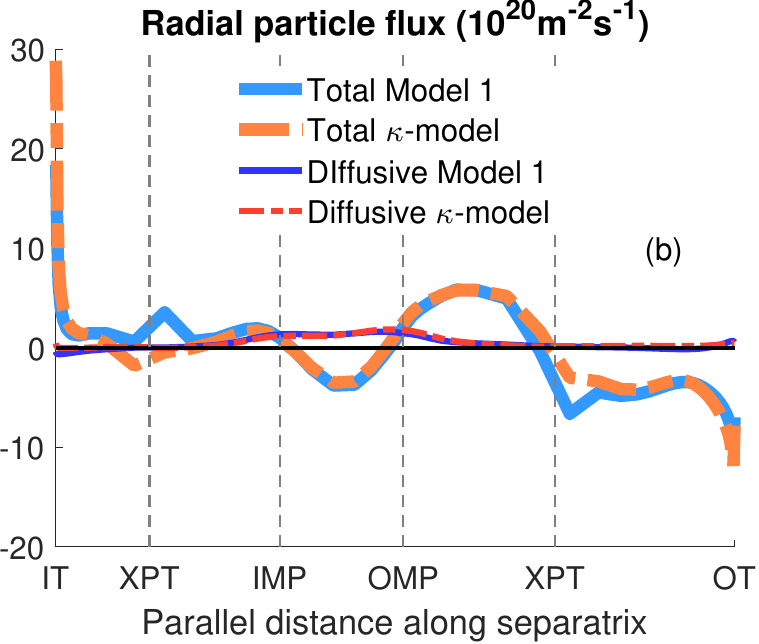}}	\subfigure{\includegraphics[scale=0.33]{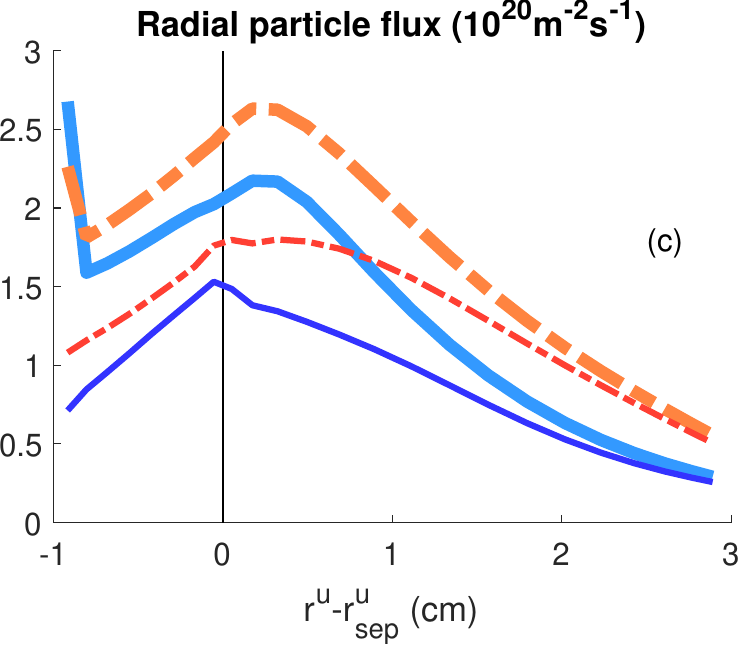}}
	\caption{Poloidal profiles in the first flux tube outside the separatrix of: (a) $D_\perp$ for Model 1 (solid blue line), $D_\perp$ for the $\kappa$-model (dash-dotted orange line) and $\kappa$ (dashed red line); (b) the total radial particle flux for Model 1 (thick solid light blue line) and the $\kappa$-model (thick dash-dotted orange line), and the diffusive radial particle flux for Model 1 (thin solid dark blue line) and the $\kappa$-model (thin dash-dotted red line). In (c) the same fluxes are plotted at the OMP as a function of the distance from the separatrix.} \label{fg:poloidal1}
\end{figure}

Given the results we showed above, one may ask themselves how Model 1 and the $\kappa$-model obtain almost identical plasma profiles despite different anomalous diffusion coefficients. From a general standpoint, we think most of this should be attributed to both models achieving the same upstream density and temperature profiles, which in a 2-point modeling setting provide a direct link to target quantities. In addition, we argue that the TCV case at hand is dominated by parallel neoclassical transport and mean-field drifts, with the perpendicular anomalous transport being less important. This is supported by ballooning being absent in case of Model 1, or negligible in case of $\kappa$-model. We can appreciate this feature from Figure \ref{fg:poloidal1}a, which shows the poloidal profile of $D_\perp$ in the first flux tube outside the separatrix for Model 1 and the $\kappa$-model, as well as the profile of $\kappa$ itself. Variations in $D_\perp$ for the $\kappa$-model are in fact present, but these are not orders of magnitude different and are related to $E\times B$ shear. Indeed, $\kappa$ itself is almost constant along the flux tube, indicating a uniform turbulence strength. As stated previously, poloidally constant diffusion coefficients are consistent with other TCV modeling efforts, where no ballooning enhancement was employed. Furthermore, being TCV a small machine and thus featuring stronger gradients, and the X21 case having specifically a small magnetic field, drift flows are the dominant cross-field transport mechanism. This can be appreciated in Figure \ref{fg:poloidal1}b and c, where the total radial particle flux is compared to the purely diffusive (anomalous) radial flux, poloidally in the first flux tube outside the separatrix (b) and radially at the OMP (c). We can clearly see that the diffusive flux is of the same order as the drift one only around the OMP/IMP, and is dominant in the far SOL. As such, it is only important in determining the upstream separatrix density and temperature. At higher densities and collisionality, turbulent transport is expected to play a more significant role, and one should expect more pronounced differences between Model 1 and the $\kappa$-model. This is indeed what we will show in Section \ref{sc:densscan}. It is still to be assessed whether this behavior is only valid for TCV or other tokamaks as well. Nevertheless, some model differences are present, which may become more relevant in other studies and experimental conditions, and we will thus discuss them in the following.

\begin{figure}[h!]
	\centering
	\subfigure{\includegraphics[scale=0.5]{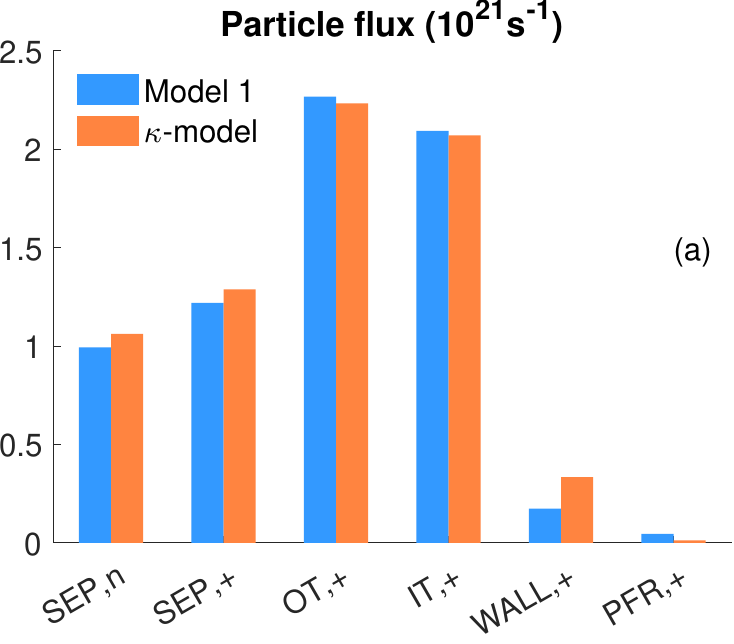}}
	\subfigure{\includegraphics[scale=0.5]{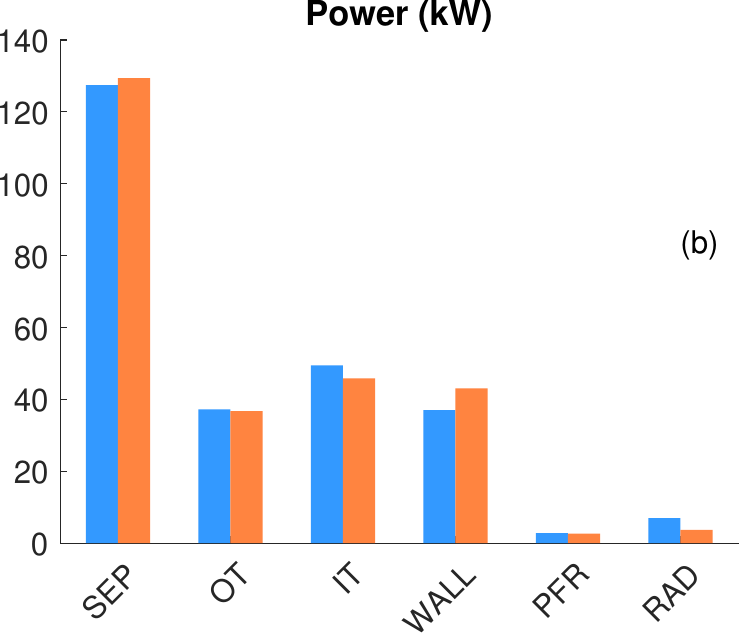}}
	\caption{Particle (a) and energy (b) balances for Model 1 (blue) and the $\kappa$-model (orange). The ``n" indicates neutral atom flux while ``+" indicates the ions flux.} \label{fg:balance1}
\end{figure}

We first analyze the particle and power balances in terms of integral fluxes to or through specific boundaries of the domain, shown in Figure \ref{fg:balance1}. While target ion particle fluxes show small differences, as a consequence of the very similar saturation current attained by both models, the particle outflux from the separatrix and that to the main chamber wall show more noticeable differences, see Figure \ref{fg:balance1}a. The $\kappa$-model has in fact a larger D$^+$ ions flux compared to Model 1, which is exactly compensated by an equally larger flux of D atoms towards the core, so that the net flux through the separatrix is the same for both models. This larger flux of neutral atoms is caused by the noticeably larger flux of ions directed towards and recycling at the main chamber wall, which is almost a factor two larger for $\kappa$-model. Indeed one should expect such behavior given the much larger $D_\perp$ in the far-SOL attained by the $\kappa$-model, see Figure \ref{fg:dperp}a, and related larger radial particle flux shown in Figure \ref{fg:poloidal1}c. This different behavior affects the neutrals density and ionization profiles, and eventually ion temperature, as shown below. This larger particle flux to the main chamber wall is an important feature of the $\kappa$-model, and is observed for a similar study carried out for the TCV-X23 case \cite{unstrpet25}. A thorough comparison with experimental data at the main chamber wall is needed in the future to verify this feature, as it significantly affects fluxes of eroded impurities and core contamination. At last, also the ions flux through the PFR boundary is slightly different among the two models, though very small compared to other boundaries. Also here the cause lies in $D_\perp$: the $\kappa$-model has a much lower diffusivity in the PFR region compared to Model 1, see again Figure \ref{fg:dperp}a, so that the particle outflux from this boundary is smaller.

Concerning the power balance, an almost identical flux through the separatrix and to the OT is visible for both models, see Figure \ref{fg:balance1}b. The $\kappa$-model shows again larger flux to the main chamber wall, caused in part by the larger particle flux discussed before, and also by the larger heat diffusivities for both electrons and ions, which follow the same trend as the $D_\perp$. This loss of power towards the main chamber wall is compensated in the $\kappa$-model by both a smaller heat flux to the inner target and radiated power.

\begin{figure}[h!]
	\centering
	\subfigure{\includegraphics[scale=0.33]{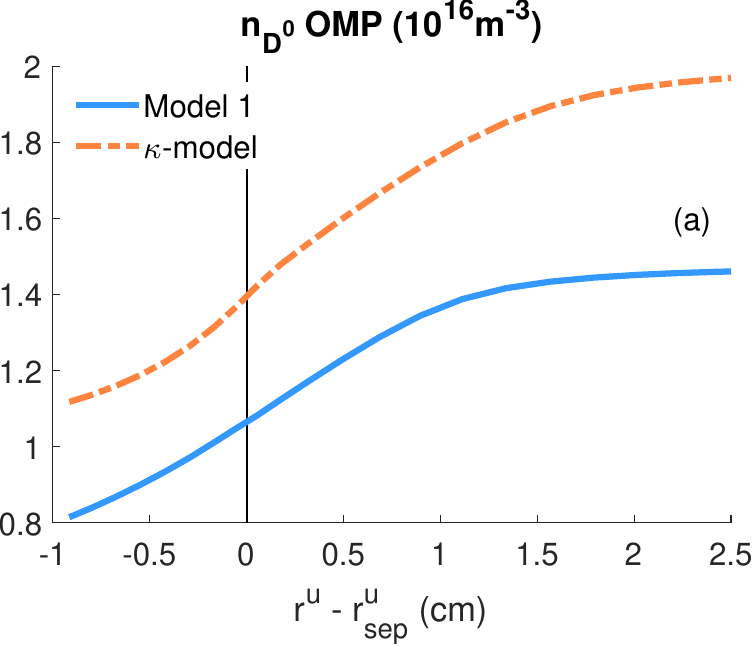}}
	\subfigure{\includegraphics[scale=0.33]{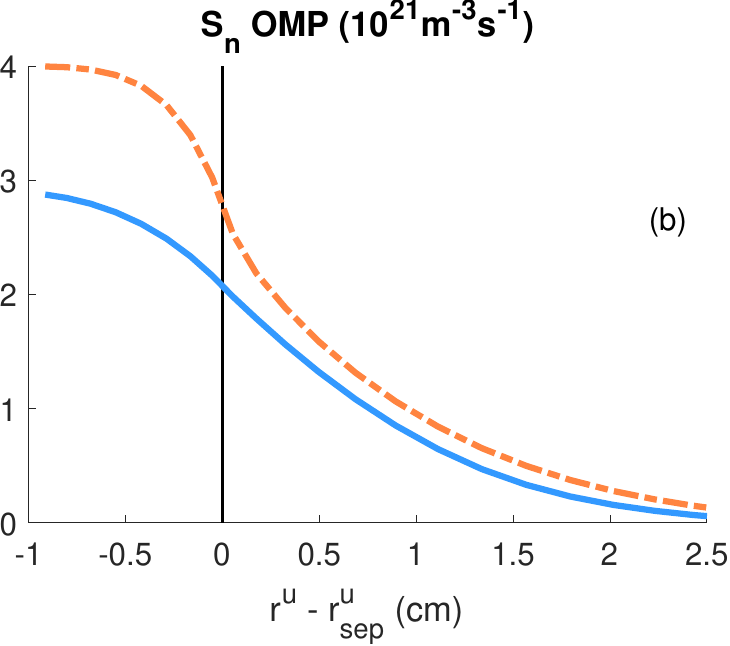}}
	\subfigure{\includegraphics[scale=0.33]{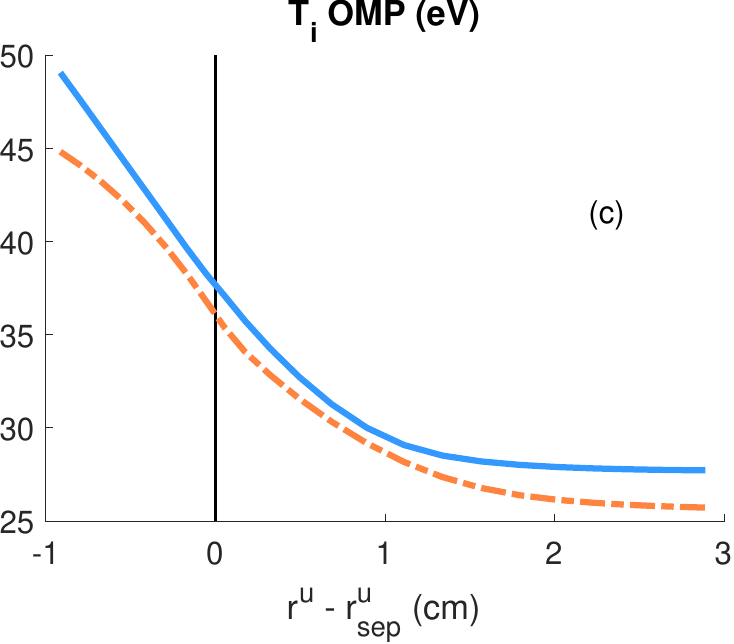}}
	\caption{Upstream profiles of D atoms density (a), particle source (b), and ion temperature (c) for the forward field case as a function of the OMP radial distance from the separatrix. Thick solid blue lines indicate Model 1 profiles, and thick dash-dotted orange lines indicate $\kappa$-model profiles. Solid vertical black lines indicate the separatrix location.} \label{fg:ompprof2}
\end{figure}

As discussed previously, the much larger ion particle flux towards the main chamber wall of the $\kappa$-model, causes a similar increase in atomic neutrals being recycled towards the main plasma. This effect can be appreciated in Figure \ref{fg:ompprof2}a, showing that indeed the D atoms density at the OMP is 50\% higher for the $\kappa$-model. This leads to increased plasma-neutrals interactions, see for example the particle source $S_n$ in Figure \ref{fg:ompprof2}b. On one side, this slightly decreases the ion temperature $T_i$, as displayed in Figure \ref{fg:ompprof2}c\footnote{We do not use a separate neutrals temperature equation in our AFN model, as this made simulations even more unstable. The lower $T_i$ can therefore be interpreted both as a consequence of temperature averaging with denser cold atomic neutrals and of increased charge exchange processes with such cold neutrals.}. As a consequence, the IT $T_i$ is $\sim$1.5 eV lower for the $\kappa$-model and of the same order of the electron one, which can explain the slightly smaller integrated heat flux to the IT seen in Figure \ref{fg:balance1}b. At the OT this difference is only $\sim$0.5 eV.

\subsection{Sensitivity assessment}\label{sc:sensitivity}
As mentioned in Section \ref{sc:setup}, we have so far employed a fluid neutral approximation, and neglected C impurities. When employing the full kinetic neutral model of Eirene, including molecules, neutral-neutral collisions, and sputtered C impurities, the results are only slightly affected for the $\kappa$-model, and show a more significant disagreement only for the saturation current in Model 1, see the results shown on Figure \ref{fg:kinetic} in Appendix \ref{app:b}. We also assess the sensitivity of our solution on the grid resolution and the anomalous conductivity $\sigma_{\perp}$, as shown on Figure \ref{fg:siggrid} in Appenidix \ref{app:c}. One can see that upstream profiles are largely unaffected, and $T_e$ at the targets is somewhat affected only by a change in $\sigma_{\perp}$ case. The largest difference is instead visible for the $j_{sat}$ peak magnitude at both targets, while the profile shape is only partially modified. With this we can conclude that, at least for the plasma scenario studied in this paper, the model parameters obtained with our simplifying assumptions do not change drastically when such assumptions are lifted.

To conclude this section, we assess the sensitivity of the obtained solution on the model parameters through the Hessian matrix, reported in Table \ref{tb:hess1} for Model 1 and Table \ref{tb:hess2} for the $\kappa$-model. These Hessian matrices have been obtained through tangent-over-tangent AD \cite{griewank} at the optimum $\hat\theta$, and normalized with respect to $\hat{\theta}$ and $\hat{\mathcal{J}}(\hat{\theta})$ as:
\begin{equation}
	\left.\begin{pmatrix}
		\frac{\partial^2 \hat{\mathcal{J}}}{\partial \theta_1^2} \frac{\theta_1^2}{\hat{\mathcal{J}}} & \frac{\partial^2 \hat{\mathcal{J}}}{\partial \theta_1 \partial\theta_2} \frac{\theta_1\theta_2}{\hat{\mathcal{J}}} & \cdots & \cdots & \frac{\partial^2 \hat{\mathcal{J}}}{\partial \theta_1 \partial\theta_n} \frac{\theta_1\theta_n}{\hat{\mathcal{J}}} \\
		
		& \frac{\partial^2 \hat{\mathcal{J}}}{\partial \theta_2^2} \frac{\theta_2^2}{\hat{\mathcal{J}}} & \frac{\partial^2 \hat{\mathcal{J}}}{\partial \theta_2 \partial\theta_3} \frac{\theta_2\theta_3}{\hat{\mathcal{J}}} & \cdots & \frac{\partial^2 \hat{\mathcal{J}}}{\partial \theta_2 \partial\theta_n} \frac{\theta_2\theta_n}{\hat{\mathcal{J}}} \\
		
		& & & \ddots & \vdots\\
		
		&  &  & & \frac{\partial^2 \hat{\mathcal{J}}}{\partial \theta_n^2} \frac{\theta_n^2}{\hat{\mathcal{J}}} \\
	\end{pmatrix}\right|_{\hat{\theta}}.
\end{equation}

We can observe that both models are very sensitive to the core density BC $n_{e,core}$, for which the curvature is largest among all parameters. Indeed, one expects such behavior because, as already mentioned, target quantities vary with power two and three of the upstream density in a 2-point model setting. For the same reason but to a lesser extent, Model 1 is sensitive also to the anomalous particle diffusion coefficient $D_\perp$, which determines the upstream density profile. In a similar way the $\kappa$-model shows sensitivity also to the core $\kappa$ dissipation coefficient $C_{\sigma_{\parallel},2,core}$, see Table \ref{tb:hess2}. This may indicate that such parameter is not more universal than the ad-hoc diffusion coefficients themselves. We can expect that, the solution being quite sensitive to these parameters, they are the \textit{Achilles heel} of the model when predictions are concerned, see Section \ref{sc:prediction}.

On the other hand, it appears that the anomalous ion heat diffusivity has a smaller influence on the results in both models. Likely, this is due to the absence of explicit ion temperature information in the cost function, while it implicitly contributes through the parallel heat flux $q_\parallel$ at the OT, and the soundspeed in the saturation current $j_{sat}$ at both targets. 

The sensitivity with respect to the anomalous electron heat diffusivity is instead situated between that of the particles and ion heat. In this case the electron temperature does enter the cost function explicitly, which explains the difference compared to the ion heat diffusivity.

Finally, for the $\kappa$-model the two parallel dissipation constants $C_{\sigma_{\parallel},1}$ and $C_{\sigma_{\parallel},2}$, as well as $C_\nu$ and $C_{ie}$ have negligible influence, wile the $\kappa$ core BC $\kappa_{core}$ and diffusivity coefficient $C_{\kappa}$ have a comparable effect to that of the electron heat diffusivity coefficient $C_{\chi_{e}}$.

\begin{table}
	\caption{Normalized Hessian for Model 1} \label{tb:hess1}
	\centering
	\begin{tabular}{c | c c c c c}
		\hline
		$\theta$ & $D_\perp$ & $\chi_{e,\perp}$ & $\chi_{i,\perp}$ & $C_b$ & $n_{e,core}$ \\
		\hline
		$D_\perp$        & 15 & 0.7 & -0.5   & 0 & 29 \\
		$\chi_{e,\perp}$ &    & 0.8 & -0.005 & 0 & 1.3 \\
		$\chi_{i,\perp}$ &    &     &  0.02  & 0 & 1.1 \\
		$C_b$            &    &     &        & 0 & 0 \\
		$n_{e,core}$     &    &     &        &   & 63 \\
		\hline
	\end{tabular}
\end{table}

\begin{table}
	\caption{Normalized Hessian for the $\kappa$-model} \label{tb:hess2}
	\centering
	\begin{tabular}{c | ccccccccccc}
		\hline
		$\theta$ & $C_{\chi_e}$ & $C_{\chi_i}$ & $C_{\sigma_{\parallel},1}$ & $C_{\sigma_{\parallel},2}$ & $C_{\sigma_{\parallel},2,core}$ &  $C_S$ & $C_{\nu}$ & $C_\kappa$ & $C_{ie}$ & $\kappa_{core}$ & $n_{e,core}$ \\
		\hline
		$C_{\chi_e}$                   &8.1& 0.5  & -0.8& -0.05 & -12 & -3.1& 0.04  & 9.5 & 0.4   &  7.9 & 33 \\
		$C_{\chi_i}$                   & ~ & 0.04 &-0.05&-0.003 &-0.7 & -0.2& 0.003 & 0.5 & 0.03  &  0.4 & 1.7 \\
		$C_{\sigma_{\parallel},1}$     & ~ & ~    & 0.1 & 0.005 & 1.3 & 0.4 &-0.004 &-1.1 & -0.05 & -0.9 & -3.6 \\
		$C_{\sigma_{\parallel},2}$     & ~ & ~    & ~   & 0.0003&0.07 & 0.02&-0.003 &-0.06& -0.003&-0.05 & -0.2 \\
		$C_{\sigma_{\parallel},2,core}$& ~ & ~    & ~   & ~     & 20  & 5.3& -0.06  & -16 & -0.7  & -14  & -56 \\
		$C_S$                          & ~ & ~    & ~   & ~     & ~   & 1.4& -0.01  &-4.3 & -0.2  & -3.6 & -15 \\
		$C_{\nu}$                      & ~ & ~    & ~   & ~     & ~   &   ~  & -1e-4  &0.04 & 0.003 & 0.04 & 0.2 \\
		$C_\kappa$                     & ~ & ~    & ~   & ~     & ~   &   ~  &  ~     & 14  & 0.5   & 11   & 46 \\
		$C_{ie}$                       & ~ & ~    & ~   & ~     & ~   &   ~  &  ~     & ~   & 0.03  & 0.5  & 1.9 \\
		$\kappa_{core}$                & ~ & ~    & ~   & ~     & ~   &   ~  &  ~     & ~   & ~     & 9.6  & 39 \\
		$n_{e,core}$                   & ~ & ~    & ~   & ~     & ~   &   ~  &  ~     & ~   & ~     & ~ & 161 \\
		\hline
	\end{tabular}
\end{table}

\section{Model predictions}\label{sc:prediction}
In this section we assess the predicative capabilities of the calibrated models. We start in \ref{sc:revfield} by modeling the same experimental scenario in reversed field configuration. As will be shown, this configuration exhibits similar upstream plasma profiles compared to the forward field case, which results in small model differences. Therefore, we perform a density scan in \ref{sc:densscan}, highlighting model differences that are not apparent close to the calibration point.

\subsection{Prediction on reversed field data}\label{sc:revfield}
\begin{figure}[ht]
	\centering
	\subfigure{\includegraphics[scale=0.33]{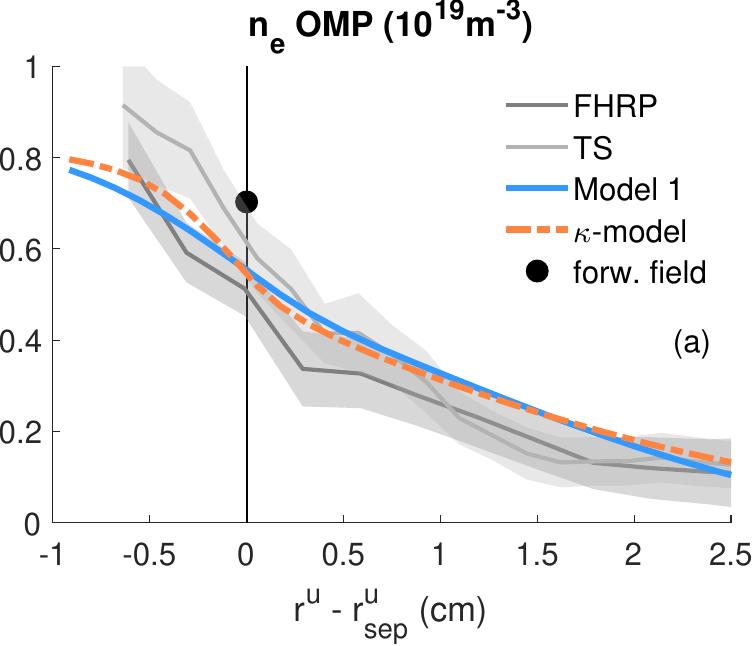}}
	\subfigure{\includegraphics[scale=0.33]{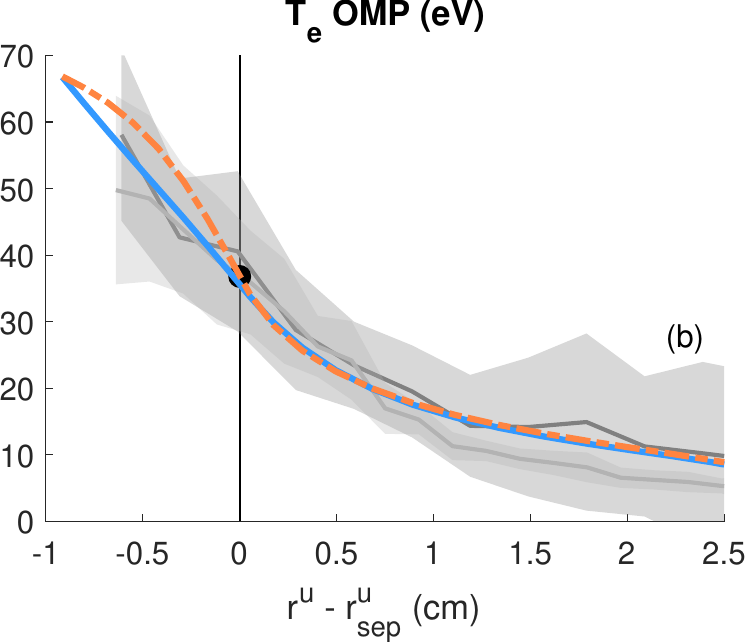}}
	\subfigure{\includegraphics[scale=0.33]{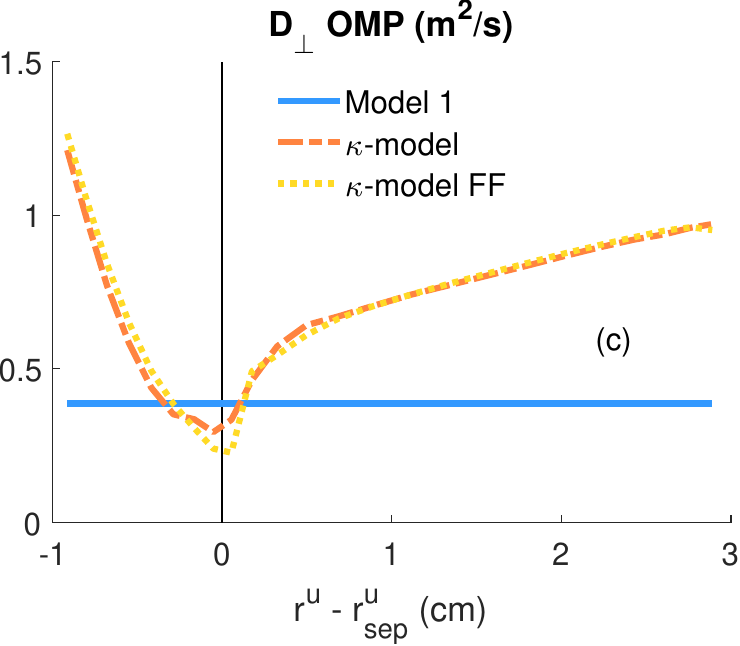}}
	\caption{ Upstream profiles of electron density (a), electron temperature (b), and anomalous particle diffusion coefficient (c) for the reversed field case as a function of the OMP radial distance from the separatrix. Thin solid grey lines indicate FHRP and TS data, with shaded regions indicating experimental uncertainty. Thick solid blue lines indicate Model 1 profiles, thick dash-dotted orange lines indicate $\kappa$-model profiles, and the thick dotted yellow line in (c) indicates $\kappa$-model profiles for the forward field case. The black solid circle indicates the mean experimental data at the separatrix in forward field. Solid vertical black lines indicate the separatrix location.} \label{fg:omprev}
\end{figure}

For this reversed field configuration we do not carry out a new calibration, but instead employ the same simulation setup and model parameters found previously, only reversing the magnetic field. The experimental upstream separatrix electron density in reversed field is slightly lower compared to that of the forward field, as visible in Figure \ref{fg:omprev}. Considering how sensitive the simulation are to this parameter, as discussed at the end of the previous section, employing the calibrated $n_{e,core}$ BC results in separatrix densities outside of the experimental range, and consequently completely different target profiles. Hence, we adjust this parameter for all models to obtain a separatrix density corresponding to the mean experimental value. The upstream electron temperature is instead well in line with the forward field data. For $\kappa_{core}$ we assume that its value does not change in reversed field. This may not be the case in reality, especially when a completely different experimental scenario is addressed. 

Before analyzing the plasma profiles, we assess the performance of the calibrated models in describing the reversed field data by recalculating the cost function. The results are listed in Table \ref{tb:cfvaluesrev} for all models considered, including the difference with respect to Model 1, and with respect to the cost function value in the forward field case. One can clearly notice how Model 1 is the best performing one, while Model 2 and 3 are performing much worse, with an at least 50\% larger cost function value. As discussed previously, this can be linked to  overfitting, especially for Model 3. The $\kappa$-model performance on the reversed field case is essentially the same as on the forward field, worse than Model 1 and only slightly better than Model 2. Moreover, one notices that the calculated cost function value for Model 1 is actually lower compared to the calibration case, indicating that a better agreement with the experimental data should be expected. As in the previous section, we only show the comparison of Model 1 and the $\kappa$-model, and leave the results of Model 2 and 3 in Appendix \ref{app:a}.

The obtained upstream plasma density profiles and anomalous diffusion coefficient are shown in Figure \ref{fg:omprev}. The same qualitative and quantitative comparison observed for the forward field case is valid also here, with the electron temperature profile which is well captured by both models, while the density one shows slightly worse agreement. Moreover, compared to the forward field case the anomalous particle diffusivity of the $\kappa$-model shows a slightly less pronounced dip around the separatrix, driven by a reduced  $E\times B$ flow shear, the rest of the profile being largely the same. The estimated near-SOL density and temperature decay lengths at the OMP are in this case well in line with the experimental ones, as listed in Table \ref{tb:lambdasrev}. There is instead a larger discrepancy at DE, where both models predict similar decay lengths as at the OMP, while the experimentally estimated ones decrease.

\begin{table}
	\caption{Cost function recalculated on the reversed field data for different models and its difference compared to Model 1 and the forward field}\label{tb:cfvaluesrev}
	\centering
	\begin{tabular}{c c c c c}
		\hline
		& Model 1 & Model 2 & Model 3 & $\kappa$-model \\
		\hline
		$\hat{\mathcal{J}}$ & 0.327 & 0.476 & 0.983 & 0.465 \\
		Differences wrt:          &       &       &       & \\
		Model 1          & 0     & +46\% & +201\% & +42\% \\
		Forward field    & -31\% & +62\% & +318\% & +2\% \\
		\hline
		& 
	\end{tabular}
\end{table}

\begin{table}[h!]
	\caption{Near-SOL decay lengths (in cm) at the OMP and DE for electron density $\lambda_n$ and temperature $\lambda_{T_e}$, and OT parallel heat flux decay length ($\lambda_q$) and spreading factor ($S$) (in mm), as obtained from experimental data (Exp.) \cite{x21paper} and with the different models, for the reversed field case} \label{tb:lambdasrev}
	\centering
	\begin{tabular}{l c c c}
		\hline
		& Exp. & Model 1 & $\kappa$-model\\
		\hline
		$\lambda_{n,OMP}$&   2.0$\pm$1.1 & 1.9 & 1.9 \\
		$\lambda_{n,DE}$&    1.1$\pm$0.2 & 1.9 & 1.8 \\
		$\lambda_{T_e,OMP}$& 1.5$\pm$1.1 & 1.4 & 1.5 \\
		$\lambda_{T_e,DE}$&  1.0$\pm$0.1 & 1.5 & 1.5 \\
		$\lambda_{q}$&       4.0$\pm$0.1 & 3.4 & 2.7 \\
		$S$ &                1.8$\pm$0.0 & 1.1 & 1.2\\
		\hline
	\end{tabular}
\end{table}

\begin{figure}[ht]
	\centering
	\subfigure{\includegraphics[scale=0.33]{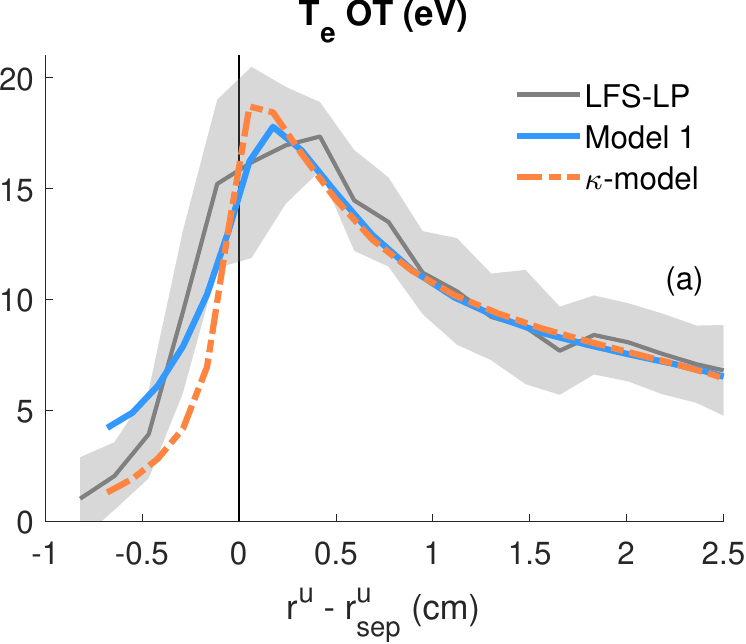}}
	\subfigure{\includegraphics[scale=0.33]{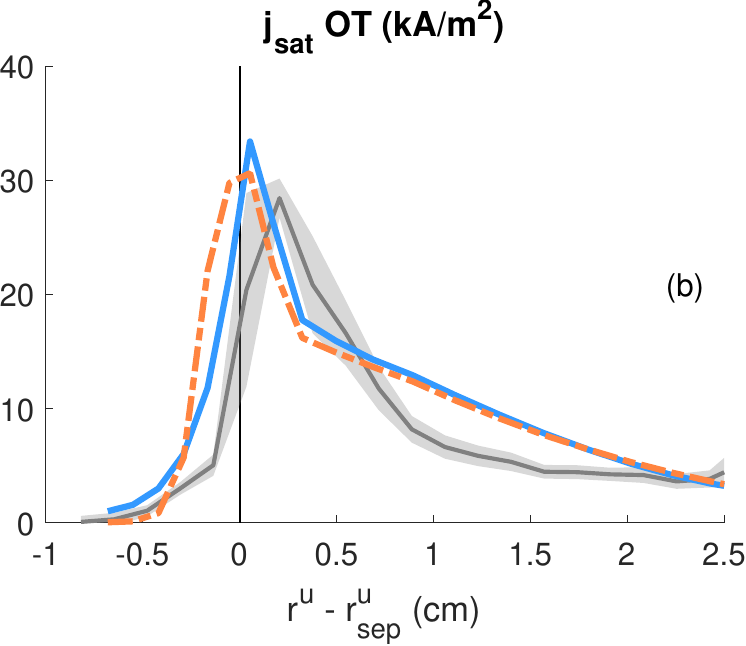}}
	\subfigure{\includegraphics[scale=0.33]{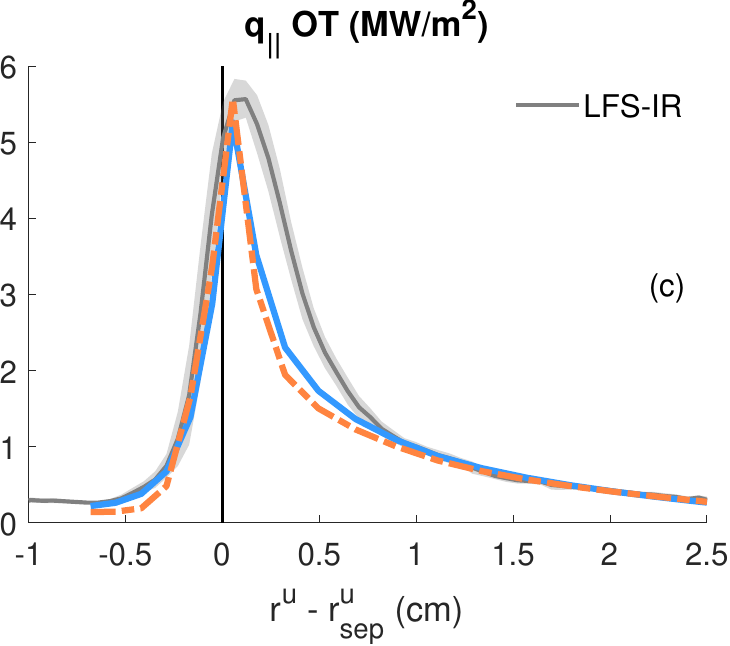}}
	\subfigure{\includegraphics[scale=0.33]{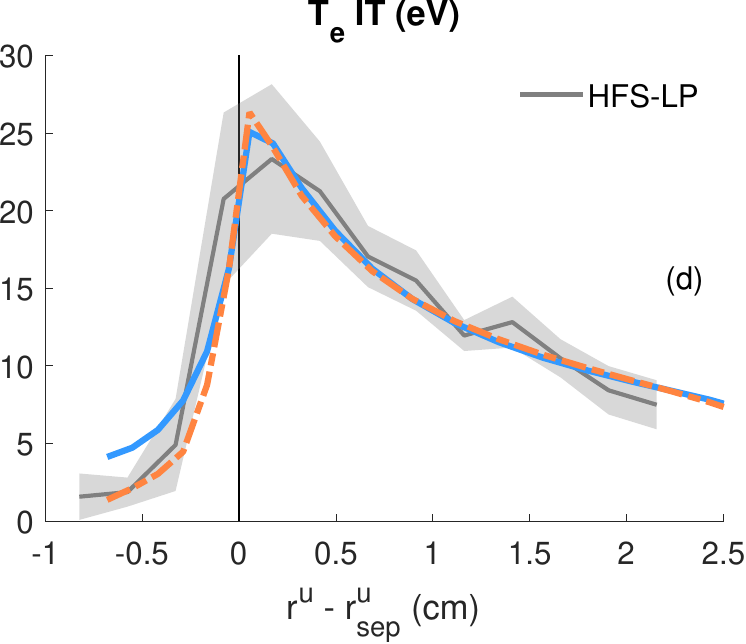}}
	\subfigure{\includegraphics[scale=0.33]{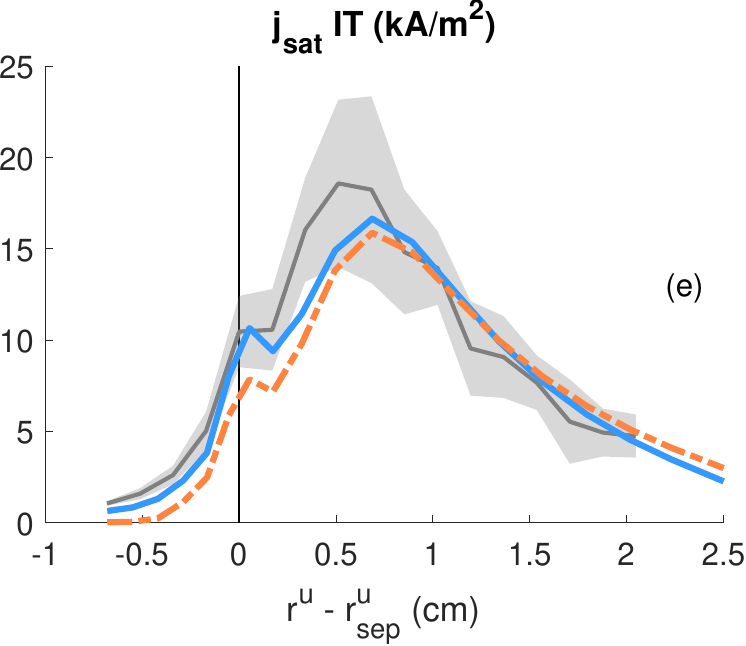}}
	\caption{Reversed field case profiles of: OT electron temperature (a), saturation current (b), and parallel heat flux (c); IT electron temperature (d), and saturation current (e). All quantities are mapped to the upstream (OMP) radial distance from the separatrix. Thin solid grey lines indicate LP and IR data, with shaded regions indicating the experimental uncertainty. Thick solid blue lines indicate Model 1 profiles, and thick dash-dotted orange lines indicate $\kappa$-model profiles. Solid vertical black lines indicate the separatrix location} \label{fg:targetrev}
\end{figure}

We now focus on target plasma profiles, which are reported in Figure \ref{fg:targetrev}, and observe that an excellent agreement between simulations and experiment is found. $T_e$ is very well captured at both targets, with the $\kappa$-model also reproducing the temperature slope in the PFR. The OT $j_{sat}$ profile shows the largest disagreement among target quantities, with the shape qualitatively captured: both models are shifted towards the PFR with respect to the experimental profile, despite the correct peak magnitude. Similarly to the forward field case, the predicted OT parallel heat flux has a narrower peak compared to experiment, but now the PFR side of the shape is better reproduced compared to the SOL side, which is steeper. This is indeed confirmed by the estimated $\lambda_{q}$ and $S$ reported in Table \ref{tb:lambdasrev}. Finally, the IT $j_{sat}$ profile is qualitatively and quantitatively reproduced by both models, with the peak slightly shifted towards the SOL side.

Concerning power and particle balances, the same qualitative picture is valid here as for the forward field case, and we do not show it. The two models are therefore capable of predicting the reversed field data with good accuracy, using the parameters calibrated on the forward field data. Still, the plasma upstream conditions were quite similar for both magnetic field configurations, and one can argue that this plays a significant role in the quality of the predicted results. Experimental data at higher density for the same scenario conditions were not yet available at the time of writing, and we leave modeling of a completely different discharge for future work, noting that modeling of the TCV-X23 case using the same parameters found in this study has already started \cite{unstrpet25}. Therefore, in the next section we assess possible model differences in predictive situations by performing a density scan.

\subsection{Prediction on density scan}\label{sc:densscan}
\begin{figure}[ht]
	\centering
	\subfigure{\includegraphics[scale=0.5]{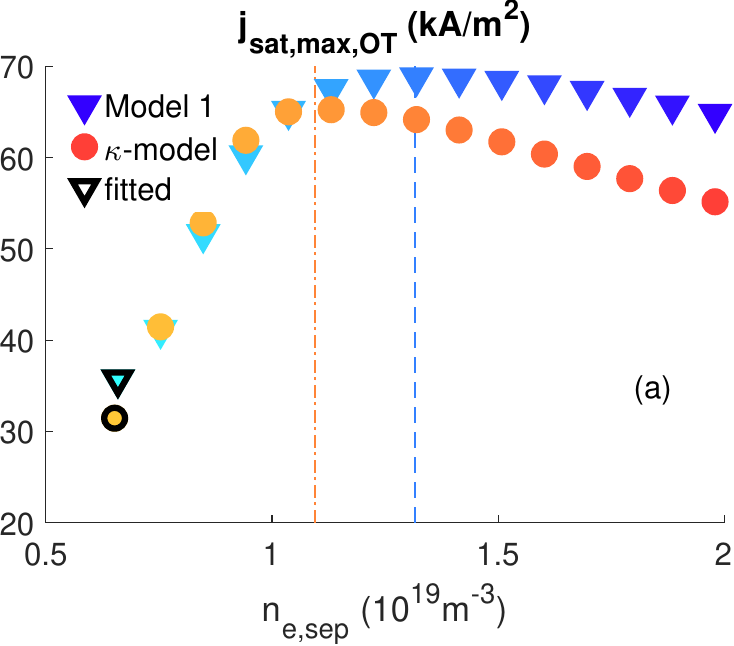}}
	\subfigure{\includegraphics[scale=0.5]{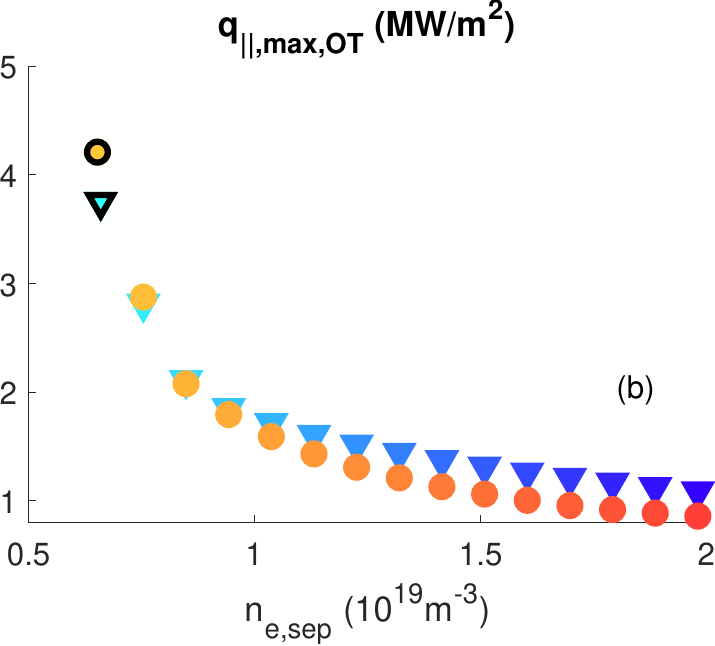}}
	\subfigure{\includegraphics[scale=0.5]{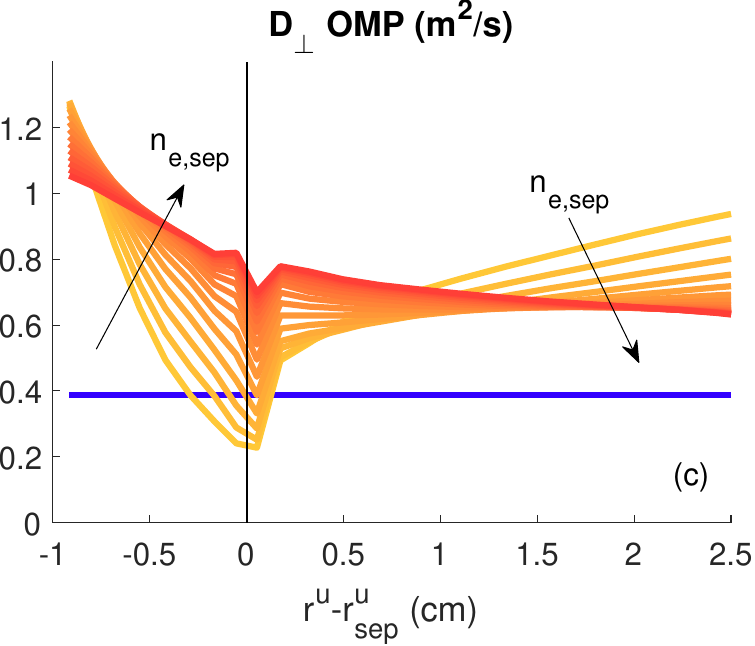}}
	\subfigure{\includegraphics[scale=0.5]{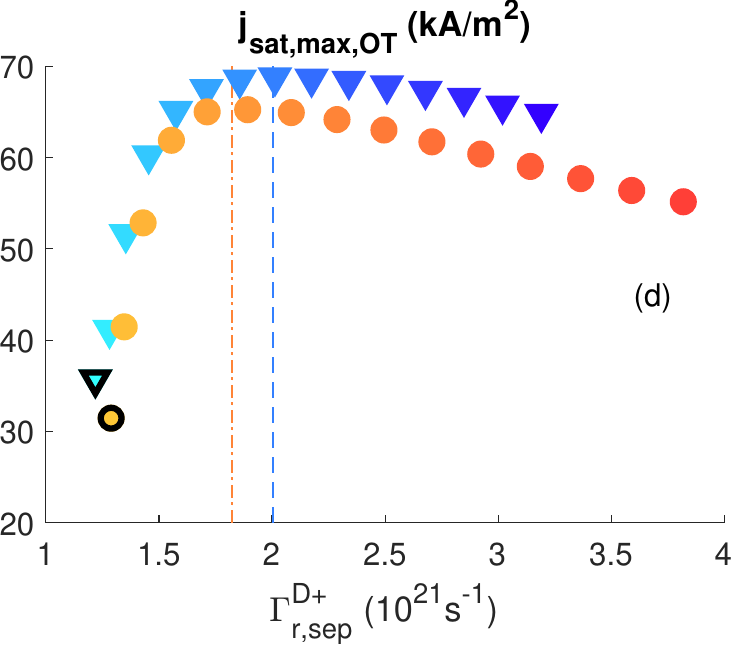}}
	\subfigure{\includegraphics[scale=0.5]{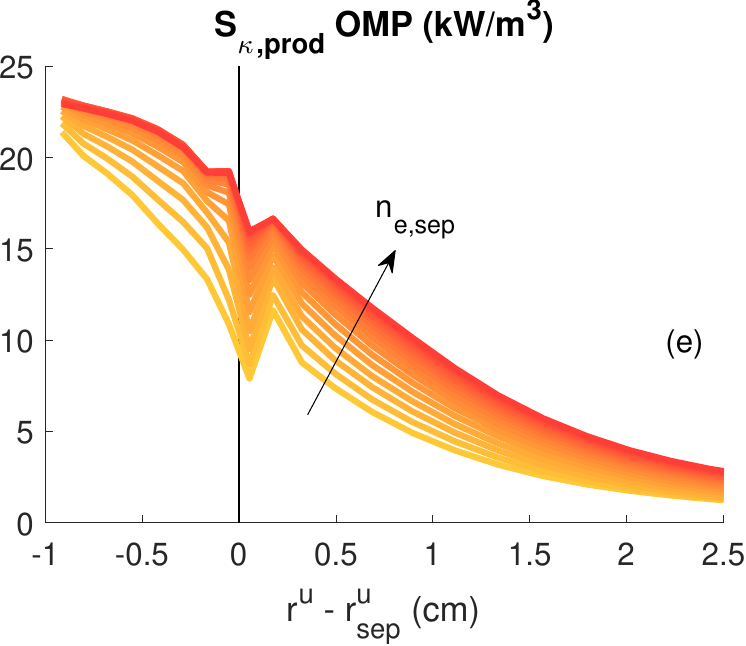}}
	\subfigure{\includegraphics[scale=0.5]{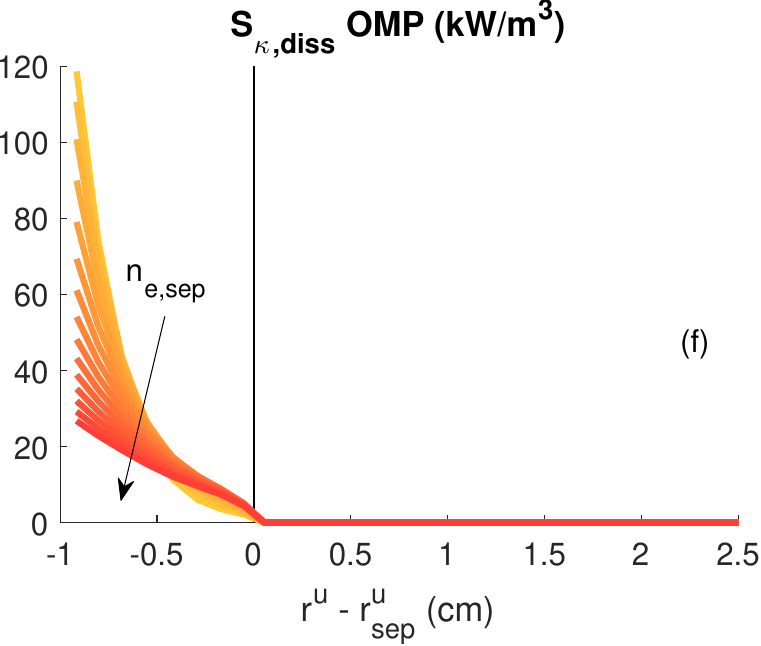}}
	\caption{Maximum OT saturation current (a) and parallel heat flux (b) as a function of the OMP separatrix density. (c) Anomalous diffusion coefficient at the OMP as a function of the distance from the separatrix. (d) Maximum OT saturation current as a function of the D ions outflux from the separatrix. $\kappa$ production (e) and dissipation (f) at the OMP as a function of the distance from the separatrix. Shades of blue indicate Model 1 results while shades of orange indicate $\kappa$-model results. Darker colors indicate higher density. The vertical dashed blue lines indicate the rollover point for Model 1 and vertical dash-dotted orange lines indicate the rollover point for the $\kappa$-model. The thick black contours on markers indicate the calibration point.} \label{fg:dscan1}
\end{figure}

The density scan study is conducted by keeping  $P_{core}$ and the calibrated model parameters fixed, except for $n_{e,core}$, which is increased for both Model 1 and the $\kappa$-model to attain the same OMP separatrix value $n_{e,sep}$. In the following figures, the point in the density scan at which the calibration has been performed is indicated by a thick black contour of the markers, while darker marker colors indicate higher density. We first show on Figure \ref{fg:dscan1}a and b the OT rollover curves of the peak saturation current and parallel heat flux, respectively. The former shows a significant difference between the two models, with the $\kappa$-model anticipating rollover and displaying a more pronounced maximum $j_{sat}$ decrease afterwards. The heat flux instead does not show any rollover, and the behavior of the two models is qualitatively the same. Similarly, we do not see a maximum temperature rollover in both models (not shown). This noticeable difference in detachment behavior can be attributed to the self-consistent evolution of $D_\perp$ for the $\kappa$-model throughout the density scan, which can be appreciated in Figure \ref{fg:dscan1}c. There, we show how the diffusion coefficient around the separatrix increases for increasing $n_{e,sep}$, and flattens in the far-SOL. Consequently, the ions radial outflux $\Gamma^{D^+}_{r,sep}$ from the core towards the SOL is higher for the $\kappa$-model compared to Model 1, at the same $n_{e,sep}$. Hence, this acts as effectively having more particles ``available'' in the divertor for detachment, for a given upstream density. If we now plot the maximum OT $j_{sat}$ as a function of $\Gamma^{D^+}_{r,sep}$, we see that the rollover point of the $\kappa$-model is more closely located to that of Model 1, see Figure \ref{fg:dscan1}d. Therefore, to a large extent this model difference can be explained by the larger particle outflux predicted by the $\kappa$-model at a given upstream density. It remains to be assessed whether this difference in rollover point can be experimentally resolved and verified.

Returning to the $D_\perp$ evolution in the density scan, its behavior can be explained by considering the local $\kappa$ production and dissipation balance, and the $E\times B$ flow shear. In particular, the interchange source term $S_{\kappa,prod}$ of Eq. \ref{eq:skprod} increases with density, as can be seen in Figure \ref{fg:dscan1}e, because the density and density gradient increase overcome the decrease in temperature necessary to sustain the same upstream energy content (in pressure). At the same time, this temperature decrease causes a reduction of the parallel conductivity $\sigma_{\parallel}$, which in turns decreases both the dissipation term $S_{\kappa,diss}$ of Eq. \ref{eq:kdiss} as well as the parallel transport term in the $\kappa$ flux of Eq. \ref{eq:kflux}. Since dissipation of $\kappa$ is dominant over the source in the core region, see Figure \ref{fg:dscan1}f, its strong decrease with density makes the net source of $\kappa$ increase, hence the growth in $D_\perp$ in the core region and around the separatrix. Furthermore, also the $E \times B$ flow shear around the separatrix decreases with density, as such reducing the $D_\perp$ dip in this region. Finally, in the far-SOL we notice a decrease in $D_\perp$, which seems contradicting the local production-dissipation balance. However, when looking at the actual conserved quantity $n_i\kappa$, this consistently increases throughout the radial OMP domain with increasing $n_{e,sep}$ (not shown): the far SOL density increases, while temperature, hence parallel conductivity and $\kappa$ parallel dissipation, drops, requiring $\kappa$ itself to decrease.

\begin{figure}[ht]
	\centering
	\subfigure{\includegraphics[scale=0.5]{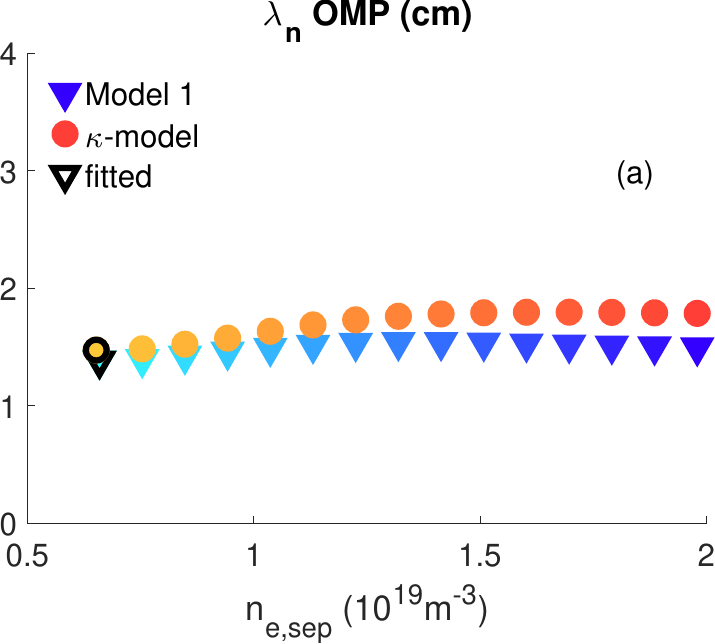}}
	\subfigure{\includegraphics[scale=0.5]{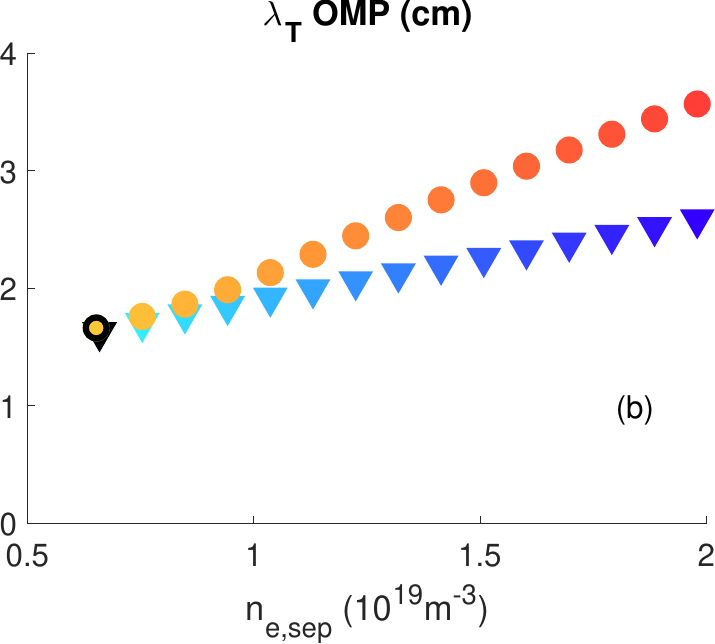}}
	\subfigure{\includegraphics[scale=0.5]{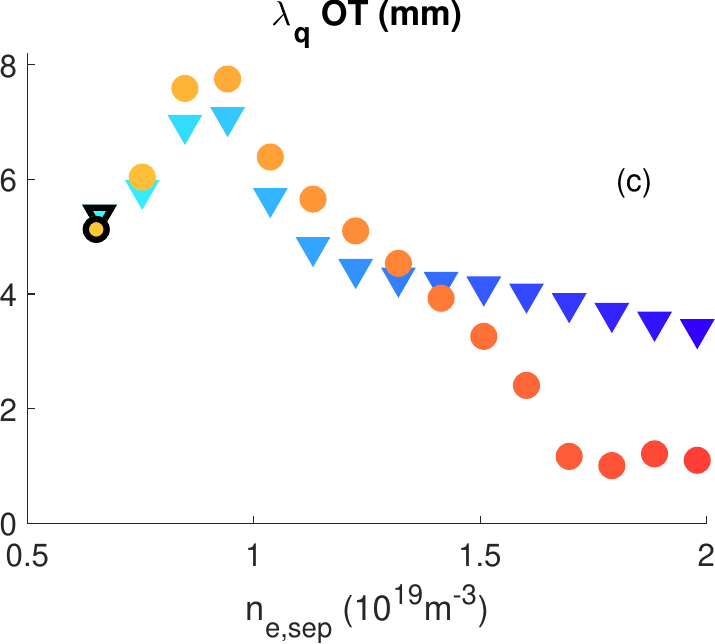}}
	\subfigure{\includegraphics[scale=0.5]{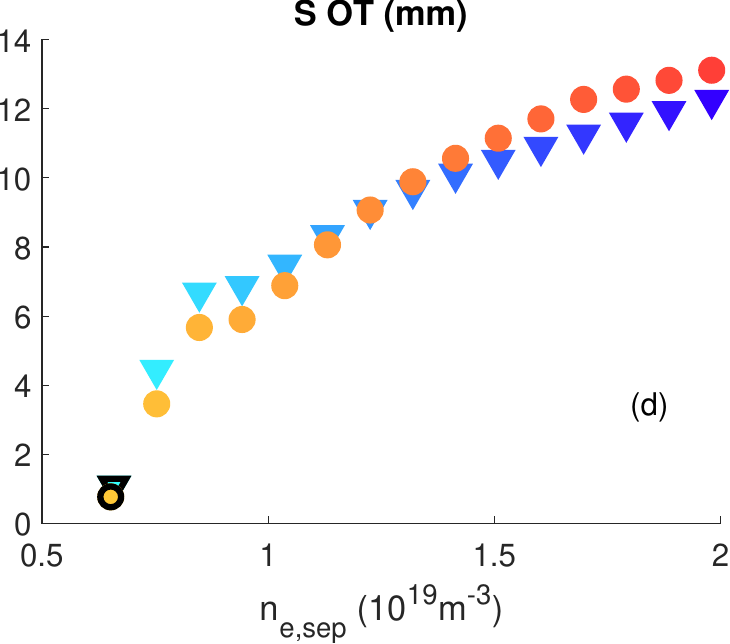}}
	\caption{Near-SOL decay lengths for density (a) and temperature (b) at the OMP, and OT heat flux decay length (c) and spreading factor (d), as a function of the OMP separatrix density. Triangles in shades of blue indicate Model 1 results while circles in shades of orange indicate $\kappa$-model results. Darker colors indicate higher density. The thick black contours on markers indicates the calibration point.} \label{fg:dscan2}
\end{figure}

We conclude this section by showing how the model differences previously analyzed, in particular the increase of anomalous transport around the separatrix predicted by the $\kappa$-model, impact upstream and downstream near-SOL decay lengths, which we plot in Figure \ref{fg:dscan2} as a function of $n_{e,sep}$. We can clearly see how the enhanced particles and heat diffusivity in this region drives larger decay lengths and broader profiles for the $\kappa$-model compared to the constant diffusivity of Model 1. On the other hand, $\lambda_{q}$ at the outer target and spreading factor $S$ do not show significant differences until $n_{e,sep}\sim1.5\cdot10^{19}$ m$^{-3}$, where $\lambda_{q}$ predicted by Model 1 flattens, while the $\kappa$-model predicts a further decrease. However, at that point we notice the confidence bounds of the MATLAB fitting algorithm become very large, the heat flux profile being almost flat, and the fitting results are thus questionable for these high density points.

\clearpage
\section{Conclusions and outlook}\label{sc:conclusion}

In this work we reviewed the currently available model calibration strategies for SOLPS-ITER and other plasma-edge codes, emphasizing the advantages of the optimization-based approach in case of highly nonlinear and expensive-to-evaluate models. We then applied this framework to the case of cross-field transport models, a routine model calibration step necessary for interpretation of experiments with mean-field plasma edge codes. We calibrated and assessed different cross-field transport models in SOLPS-ITER using the TCV-X21 reference case. The considered models ranged from conventional ad-hoc diffusivity descriptions, with different levels of parametric freedom, to the self-consistent $\kappa$-model. Model parameters were determined against a combination of upstream and target measurements using forward field data.

We showed that within the first 10 iterations of the parameter optimization loop the performance indicator achieves 99\% of its total reduction, indicating that the optimization could already be stopped there. At that point however, the true optimum has not been reached yet. The obtained model parameters were in line with other studies from literature and highlighted that no or negligible ballooning of the anomalous transport coefficients was necessary to describe this experimental scenario. The calibration results showed that increasing model flexibility substantially improves agreement with the calibration dataset. In particular, models with more parametric freedom, i.e. radially varying transport coefficients, achieved reductions of up to 50\% in the cost function compared to the baseline constant-diffusivity model. However, when applied without recalibration to the reversed field configuration, these models exhibited significantly degraded performance. This indicates that the additional degrees of freedom primarily captured features specific to the calibration case rather than improving the underlying predictive capability, the well-known overfitting problem. The simplest constant-diffusivity model provided the best overall predictive performance on the reversed field dataset while requiring only a limited number of fitted parameters. The $\kappa$-model achieved a calibration quality comparable to this baseline model and reproduced the experimental profiles with similar accuracy in both magnetic field configurations.

We observed small differences between the simplest constant-diffusivity model and the $\kappa$-model in both magnetic field configurations, despite the large difference in diffusion coefficients. We linked this to two main aspects, the most important one being that both models attained the same upstream separatrix density and temperature, which uniquely determines target quantities in a 2-point modeling setting. The second aspect is that for this TCV case the dominant transport mechanisms seemed parallel neoclassical and drifts transport, and not the anomalous one, corroborated by the negligible or absent ballooning enhancement found here and in other TCV modeling efforts. We point out once again that, while for the reversed field configuration the same upstream separatrix values in both models were user-imposed, in the calibration step this was a consequence of the optimization. We also showed the importance of including both upstream and target experimental data in the cost function, providing the best overall fit and consistency of the plasma quantities, compared to including only upstream data, as is often done in current practice. We think this is a remarkable result which opens a new way of interpretive modeling with SOLPS-ITER: the optimization approach allows finding model parameters and separatrix values providing the highest consistency between upstream and targets. One noticeable difference in the two models concerns the particle flux to the vessel wall, which is almost a factor two larger for the $\kappa$-model, due to the increased diffusivity in the far-SOL. This aspect is of particular importance for impurities erosion and plasma contamination, and needs further confirmation from experiments.

We analyzed the sensitivity of our results on model parameters based on the Hessian matrix, which showed that the most sensitive parameter was the core density BC for both the constant-diffusivity and $\kappa$ models, the latter also having high sensitivity to the $\kappa$ parallel dissipation constant in the core. This means that such parameters are likely less universal, which has been confirmed in this study on the reversed field configuration, and in the ongoing modeling of the TCV-X23 case \cite{unstrpet25}.

The density-scan study revealed differences between the two models that were not apparent close to the calibration point. While both models predicted similar peak heat flux behavior, the $\kappa$-model exhibited increasing transport levels around the separatrix with increasing density, originating from the self-consistent evolution of the transport coefficients. This resulted in larger particle outflux from the core for the same separatrix density, broader upstream profiles and an earlier onset of rollover in saturation current.

The present work demonstrates that optimization-based model calibration provides a powerful framework for systematic use of plasma edge codes as interpretive tools. In addition, it provides the community with a first set of calibrated parameters for the $\kappa$-model.

Future work should focus on different trajectories. Current efforts are being put into extending the $\kappa$-model itself, improving the closure models for the various terms and incorporating new terms or improved descriptions by analyzing 3D turbulence data. In parallel to this, the validation database for the $\kappa$-model should be extended to substantially different operating conditions, including ongoing studies of the TCV-X23 scenario, in order to further evaluate the model predictive capabilities and confirm its peculiar features of higher main chamber fluxes and earlier rollover onset. In addition, reliable estimations of $\kappa$ from experimental data or turbulence codes will further constrain the calibration process and allow including the $C_D$ parameter in the closure of the anomalous particle diffusion coefficient Eq.\ref{eq:kdanml}, which is at the moment arbitrarily fixed. Furthermore, first of a kind uncertainty estimates for the calibrated parameters are well within reach: employing the already available Bayesian MAP \cite{carlibayes} and Hessian matrix information, the Laplace approximation allows Guassian uncertainty estimates which are computationally cheap compared to sampling-based methods \cite{laplace, papadim}. Additionally, the Laplace approximation provides an estimate of the Bayesian evidence, a formal tool to carry-out an objective model comparison not solely based on the performance indicator, i.e. the cost function or likelihood in Bayesian terms, but also incorporates information on model complexity, thereby alleviating the overfitting problem. In this context, model parameters that we treated as fixed in this study, such as the core-boundary input power $P_{core}$ or the recycling coefficient $R_c$, should be included in the estimation and UQ framework, even if their assumed values are well informed. This will pave the way for a comprehensive framework for model comparison, Uncertainty Quantification, and model validation for plasma edge codes.

\section*{Acknowledgments}
Stefano Carli is a postdoctoral research fellow of the Research Foundation-Flanders (FWO) under grant number 12E3623N. Parts of the work are supported by the Research Foundation Flanders (FWO) under project grant G085922N. This work has been carried out within the framework of the EUROfusion Consortium, partially funded by the European Union via the Euratom Research and Training Programme (Grant Agreement No 101052200 — EUROfusion). The Swiss contribution to this work has been funded by the Swiss State Secretariat for Education, Research and Innovation (SERI). Views and opinions expressed are however those of the author(s) only and do not  necessarily reflect those of the European Union, the European Commission or SERI. Neither the European Union nor the European Commission nor SERI can be held responsible for them. The resources and services used in this work were provided by the VSC (Flemish Supercomputer Center), funded by the Research Foundation - Flanders (FWO) and the Flemish Government. This work made use of the version 2.0 of the TCV-X21 experimental dataset \cite{x21data} provided by the Experimental and Outboard Midplane Probe Data teams and publicly released under the CC-BY 4.0 license.

\section*{Author contributions}
S. Carli: Conceptualization, Data Curation, Formal Analysis, Funding Acquisition, Investigation, Methodology, Project Administration, Software, Resources, Validation, Visualization, Writing – original draft.
R. Coosemans: Methodology, Writing – review \& editing.
C. Colandrea: Data Curation, Writing – review \& editing.
W. Dekeyser: Conceptualization, Funding Acquisition, Methodology, Project Administration, Resources, Software, Supervision, Writing – review \& editing.

\section*{Data Availability}
The results presented in this work have been obtained with SOLPS-ITER version \href{https://git.iter.org/projects/BND/repos/solps-iter/commits/6e5200901c437728493701ba9a48661d9c2831c0}{6e5200901c4}, with B2.5 version \href{https://git.iter.org/projects/BND/repos/b2.5/commits/d9590892374e344b18c0deaa9a7286d0f03a964a}{d9590892374}, and TAPENADE version \href{https://gitlab.inria.fr/tapenade/tapenade/-/commit/686e29d1a492ce72f4f7f61d94a47066695c9da8}{3.16-v2-120-g686e29d1a}.
The data that support the findings of this study are available in an open source online repository \cite{repodata}. The scripts to reproduce the figures of this paper using such data are also available online \cite{reposcript}.

\appendix

\section{Strategies to avoid instabilities in optimization} \label{app:d}
This appendix aims at giving an overview of issues encountered when performing the optimizations described in this paper, and strategies devised to overcome them.

The main source of issues is related to the linesearch described in Section \ref{sc:adjointop}. Specifically, for the linesearch to be efficient the first stepzise attempted is $\alpha_k=1$, which therefore may lead to large changes in model parameters $\theta^{k+1}$, depending on the gradient. SOLPS-ITER cases with electromagnetic drifts being prone to numerical instabilities, such large parameter changes often lead to divergence and crash of the simulation. To avoid these instabilities, three strategies can be adopted, in any combination:
\begin{enumerate}
	\item Reduce the initial stepsize $\alpha_k$ attempted by the linesearch. This can be easily adopted by specifying an additional command line argument for PETSc/TAO, namely \verb|-tao_ls_stepinit X|, where \verb|X| is a real number between 0 and 1. Of course, the smaller it is, the more optimization iterations are required to reach the optimum.
	
	\item Reduce the strength of electromagnetic drifts at each linesearch attempt, and gradually increase it back to 100\%. This strategy relies on the already available capability of SOLPS-ITER of gradually increasing the strength of drift velocities during simulations. For the optimization, users can re-set the initial drift strength to X\%, where X is defined by the switch \verb|b2optim_reset_drift|. Then, drifts will be gradually increased to 100\% in N iterations, where N is defined through the switch \verb|b2optim_reset_drift_iter| such that  N=\verb|b2mndr_ntim|/\verb|b2optim_reset_drift_iter|. The consequence is that simulations will likely require more iterations to converge, and one should thus increase \verb|b2mndr_ntim| to accommodate this.
	
	\item Gradual increase of the model parameters at each linesearch attempt. The philosophy here is essentially the same as for the drifts, but applied to $\theta$: the initial value  $\theta^k$ is gradually increased (or decreased) until $\theta^{k+1}$ in N iterations, where N is now defined through the switch \verb|b2optim_reset_param_iter| such that N=\verb|b2mndr_ntim|/\verb|b2optim_reset_param_iter|.
\end{enumerate} 

Simulations without electromagnetic drifts activated seem less sensitive to abrupt changes in model parameters and in most cases do not need the above safeguards to be activated.

Another difficulty arised when an optimization is continued after interruption. As mentioned in the main text, in this case the gradient and cost function are automatically normalized by PETSc/TAO using values of the new first iteration, meaning that convergence criteria essentially restart from scratch. The only solution possible at the moment is to re-calculate the optimization history in postprocessing, as explained in Section \ref{sc:optim_evol}. In addition, we often observed that the linesearch attempted too large steps in the first optimization iteration after restart. It is unclear whether this is related to the gradient normalization or to the information on the estimated BFGS Hessian $B_k$ being reinitialized. The effective solution employed was to limit the initial stepsize of the linesearch as per bullet 1 above.

\section{Results of Model 2 and Model 3} \label{app:a}
\begin{figure}[ht]
	\centering
	\subfigure{\includegraphics[scale=0.33]{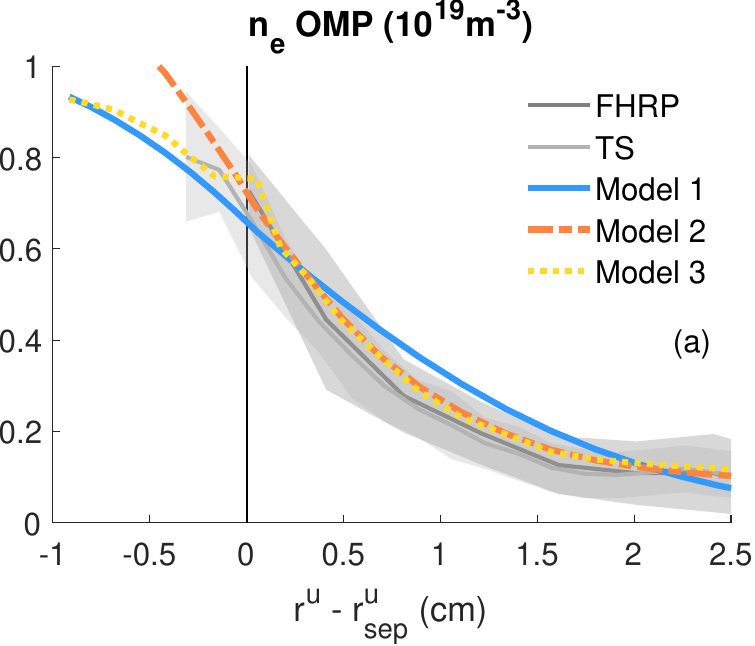}}
	\subfigure{\includegraphics[scale=0.33]{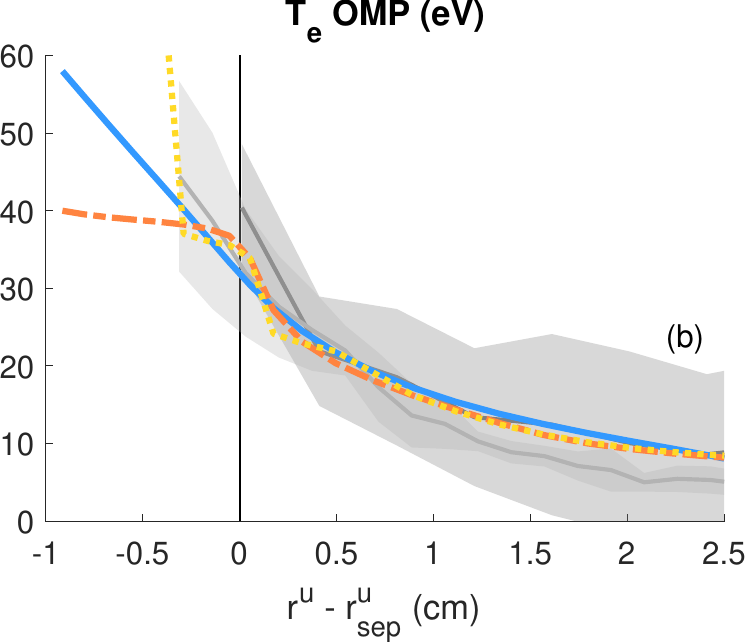}}
	\subfigure{\includegraphics[scale=0.33]{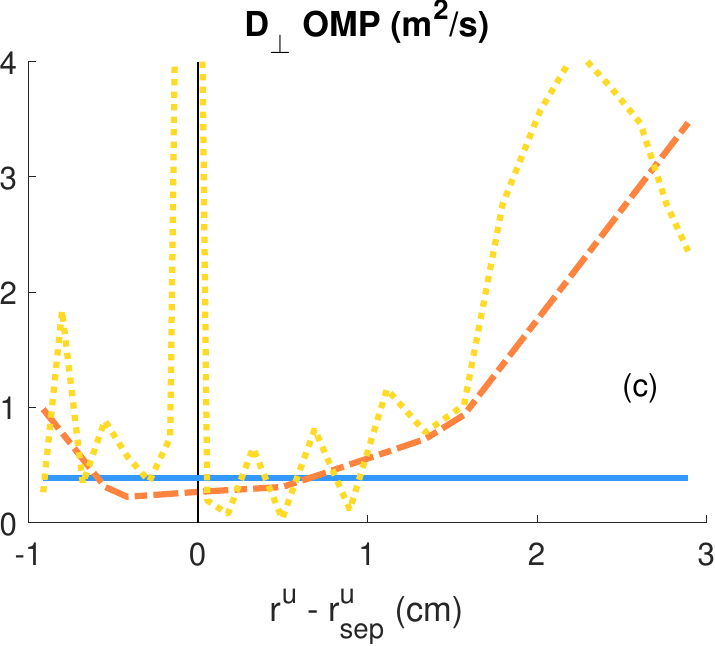}}
	\caption{ Upstream profiles of electron density (a), electron temperature (b), and anomalous particle diffusion coefficient (c) for the forward field case as a function of OMP radial distance from the separatrix. Thin solid grey lines indicate FHRP and TS data, with shaded regions indicating experimental uncertainty. Thick solid blue lines indicate Model 1 profiles, thick dash-dotted orange lines indicate Model 2 profiles, and thick dotted yellow lines indicate Model 3 profiles. Solid vertical black lines indicate the separatrix location.} \label{fg:ompradials}
\end{figure}

\begin{figure}[ht]
	\centering
	\subfigure{\includegraphics[scale=0.33]{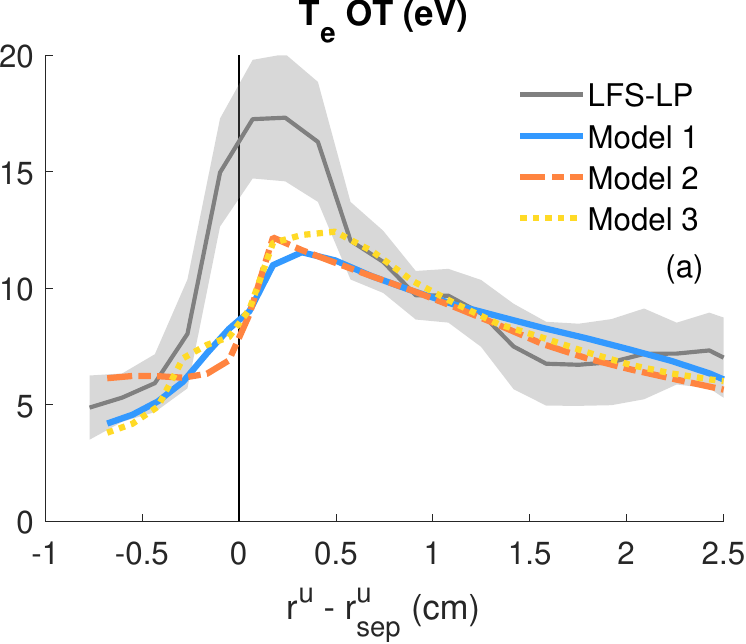}}
	\subfigure{\includegraphics[scale=0.33]{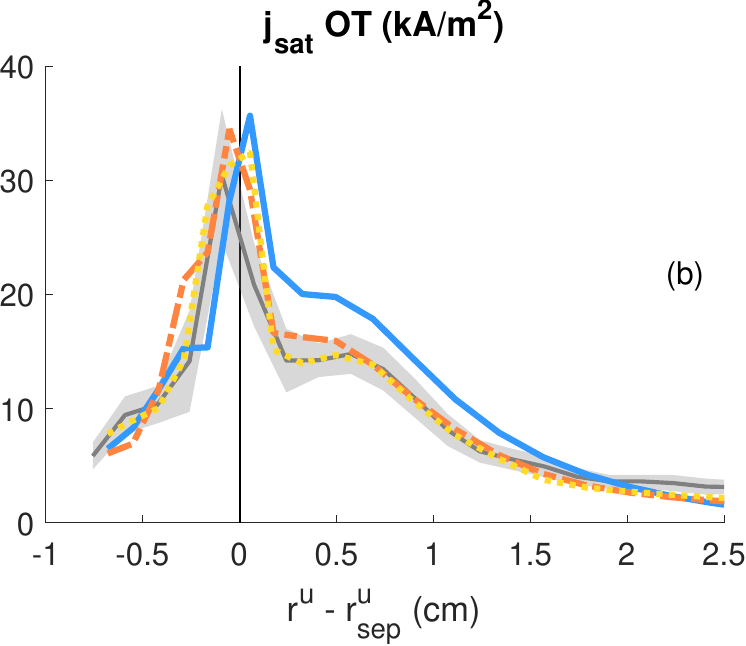}}
	\subfigure{\includegraphics[scale=0.33]{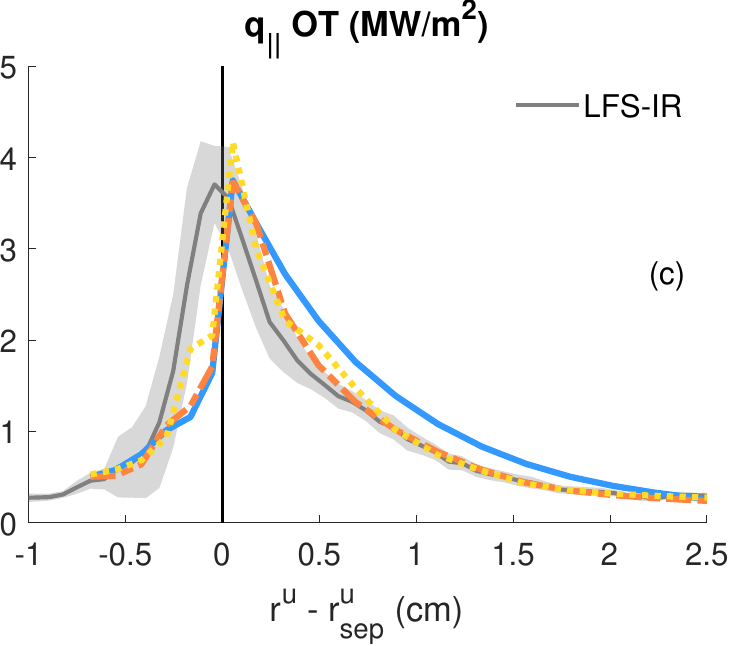}}
	\subfigure{\includegraphics[scale=0.33]{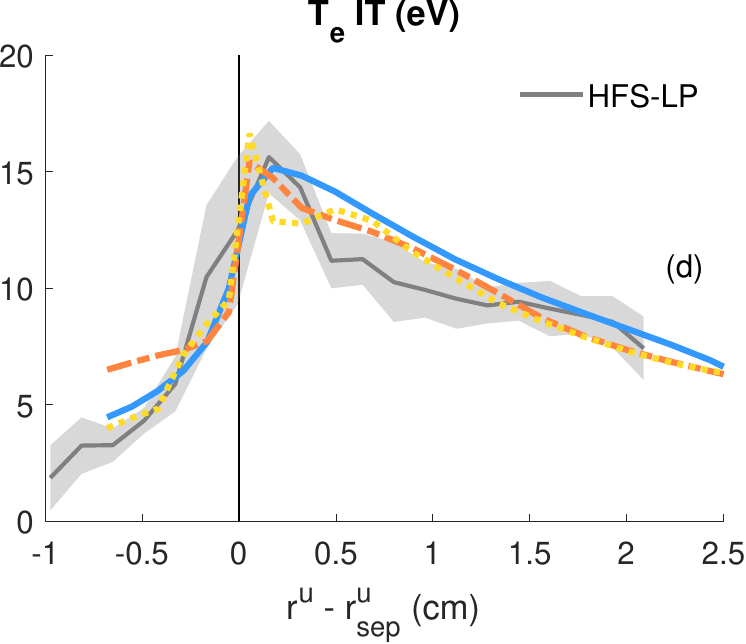}}
	\subfigure{\includegraphics[scale=0.33]{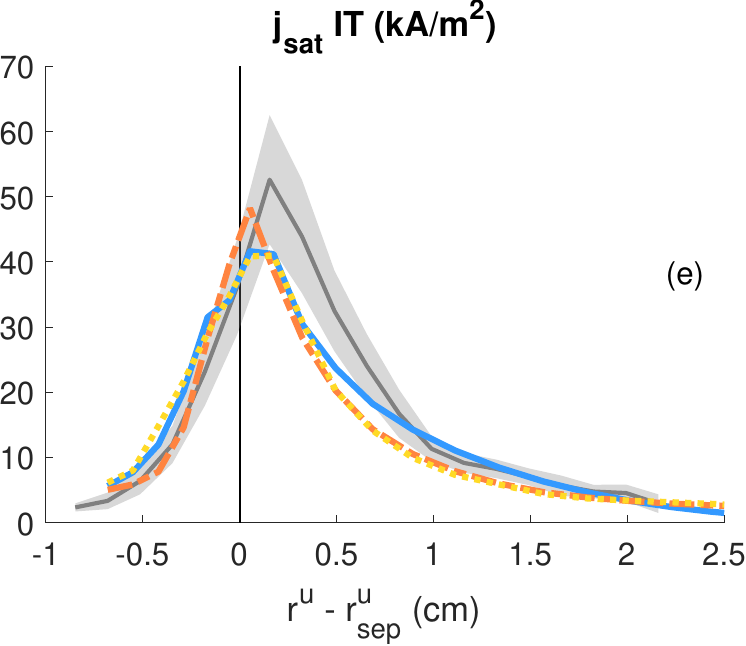}}
	\caption{Forward field case profiles of: OT electron temperature (a), saturation current (b), and parallel heat flux (c); IT electron temperature (d), and saturation current (e). All quantities are mapped to the OMP radial distance from the separatrix. Thin solid grey lines indicate LP and IR data, with shaded regions indicating the experimental uncertainty. Thick solid blue lines indicate Model 1 profiles, thick dash-dotted orange lines indicate Model 2 profiles, and thick dotted yellow lines indicate Model 3 profiles. Solid vertical black lines indicate the separatrix location} \label{fg:targetsradial}
\end{figure}

In this appendix we first display the model calibration results for Model 2 and 3, comparing them with Model 1 already shown in the main text. The OMP plasma profiles are visible in Figure \ref{fg:ompradials}a and b, with Figure \ref{fg:ompradials}c showing the obtained particle diffusion coefficients, and Figure \ref{fg:targetsradial} the target plasma profiles. Note that the ballooning coefficient is $C_b=0.3$ for Model 2 and $C_b=1.25$ for Model 3. From these figures it is clear that the improvement in cost function value listed in Table \ref{tb:cfvalues1} for Model 2 and 3 is attained at the expense of somewhat unrealistic behavior. This is especially true for Model 3, which shows an nonphysical sharp increase in $T_e$ at the OMP core side in Figure \ref{fg:ompradials}b, which, being outside the radial range of experimental data, does not contribute to the cost function. Furthermore, its diffusion coefficients are oscillating, with an abrupt peak around the separatrix, see Figure \ref{fg:ompradials}c. This feature points toward having either a better parametrization of these profiles, or including a penalty term in the cost function that punishes oscillations. Suitable priors are available in this sense, and can be employed in a MAP estimation. Model 2 provides results which are physically more reasonable and a $D_\perp$ profile increasing near the core and in the far-SOL resembling that of the $\kappa$-model, though the high electron density and flat electron temperature in the core seem inconsistent with the extrapolated measurements in this region. As stated in the main text, we expect that a formal model comparison through Bayesian evidence will show that their improvement in the performance indicator does not compensate the increase in model complexity.

\begin{figure}[ht]
	\centering
	\subfigure{\includegraphics[scale=0.5]{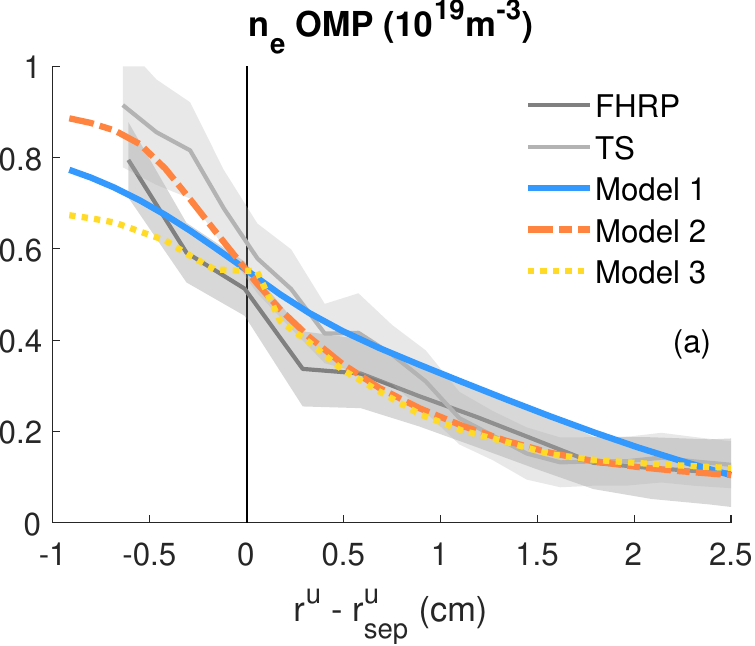}}
	\subfigure{\includegraphics[scale=0.5]{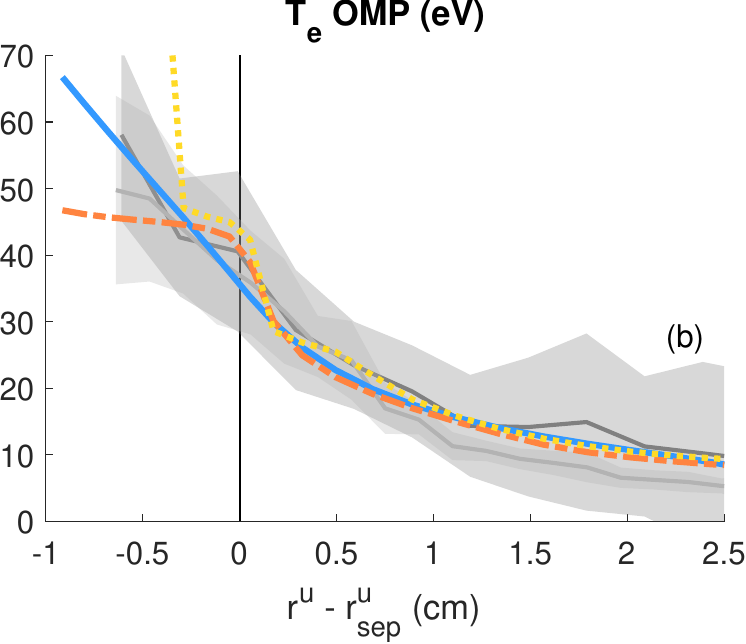}}
	\subfigure{\includegraphics[scale=0.5]{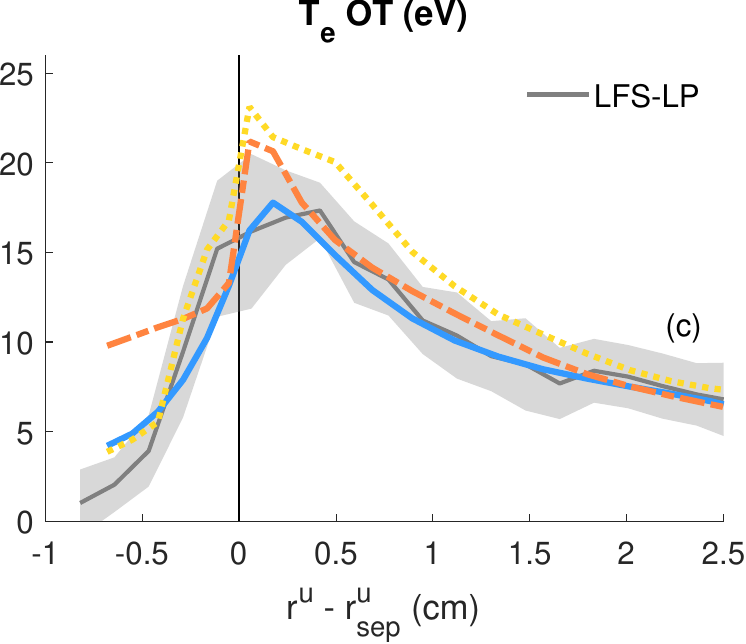}}
	\subfigure{\includegraphics[scale=0.5]{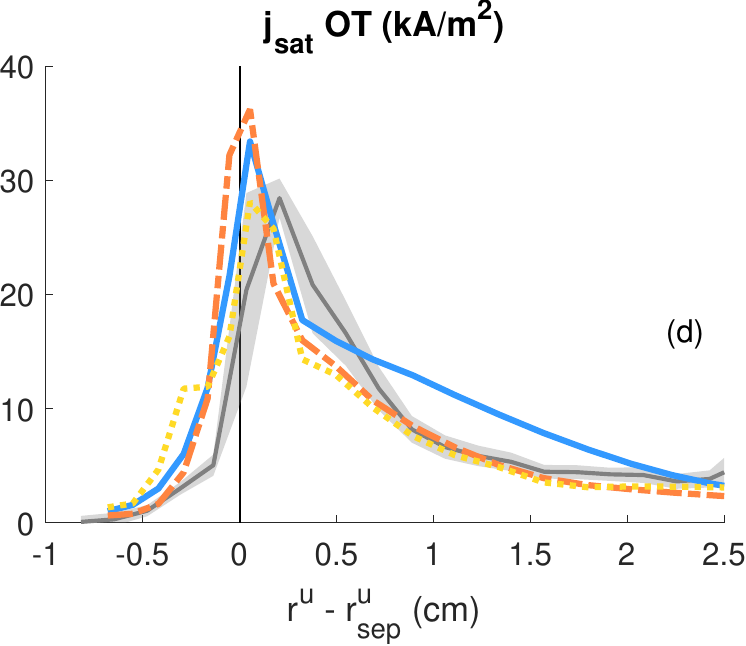}}
	\subfigure{\includegraphics[scale=0.5]{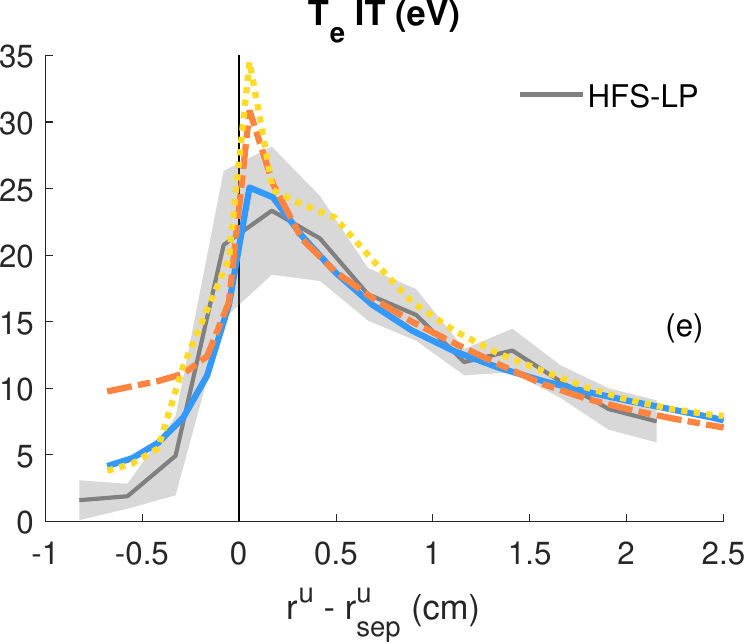}}
	\subfigure{\includegraphics[scale=0.5]{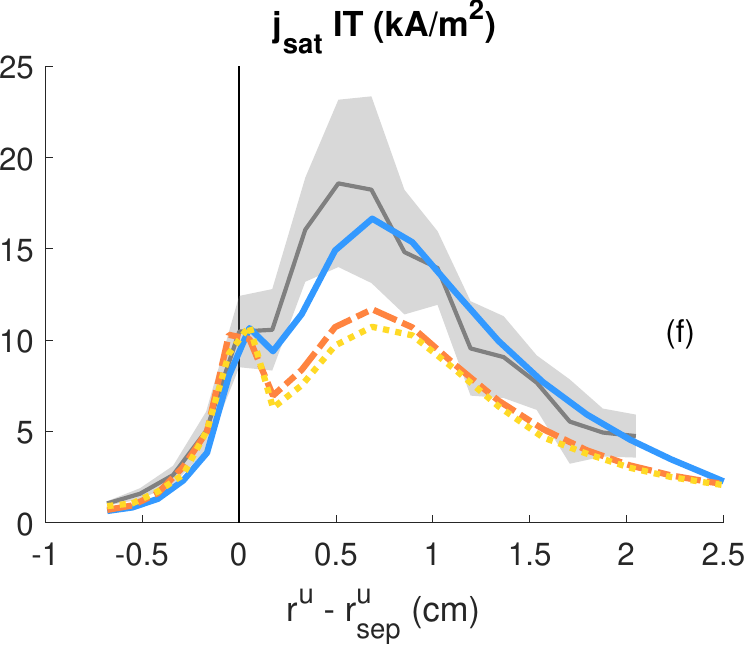}}
	\caption{Reversed field case profiles as a function of the OMP radial distance from the separatrix of: uspstream electron density (a) and temperature (b); OT electron temperature (c) and saturation current (d); IT electron temperature (e) and saturation current (f). Thin solid grey lines indicate experimental data, with shaded regions indicating experimental uncertainty. Thick solid blue lines indicate Model 1, thick dash-dotted orange lines indicate Model 2, and thick dotted yellow lines indicate Model 3. Solid vertical black lines indicate the separatrix location.} \label{fg:radialrev}
\end{figure}

A further proof of the weaknesses of Model 2 and 3, which we link to overfitting, can be appreciated from Figure \ref{fg:radialrev}. In there we show the results for the reversed field configuration obtained in the same way as described in Section \ref{sc:revfield}, namely adjusting $n_{e,core}$ to get the mean OMP experimental value at the separatrix. Clearly, the worse performance of Model 2 and 3 listed in Table \ref{tb:cfvaluesrev} is confirmed by the plasma profiles shown here, giving significantly different results compared to Model 1.

\section{Results with kinetic neutrals and impurities} \label{app:b}
In this appendix we report the results obtained by employing the EIRENE kinetic model for neutral particles as well as carbon impurities. The case setup is based on previous modeling efforts of TCV with kinetic neutrals and C impurities \cite{wensing21, wangx21, tonello2024, carpita2025} and features: D atoms and molecules, including neutral-neutral collisions; physical and chemical sputtering of C from the walls, with chemical sputtering yield $Y_{chem}$=3.5\%; particle recycling $R_D=0.99$ (same as in the AFN model) and $R_C=0.0$.

\begin{figure}[ht]
\centering
\subfigure{\includegraphics[scale=0.5]{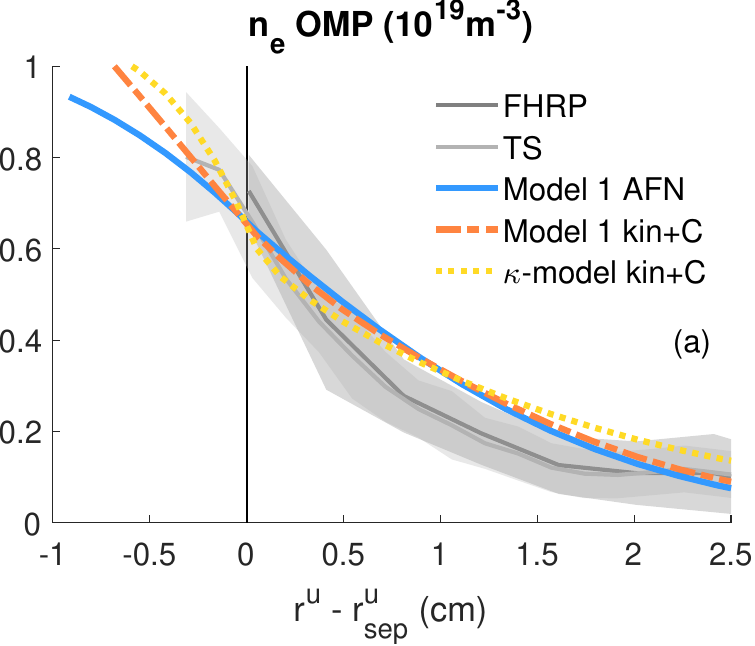}}
\subfigure{\includegraphics[scale=0.5]{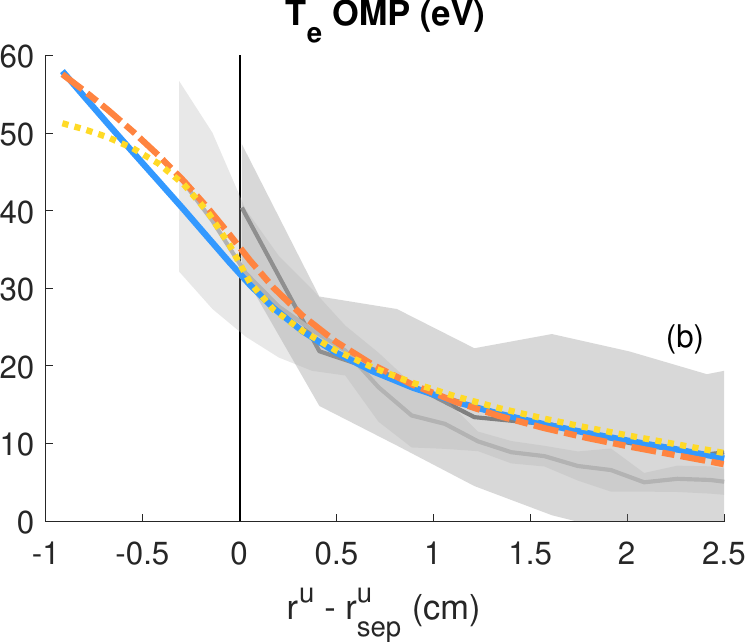}}
\subfigure{\includegraphics[scale=0.5]{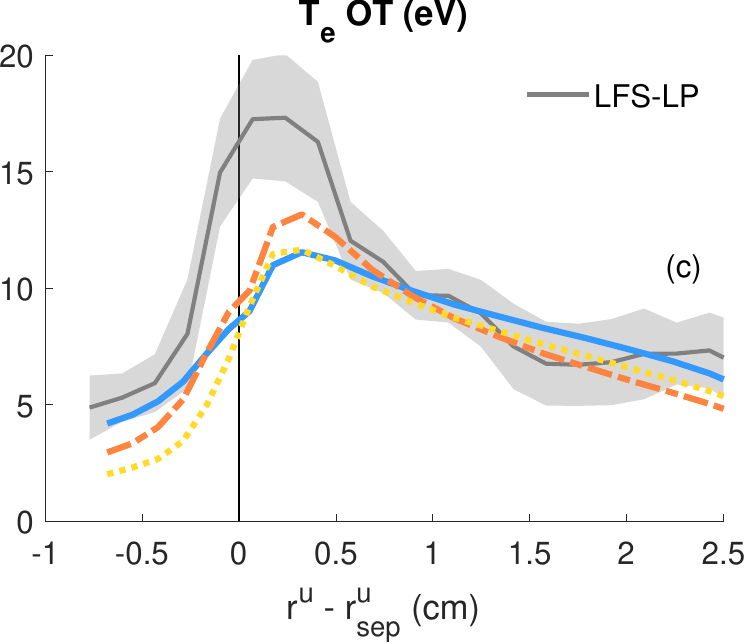}}
\subfigure{\includegraphics[scale=0.5]{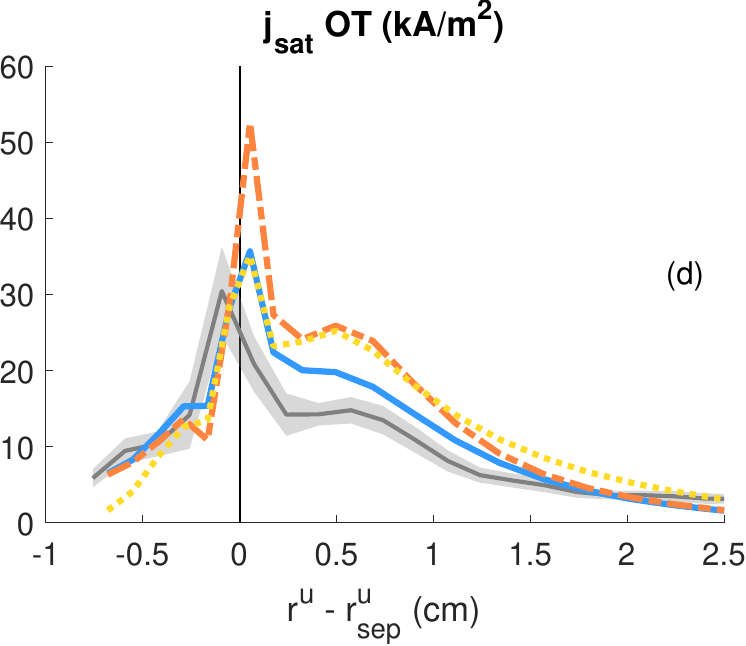}}
\subfigure{\includegraphics[scale=0.5]{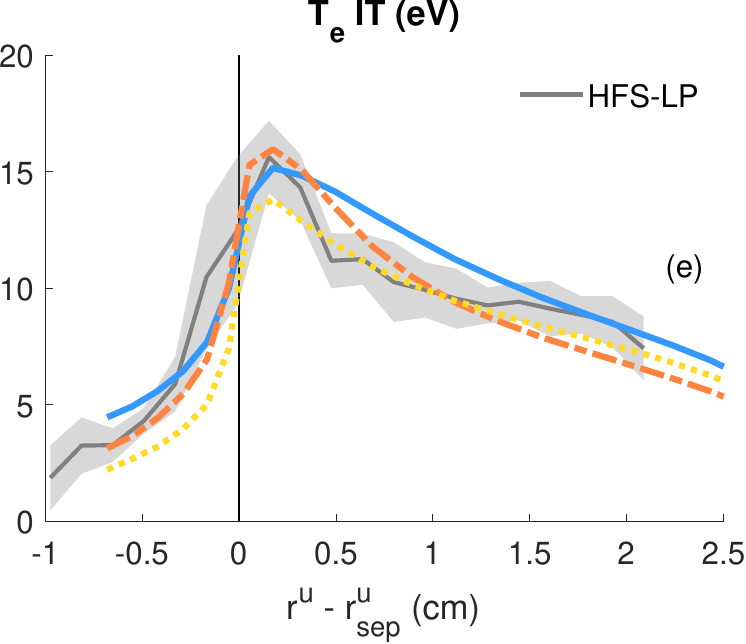}}
\subfigure{\includegraphics[scale=0.5]{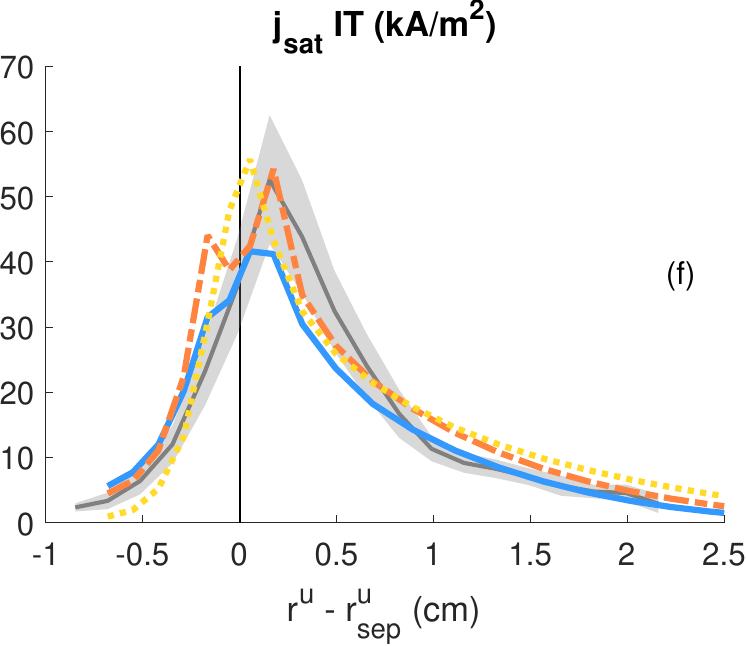}}
\caption{Forward field case profiles as a function of the OMP radial distance from the separatrix of: uspstream electron density (a) and temperature (b); OT electron temperature (c) and saturation current (d); IT electron temperature (e) and saturation current (f). Thin solid grey lines indicate experimental data, with shaded regions indicating experimental uncertainty. Thick solid blue lines indicate Model 1 with AFN, thick dash-dotted orange lines indicate Model 1 with kinetic neutrals and C, and thick dotted yellow lines indicate $\kappa$-model with kinetic neutrals and C. Solid vertical black lines indicate the separatrix location.} \label{fg:kinetic}
\end{figure}

With this setup for the neutral particles, we adopt the calibrated parameters for Model 1 and the $\kappa$-model, and compare the results against Model 1 with AFNs, see Figure \ref{fg:kinetic}. Note that similarly as for the reversed field scenario, we adjusted $n_{e,core}$ of the kinetic neutrals cases to attain the same OMP separatrix density of the AFN ones, otherwise target plasma profiles would have been considerably off. Instead, with this approach the kinetic neutral cases provide very similar results compared to AFNs, despite having introduced two significant changes, namely the kinetic neutral model and the impurities. The largest difference is visible for Model 1 OT saturation current, which overestimates the peak by $\sim$25\%, while the profile shape is qualitatively the same, see Figure \ref{fg:kinetic}d. Model 1 exhibits also a double peak in the IT saturation current, see Figure \ref{fg:kinetic}f, which has been found in other TCV studies as well \cite{wensing21, colandrea}. Likely, one could have kept $n_{e,core}$ constant and tuned the other model parameters to obtain this same result, but we argue that, at least for the case at hand, employing AFNs and neglecting C impurity in the optimization are good enough approximations, and that to a large extent the calibrated model parameters do not vary when employing kinetic neutrals.

\section{Results with finer grid and smaller $\sigma_\perp$} \label{app:c}
\begin{figure}[ht]
	\centering
	\subfigure{\includegraphics[scale=0.5]{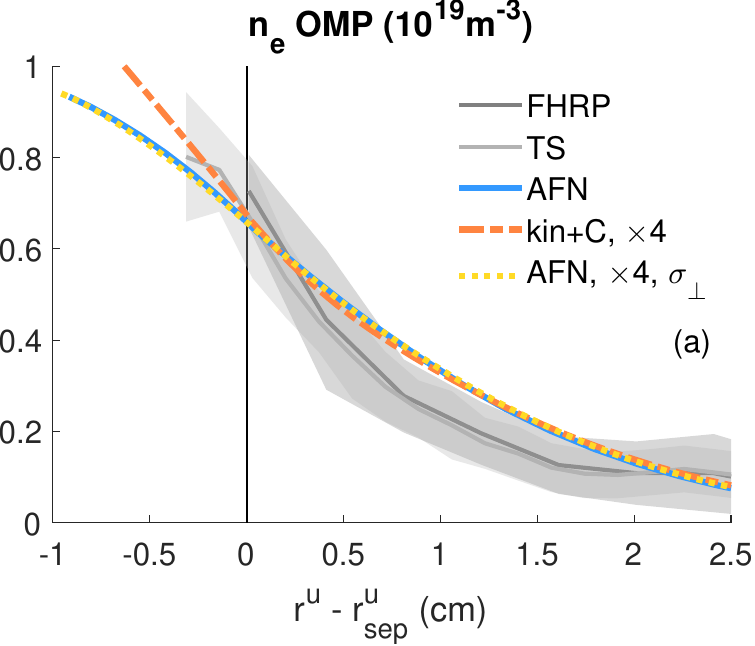}}
	\subfigure{\includegraphics[scale=0.5]{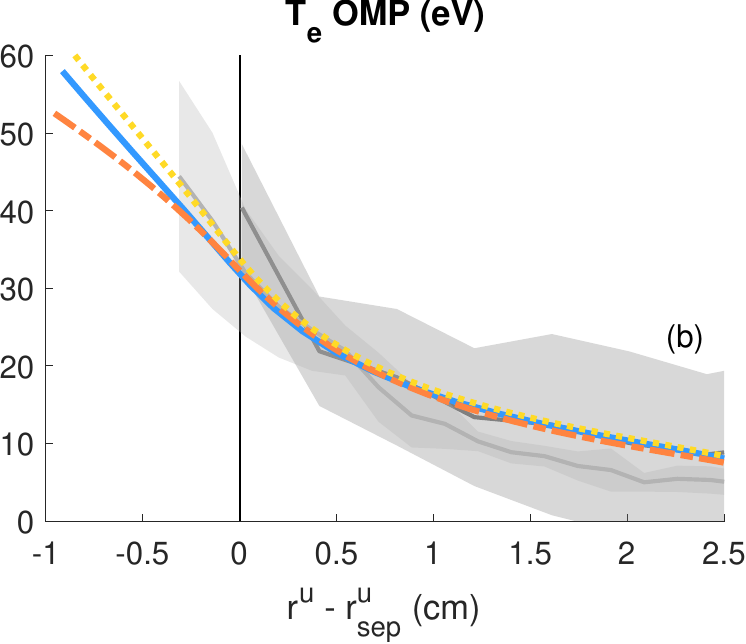}}
	\subfigure{\includegraphics[scale=0.5]{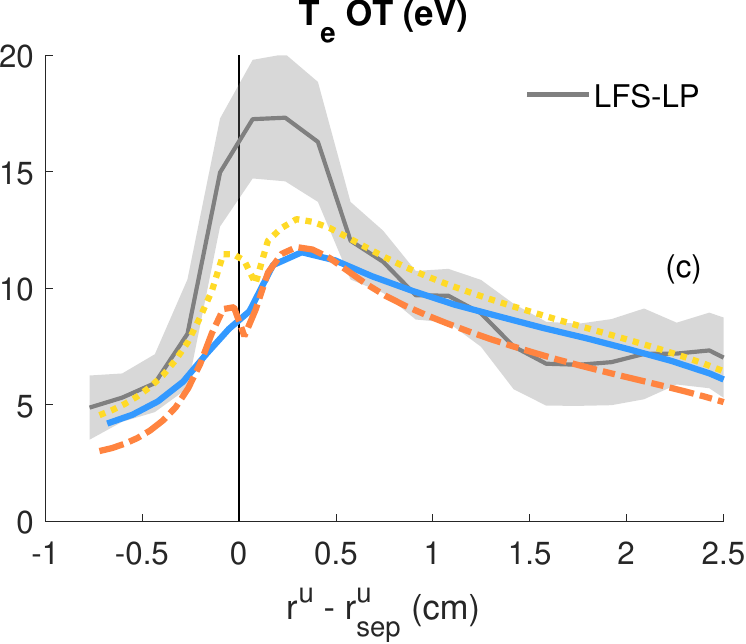}}
	\subfigure{\includegraphics[scale=0.5]{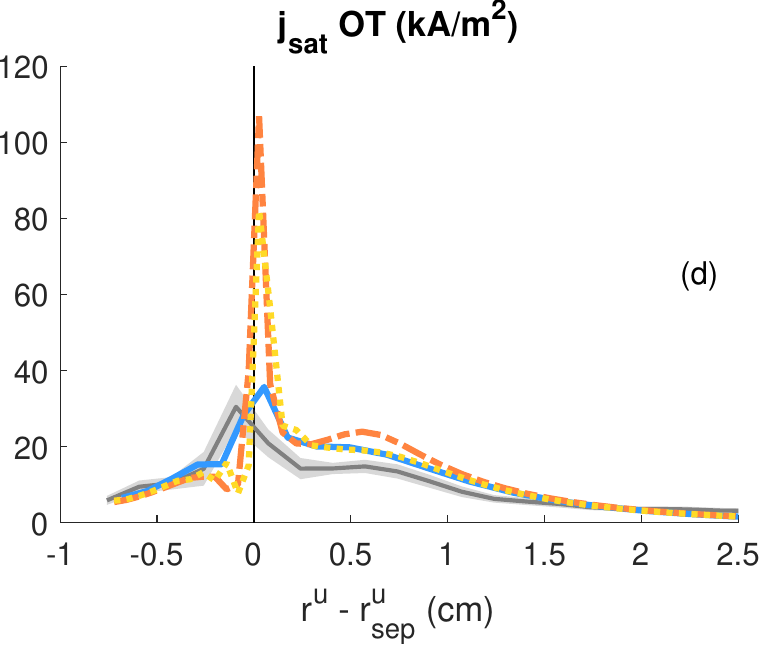}}
	\subfigure{\includegraphics[scale=0.5]{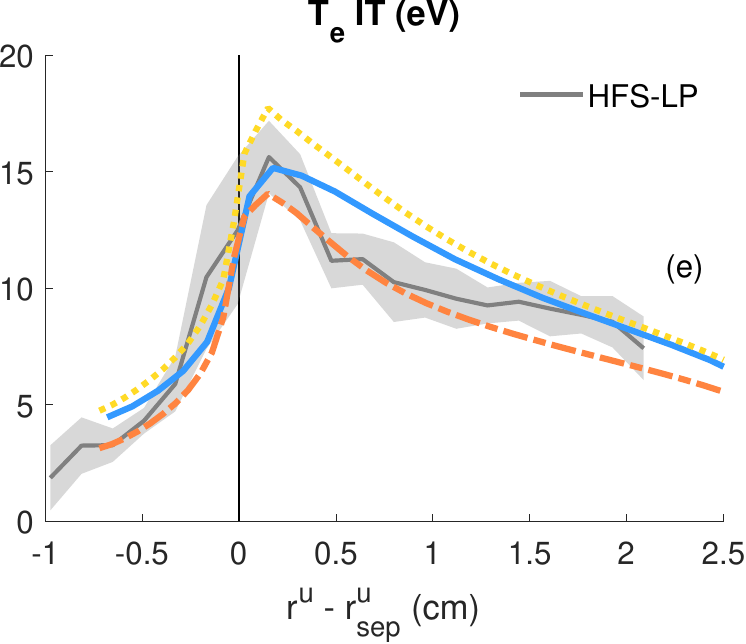}}
	\subfigure{\includegraphics[scale=0.5]{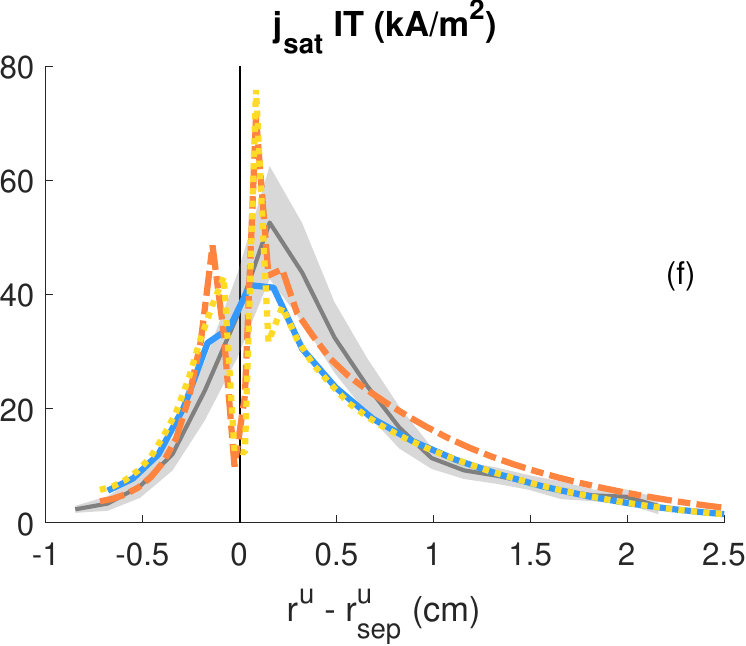}}
	\caption{Forward field case profiles as a function of the OMP radial distance from the separatrix of: uspstream electron density (a) and temperature (b); OT electron temperature (c) and saturation current (d); IT electron temperature (e) and saturation current (f). Thin solid grey lines indicate experimental data, with shaded regions indicating experimental uncertainty. Thick solid blue lines indicate Model 1 with AFN, thick dash-dotted orange lines indicate Model 1 with kinetic neutrals, C and $\times4$ finer mesh, and thick dotted yellow lines indicate Model 1 with AFN, a $\times4$ finer mesh and $\times50$ smaller $\sigma_{\perp}$. Solid vertical black lines indicate the separatrix location.} \label{fg:siggrid}
\end{figure}

In this appendix we assess the sensitivity of our simulation results to refinements in mesh size and on the perpendicular anomalous conductivity $\sigma_{\perp}$. The former is tackled by doubling the number of poloidal and radial points in which the computational domain is discretized, effectively obtaining a $\times4$ increase in the number of grid cells. For this finer mesh case we also employ the kinetic neutral model with carbon impurities described in Appendix \ref{app:b}. This way we aim at giving the total effect of our modeling assumptions, and not each of them separately. However, decreasing $\sigma_{\perp}$ in this finer grid with kinetic neutrals has proved impossible, and we therefore fell back to an AFN approximation. With this, we decreased the anomalous conductivity from $\sigma_{\perp}=2.5\cdot10^{-4}en_e$ S/m to $\sigma_{\perp}=5\cdot10^{-6}en_e$ S/m, a factor 50 reduction. The results are displayed on Figure \ref{fg:siggrid} for Model 1 only, and show that both using a finer grid and a smaller $\sigma_{\perp}$ have huge impact on the target saturation current, with OMP quantities and target $T_e$ less affected. In particular, one can notice the OT $j_{sat}$ peak increase by almost a factor three, while its location and the rest of the profile remains qualitatively the same. The double peak at the IT $j_{sat}$ seen in Figure \ref{fg:kinetic}f is further enhanced here. However, neither of the two effects is visible in the experimental measurements. While such data is an average of different shots, each of them being an average over multiple time instants, it is unlikely for such peaks to be completely smoothed out by the averaging procedure. Therefore, these may be of numerical nature. In absence of further evidence, we conclude that target $j_{sat}$ seems quite sensitive to these model assumptions, and that the calibrated parameters provide the same ballpark of results also in this case.

\clearpage

\bibliographystyle{ieeetr}
\bibliography{refs}

\begin{thebibliography}{100}

\bibitem{roadmap}
A.~Donn\'{e}, ``The {E}uropean roadmap towards fusion electricity,'' {\em
  Philosophical Transactions of the Royal Society A}, vol.~377, p.~20170432,
  2019.

\bibitem{grillix}
A.~Stegmeir, A.~Ross, T.~Body, M.~Francisquez, W.~Zholobenko, D.~Coster,
  O.~Maj, P.~Manz, F.~Jenko, B.~N. Rogers, and K.~S. Kang, ``Global turbulence
  simulations of the tokamak edge region with grillix,'' {\em Physics of
  Plasmas}, vol.~26, p.~052517, 05 2019.

\bibitem{soledge3x}
H.~Bufferand, J.~Bucalossi, G.~Ciraolo, G.~Falchetto, A.~Gallo, P.~Ghendrih,
  N.~Rivals, P.~Tamain, H.~Yang, G.~Giorgiani, F.~Schwander,
  M.~Scotto~d’Abusco, E.~Serre, Y.~Marandet, M.~Raghunathan, W.~Team, and the
  JET~Team, ``Progress in edge plasma turbulence modelling—hierarchy of
  models from 2d transport application to 3d fluid simulations in realistic
  tokamak geometry,'' {\em Nuclear Fusion}, vol.~61, p.~116052, oct 2021.

\bibitem{gbs}
P.~Ricci, F.~D. Halpern, S.~Jolliet, J.~Loizu, A.~Mosetto, A.~Fasoli, I.~Furno,
  and C.~Theiler, ``Simulation of plasma turbulence in scrape-off layer
  conditions: the gbs code, simulation results and code validation,'' {\em
  Plasma Physics and Controlled Fusion}, vol.~54, p.~124047, nov 2012.

\bibitem{detachment1}
A.~W. Leonard, ``Plasma detachment in divertor tokamaks,'' {\em Plasma Physics
  and Controlled Fusion}, vol.~60, p.~044001, feb 2018.

\bibitem{detachment2}
S.~I. Krasheninnikov and A.~S. Kukushkin, ``Physics of ultimate detachment of a
  tokamak divertor plasma,'' {\em Journal of Plasma Physics}, vol.~83, no.~5,
  p.~155830501, 2017.

\bibitem{detachment3}
P.~C. Stangeby, ``Basic physical processes and reduced models for plasma
  detachment,'' {\em Plasma Physics and Controlled Fusion}, vol.~60, p.~044022,
  mar 2018.

\bibitem{soledgeneut}
H.~Bufferand, G.~Ciraolo, R.~Düll, G.~Falchetto, N.~Fedorczak, Y.~Marandet,
  V.~Quadri, M.~Raghunathan, N.~Rivals, F.~Schwander, E.~Serre, S.~Sureshkumar,
  P.~Tamain, and N.~Varadarajan, ``Global 3d full-scale turbulence simulations
  of tcv-x21 experiments with soledge3x,'' {\em Nuclear Materials and Energy},
  vol.~41, p.~101824, 2024.

\bibitem{grillixneut}
K.~Eder, A.~Stegmeir, W.~Zholobenko, J.~Pfennig, D.~Brida, G.~Grenfell,
  F.~Jenko, and the ASDEX Upgrade~Team, ``Self-consistent plasma-neutrals fluid
  modeling of edge and scrape-off layer turbulence in diverted tokamaks,'' {\em
  Plasma Physics and Controlled Fusion}, vol.~67, p.~065034, jun 2025.

\bibitem{gbsneut}
M.~Giacomin, P.~Ricci, A.~Coroado, G.~Fourestey, D.~Galassi, E.~Lanti,
  D.~Mancini, N.~Richart, L.~Stenger, and N.~Varini, ``The gbs code for the
  self-consistent simulation of plasma turbulence and kinetic neutral dynamics
  in the tokamak boundary,'' {\em Journal of Computational Physics}, vol.~463,
  p.~111294, 2022.

\bibitem{reiter}
D.~Reiter, M.~Baelmans, and P.~Börner, ``The {EIRENE} and {B2-EIRENE} codes,''
  {\em Fusion Science and Technology}, vol.~47, no.~2, pp.~172--186, 2005.

\bibitem{closure1}
A.~L. Moser, L.~Casali, B.~M. Covele, A.~W. Leonard, A.~G. McLean, M.~W.
  Shafer, H.~Q. Wang, and J.~G. Watkins, ``Separating divertor closure effects
  on divertor detachment and pedestal shape in diii-d,'' {\em Physics of
  Plasmas}, vol.~27, p.~032506, 03 2020.

\bibitem{closure2}
O.~Février, H.~Reimerdes, C.~Theiler, D.~Brida, C.~Colandrea, H.~{De
  Oliveira}, B.~Duval, D.~Galassi, S.~Gorno, S.~Henderson, M.~Komm, B.~Labit,
  B.~Linehan, L.~Martinelli, A.~Perek, H.~Raj, U.~Sheikh, C.~Tsui, and
  M.~Wensing, ``Divertor closure effects on the tcv boundary plasma,'' {\em
  Nuclear Materials and Energy}, vol.~27, p.~100977, 2021.

\bibitem{closure3}
C.~Cowley, D.~Moulton, and B.~Lipschultz, ``Isolating the impacts of divertor
  closure on fueling and detachment with simplified solps-iter simulations of
  mast-u,'' {\em Physics of Plasmas}, vol.~33, p.~022508, 02 2026.

\bibitem{radiation1}
A.~Kallenbach, M.~Bernert, R.~Dux, L.~Casali, T.~Eich, L.~Giannone,
  A.~Herrmann, R.~McDermott, A.~Mlynek, H.~W. Müller, F.~Reimold,
  J.~Schweinzer, M.~Sertoli, G.~Tardini, W.~Treutterer, E.~Viezzer,
  R.~Wenninger, M.~Wischmeier, and the ASDEX Upgrade~Team, ``Impurity seeding
  for tokamak power exhaust: from present devices via iter to demo,'' {\em
  Plasma Physics and Controlled Fusion}, vol.~55, p.~124041, nov 2013.

\bibitem{radiation2}
M.~Bernert, M.~Wischmeier, A.~Huber, F.~Reimold, B.~Lipschultz, C.~Lowry,
  S.~Brezinsek, R.~Dux, T.~Eich, A.~Kallenbach, A.~Lebschy, C.~Maggi,
  R.~McDermott, T.~Pütterich, and S.~Wiesen, ``Power exhaust by sol and
  pedestal radiation at asdex upgrade and jet,'' {\em Nuclear Materials and
  Energy}, vol.~12, pp.~111--118, 2017.
\newblock Proceedings of the 22nd International Conference on Plasma Surface
  Interactions 2016, 22nd PSI.

\bibitem{pump1}
J.~Roth, K.~Krieger, and G.~Fussmann, ``Divertor retention for recycling
  impurities,'' {\em Nuclear Fusion}, vol.~32, p.~1835, oct 1992.

\bibitem{pump2}
A.~Kukushkin, H.~Pacher, V.~Kotov, D.~Reiter, D.~Coster, and G.~Pacher,
  ``Effect of the dome on divertor performance in iter,'' {\em Journal of
  Nuclear Materials}, vol.~363-365, pp.~308--313, 2007.
\newblock Plasma-Surface Interactions-17.

\bibitem{soledgezhdanov}
H.~Bufferand, J.~Balbin, S.~Baschetti, J.~Bucalossi, G.~Ciraolo, P.~Ghendrih,
  R.~Mao, N.~Rivals, P.~Tamain, H.~Yang, G.~Giorgiani, F.~Schwander,
  M.~Scotto~d’Abusco, E.~Serre, J.~Denis, Y.~Marandet, M.~Raghunathan,
  P.~Innocente, D.~Galassi, and J.~Contributors, ``Implementation of
  multi-component zhdanov closure in soledge3x,'' {\em Plasma Physics and
  Controlled Fusion}, vol.~64, p.~055001, mar 2022.

\bibitem{gbsimpurity}
A.~Coroado and P.~Ricci, ``A self-consistent multi-component model of plasma
  turbulence and kinetic neutral dynamics for the simulation of the tokamak
  boundary,'' {\em Nuclear Fusion}, vol.~62, p.~036015, mar 2022.

\bibitem{SOLPS1}
S.~Wiesen, D.~Reiter, V.~Kotov, M.~Baelmans, W.~Dekeyser, A.~Kukushkin,
  S.~Lisgo, R.~Pitts, V.~Rozhansky, G.~Saibene, I.~Veselova, and
  S.~Voskoboynikov, ``The new {SOLPS-ITER} code package,'' {\em Journal of
  Nuclear Materials}, vol.~463, pp.~480--484, 2015.

\bibitem{SOLPS2}
X.~Bonnin, W.~Dekeyser, R.~Pitts, D.~Coster, S.~S.~Voskoboyinikov, and
  S.~Wiesen, ``Presentation of the new {SOLPS-ITER} code package for tokamak
  plasma edge modelling,'' {\em Plasma and Fusion Research}, vol.~11,
  p.~1403102, 2016.

\bibitem{soledge}
H.~Bufferand, C.~Baudoin, J.~Bucalossi, G.~Ciraolo, J.~Denis, N.~Fedorczak,
  D.~Galassi, P.~Ghendrih, R.~Leybros, Y.~Marandet, N.~Mellet, J.~Morales,
  N.~Nace, E.~Serre, P.~Tamain, and M.~Valentinuzzi, ``Implementation of drift
  velocities and currents in {SOLEDGE2D–EIRENE},'' {\em Nuclear Materials and
  Energy}, vol.~12, pp.~852--857, 2017.

\bibitem{uedge}
T.~Rognlien, J.~Milovich, M.~Rensink, and G.~Porter, ``A fully implicit, time
  dependent {2-D} fluid code for modeling tokamak edge plasmas,'' {\em Journal
  of Nuclear Materials}, vol.~196-198, pp.~347--351, 1992.
\newblock Plasma-Surface Interactions in Controlled Fusion Devices.

\bibitem{edge2d}
G.~J. Radford, A.~V. Chankin, G.~Corrigan, R.~Simonini, J.~Spence, and
  A.~Taroni, ``The particle and heat drift fluxes and their implementation into
  the {EDGE2D} transport code,'' {\em Contributions to Plasma Physics},
  vol.~36, no.~2-3, pp.~187--191, 1996.

\bibitem{pitts}
R.~Pitts, X.~Bonnin, F.~Escourbiac, H.~Frerichs, J.~Gunn, T.~Hirai,
  A.~Kukushkin, E.~Kaveeva, M.~Miller, D.~Moulton, V.~Rozhansky,
  I.~Senichenkov, E.~Sytova, O.~Schmitz, P.~Stangeby, G.~{De Temmerman},
  I.~Veselova, and S.~Wiesen, ``Physics basis for the first {ITER} tungsten
  divertor,'' {\em Nuclear Materials and Energy}, vol.~20, p.~100696, 2019.

\bibitem{park}
J.-S. Park, X.~Bonnin, and R.~Pitts, ``Assessment of {ITER} divertor
  performance during early operation phases,'' {\em Nuclear Fusion}, vol.~61,
  p.~016021, 2021.

\bibitem{panasdex}
O.~Pan, M.~Bernert, T.~Lunt, M.~Cavedon, B.~Kurzan, S.~Wiesen, M.~Wischmeier,
  U.~Stroth, and t.~A. Upgrade~Team, ``Solps-iter simulations of an x-point
  radiator in the asdex upgrade tokamak,'' {\em Nuclear Fusion}, vol.~63,
  p.~016001, nov 2022.

\bibitem{horsten2025}
N.~Horsten, M.~Groth, V.-P. Rikala, B.~Lomanowski, A.~Meigs, S.~Aleiferis,
  X.~Bonnin, G.~Corrigan, W.~Dekeyser, R.~Futtersack, D.~Harting, D.~Reiter,
  V.~Solokha, B.~Thomas, S.~{Van den Kerkhof}, and N.~Vervloesem, ``Validation
  of solps-iter and edge2d-eirene simulations for h, d, and t jet iter-like
  wall low-confinement mode plasmas,'' {\em Nuclear Materials and Energy},
  vol.~42, p.~101842, 2025.

\bibitem{solpseast}
J.~Chen, Z.~Yang, D.~Coster, K.~Li, K.~Wu, Y.~Duan, L.~Wang, J.~Xu, X.~Chen,
  F.~Ding, Q.~Zang, Y.~Wang, J.~Wu, G.-N. Luo, and E.~Team, ``Experimental
  investigation and solps-iter modeling of ne-seeded radiative divertor h-modes
  plasma on east,'' {\em Physics of Plasmas}, vol.~26, p.~052501, 05 2019.

\bibitem{solpsgym}
M.~Sala, E.~Tonello, A.~Uccello, X.~Bonnin, D.~Ricci, D.~Dellasega,
  G.~Granucci, and M.~Passoni, ``Simulations of argon plasmas in the linear
  plasma device gym with the solps-iter code,'' {\em Plasma Physics and
  Controlled Fusion}, vol.~62, p.~055005, mar 2020.

\bibitem{solpsmagnumpsi}
J.~Gonzalez, E.~Westerhof, and T.~W. Morgan, ``Solps-iter simulations of a
  vapour box design for the linear device magnum-psi,'' {\em Plasma Physics and
  Controlled Fusion}, vol.~65, p.~055021, apr 2023.

\bibitem{subba}
F.~Subba, D.~Coster, M.~Moscheni, and M.~Siccinio, ``{SOLPS}-{ITER} modeling of
  divertor scenarios for {EU}-{DEMO},'' {\em Nuclear Fusion}, vol.~61,
  p.~106013, sep 2021.

\bibitem{solpssparc}
J.~D. Lore, J.-S. Park, T.~Eich, A.~Q. Kuang, M.~L. Reinke, S.~De~Pascuale,
  B.~Lomanowski, A.~Creely, and J.~M. Canik, ``Evaluation of sparc divertor
  conditions in h-mode operation using solps-iter,'' {\em Nuclear Fusion},
  vol.~64, p.~126054, oct 2024.

\bibitem{solpscfedr}
R.~DING, X.~LIU, H.~WANG, G.~XU, G.~JIA, F.~WANG, H.~SI, C.~SANG, C.~ZHANG,
  X.~ZHAO, and V.~CHAN, ``Divertor power exhaust with ar impurity seeding for
  cfedr h-mode operation,'' {\em Plasma Science and Technology}, vol.~27,
  p.~104002, oct 2025.

\bibitem{b25}
B.~Braams, ``Radiative divertor modelling for {ITER} and {TPX},'' {\em
  Contributions to Plasma Physics}, vol.~2/3, pp.~276--281, 1996.

\bibitem{braginskii}
S.~Braginskii, ``Transport processes in a plasma,'' {\em Reviews of Plasma
  Physics, edited by M. A. Leontovich, Consultants Bureau, New York, p.1},
  1965.

\bibitem{reimold}
F.~Reimold, M.~Wischmeier, M.~Bernert, S.~Potzel, D.~Coster, X.~Bonnin,
  D.~Reiter, G.~Meisl, A.~Kallenbach, L.~Aho-Mantila, and U.~Stroth,
  ``Experimental studies and modeling of complete h-mode divertor detachment in
  {ASDEX} {U}pgrade,'' {\em Journal of Nuclear Materials}, vol.~463,
  pp.~128--134, 2015.

\bibitem{solps_eq}
V.~Rozhansky, E.~Kaveeva, P.~Molchanov, I.~Veselova, S.~Voskoboynikov,
  D.~Coster, G.~Counsell, A.~Kirk, S.~Lisgo, and and, ``New b2solps5.2
  transport code for h-mode regimes in tokamaks,'' {\em Nuclear Fusion},
  vol.~49, no.~2, p.~025007, 2009.

\bibitem{paradela}
I.~Paradela~Pérez, B.~Lomanowski, J.~Lore, J.~Lovell, D.~Moulton, and M.~U.
  team, ``Power balance and divertor asymmetries in the super-x divertors of
  mast-u using solps-iter*,'' {\em Nuclear Fusion}, vol.~65, p.~066026, may
  2025.

\bibitem{solpsadc}
R.~Maurizio, A.~Leonard, J.~Yu, J.~Harrison, K.~Verhaegh, N.~Lonigro, and
  A.~McLean, ``Solps-iter modeling of divertor detachment in mast-u’s super-x
  double null configuration and comparison to experiment,'' {\em Nuclear
  Materials and Energy}, vol.~41, p.~101736, 2024.

\bibitem{unstructured}
W.~Dekeyser, P.~Boerner, S.~Voskoboynikov, V.~Rozhanksy, I.~Senichenkov,
  L.~Kaveeva, I.~Veselova, E.~Vekshina, X.~Bonnin, R.~Pitts, and M.~Baelmans,
  ``Plasma edge simulations including realistic wall geometry with
  {SOLPS-ITER},'' {\em Nuclear Materials and Energy}, vol.~27, p.~100999, 2021.

\bibitem{9point}
W.~Dekeyser, X.~Bonnin, S.~W. Lisgo, R.~A. Pitts, and B.~LaBombard,
  ``Implementation of a 9-point stencil in {SOLPS-ITER} and implications for
  {Alcator C-Mod} divertor plasma simulations,'' {\em Nuclear Materials and
  Energy}, vol.~18, pp.~125--130, 2019.

\bibitem{eireneaveraging}
M.~Baelmans, P.~Börner, K.~Ghoos, and G.~Samaey, ``Efficient code simulation
  strategies for b2-eirene,'' {\em Nuclear Materials and Energy}, vol.~12,
  pp.~858--863, 2017.
\newblock Proceedings of the 22nd International Conference on Plasma Surface
  Interactions 2016, 22nd PSI.

\bibitem{afns1}
N.~Horsten, G.~Samaey, and M.~Baelmans, ``Development and assessment of 2{D}
  fluid neutral models that include atomic databases and a microscopic
  reflection model,'' {\em Nuclear Fusion}, vol.~57, no.~11, p.~116043, 2017.

\bibitem{afns2}
W.~Van~Uytven, W.~Dekeyser, M.~Blommaert, S.~Carli, and M.~Baelmans,
  ``Assessment of advanced fluid neutral models for the neutral atoms in the
  plasma edge and application in iter geometry,'' {\em Nuclear Fusion},
  vol.~62, p.~086023, jun 2022.

\bibitem{spatiallyhybr1}
M.~Blommaert, N.~Horsten, P.~Börner, and W.~Dekeyser, ``A spatially hybrid
  fluid-kinetic neutral model for solps-iter plasma edge simulations,'' {\em
  Nuclear Materials and Energy}, vol.~19, pp.~28--33, 2019.

\bibitem{spatiallyhybr2}
N.~Horsten, M.~Groth, M.~Blommaert, W.~Dekeyser, I.~P. Pérez, and S.~Wiesen,
  ``Application of spatially hybrid fluid–kinetic neutral model on jet l-mode
  plasmas,'' {\em Nuclear Materials and Energy}, vol.~27, p.~100969, 2021.

\bibitem{makarov23}
S.~Makarov, D.~Coster, E.~Kaveeva, V.~Rozhansky, I.~Senichenkov, I.~Veselova,
  S.~Voskoboynikov, A.~Stepanenko, X.~Bonnin, and R.~Pitts, ``Implementation of
  solps-iter code with new grad–zhdanov module for d–t mixture,'' {\em
  Nuclear Fusion}, vol.~63, p.~026014, jan 2023.

\bibitem{diffusion}
B.~LaBombard, M.~Umansky, R.~Boivin, J.~Goetz, J.~Hughes, B.~Lipschultz,
  D.~Mossessian, C.~Pitcher, J.~Terry, and A.~Group, ``Cross-field plasma
  transport and main-chamber recycling in diverted plasmas on {A}lcator
  {C}-{M}od,'' {\em Nuclear Fusion}, vol.~40, pp.~2041--2060, dec 2000.

\bibitem{kmodel4}
R.~Coosemans, W.~Dekeyser, and M.~Baelmans, ``A self-consistent mean-field
  model for turbulent particle and heat transport in 2d interchange-dominated
  electrostatic exb turbulence in a sheath-limited scrape-off layer,'' {\em
  Contributions to Plasma Physics}, vol.~62, no.~5-6, p.~e202100193, 2022.

\bibitem{kmodel}
W.~Dekeyser, R.~Coosemans, S.~Carli, and M.~Baelmans, ``A self-consistent
  $k$-model for anomalous transport due to electrostatic, interchange-dominated
  {E}x{B} drift turbulence in the scrape-off layer and implementation in
  {SOLPS-ITER},'' {\em Contributions to Plasma Physics}, vol.~n/a, no.~n/a,
  p.~e202100190, 2022.

\bibitem{baschetti}
S.~Baschetti, H.~Bufferand, G.~Ciraolo, P.~Ghendrih, E.~Serre, P.~Tamain, and
  t.~WEST~Team, ``Self-consistent cross-field transport model for core and edge
  plasma transport,'' {\em Nuclear Fusion}, vol.~61, p.~106020, sep 2021.

\bibitem{kmodelcompass}
S.~Carli, W.~Dekeyser, R.~Coosemans, R.~Dejarnac, M.~Komm, M.~Dimitrova,
  J.~Adámek, P.~Bílková, and P.~Böhm, ``Interchange-turbulence-based radial
  transport model for {SOLPS-ITER}: a {COMPASS} case study,'' {\em
  Contributions to Plasma Physics}, vol.~60, no.~5-6, p.~e201900155, 2020.

\bibitem{x21paper}
D.~Oliveira, T.~Body, D.~Galassi, C.~Theiler, E.~Laribi, P.~Tamain,
  A.~Stegmeir, M.~Giacomin, W.~Zholobenko, P.~Ricci, H.~Bufferand, J.~Boedo,
  G.~Ciraolo, C.~Colandrea, D.~Coster, H.~de~Oliveira, G.~Fourestey, S.~Gorno,
  F.~Imbeaux, F.~Jenko, V.~Naulin, N.~Offeddu, H.~Reimerdes, E.~Serre, C.~Tsui,
  N.~Varini, N.~Vianello, M.~Wiesenberger, C.~Wüthrich, and the TCV~Team,
  ``Validation of edge turbulence codes against the tcv-x21 diverted l-mode
  reference case,'' {\em Nuclear Fusion}, vol.~62, p.~096001, jul 2022.

\bibitem{dudsonx21}
B.~Dudson, M.~Kryjak, H.~Muhammed, and J.~Omotani, ``Validation of hermes-3
  turbulence simulations against the tcv-x21 diverted l-mode reference case,''
  {\em Nuclear Fusion}, vol.~66, p.~036015, feb 2026.

\bibitem{jaynes}
E.~T. Jaynes, {\em Probability Theory: The Logic of Science}.
\newblock Cambridge University Press, 2003.

\bibitem{tarantola}
A.~Tarantola, {\em Inverse Problem Theory and Methods for Model Parameter
  Estimation}.
\newblock Society for Industrial and Applied Mathematics, 2005.

\bibitem{kennedyohagan}
M.~C. Kennedy and A.~O'Hagan, ``Bayesian calibration of computer models,'' {\em
  Journal of the Royal Statistical Society: Series B (Statistical
  Methodology)}, vol.~63, no.~3, pp.~425--464, 2001.

\bibitem{dose}
V.~Dose, ``Bayesian inference in physics: case studies,'' {\em Reports on
  Progress in Physics}, vol.~66, p.~1421, 2003.

\bibitem{toussaint}
U.~von Toussaint, ``Bayesian inference in physics,'' {\em Rev. Mod. Phys.},
  vol.~83, pp.~943--999, Sep 2011.

\bibitem{carlibayes}
S.~Carli, W.~Dekeyser, M.~Blommaert, R.~Coosemans, W.~Van~Uytven, and
  M.~Baelmans, ``Bayesian {MAP}-estimation of $\kappa$ turbulence model
  parameters using {A}lgorithmic {D}ifferentiation in {SOLPS-ITER},'' {\em
  Contributions to Plasma Physics}, p.~e202100184, 2021.

\bibitem{xuereb}
A.~Xuereb, M.~Groth, K.~Krieger, O.~Asunta, T.~Kurki-Suonio, J.~Likonen, and
  D.~Coster, ``{DIVIMP}-{B2}-{EIRENE} modelling of 13{C} migration and
  deposition in {ASDEX} {U}pgrade {L}-mode plasmas,'' {\em Journal of Nuclear
  Materials}, vol.~396, no.~2, pp.~228--233, 2010.

\bibitem{coster}
D.~Coster, J.~Kim, G.~Haas, B.~Kurzan, H.~Murmann, J.~Neuhauser, H.~Salzmann,
  R.~Schneider, W.~Schneider, J.~Schweinzer, and U.~Team, ``Automatic
  evaluation of edge transport coefficients with {B2-SOLPS}5.0,'' {\em
  Contributions to Plasma Physics}, vol.~40, no.~3-4, pp.~334--339, 2000.

\bibitem{kim}
J.-W. Kim and \textit{et al.}, ``Asdex-upgrade edge transport scalings from the
  two-dimensional interpretive code b2.5-i,'' {\em Journal of Nuclear
  Materials}, vol.~290-293, pp.~644--647, 2001.

\bibitem{wangx21}
Y.~Wang, C.~Colandrea, D.~Oliveira, C.~Theiler, H.~Reimerdes, T.~Body,
  D.~Galassi, L.~Martinelli, K.~Lee, and the TCV~Team, ``Validation of
  solps-iter simulations against the tcv-x21 reference case,'' {\em Nuclear
  Fusion}, vol.~64, p.~056040, apr 2024.

\bibitem{canik}
J.~Canik, R.~Maingi, V.~Soukhanovskii, R.~Bell, H.~Kugel, B.~LeBlanc, and
  T.~Osborne, ``Measurements and 2-d modeling of recycling and edge transport
  in discharges with lithium-coated pfcs in nstx,'' {\em Journal of Nuclear
  Materials}, vol.~415, no.~1, Supplement, pp.~S409--S412, 2011.
\newblock Proceedings of the 19th International Conference on Plasma-Surface
  Interactions in Controlled Fusion.

\bibitem{zito}
A.~Zito, M.~Wischmeier, D.~Carralero, P.~Manz, I.~Paradela~Pérez, M.~Passoni,
  and the ASDEX Upgrade~Team, ``Numerical modelling of an enhanced
  perpendicular transport regime in the scrape-off layer of asdex upgrade,''
  {\em Plasma Physics and Controlled Fusion}, vol.~63, p.~075003, may 2021.

\bibitem{bowman2}
C.~{Bowman} and J.~{Harrison}, ``{Automatic fitting of SOLPS-ITER predictions
  to experimental data using Bayesian optimization},'' in {\em APS Division of
  Plasma Physics Meeting Abstracts}, vol.~2022 of {\em APS Meeting Abstracts},
  p.~PO05.013, Jan. 2022.

\bibitem{hecko}
J.~{Hecko}, S.~{Soldat}, M.~{Tomes}, J.~{Seidl}, V.~{Smidl}, K.~{Hromasova},
  G.~{Verdoolaege}, and {the COMPASS team}, ``{Optimizing interpretive
  SOLPS-ITER simulations with an advanced Bayesian inference workflow}.''
  Presented at the 51st EPS Conference on Plasma Physics, July 2025.

\bibitem{fu2026}
Y.~Fu, B.~D. Dudson, X.~Chen, M.~V. Umansky, F.~Scotti, T.~D. Rognlien, and
  A.~W. Leonard, ``Statistical inference of anomalous thermal transport with
  uncertainty quantification for interpretive 2d sol models,'' {\em Plasma
  Physics and Controlled Fusion}, vol.~68, p.~015010, jan 2026.

\bibitem{bayesopt}
P.~Frazier, ``A tutorial on bayesian optimization,'' {\em ArXiv},
  vol.~abs/1807.02811, 2018.

\bibitem{baelmans2014}
M.~Baelmans, M.~Blommaert, J.~D. Schutter, W.~Dekeyser, and D.~Reiter,
  ``Efficient parameter estimation in 2{D} transport models based on an adjoint
  formalism,'' {\em Plasma Physics and Controlled Fusion}, vol.~56, p.~114009,
  oct 2014.

\bibitem{carliadjoint}
S.~Carli, L.~Hascoët, W.~Dekeyser, and M.~Blommaert, ``Algorithmic
  differentiation for adjoint sensitivity calculation in plasma edge codes,''
  {\em Journal of Computational Physics}, vol.~491, p.~112403, 2023.

\bibitem{carlipsi}
S.~Carli, M.~Blommaert, W.~Dekeyser, and M.~Baelmans, ``Sensitivity analysis of
  plasma edge code parameters through algorithmic differentiation,'' {\em
  Nuclear Materials and Energy}, vol.~18, pp.~6--11, 2019.

\bibitem{horstenad}
N.~Horsten, S.~Carli, and W.~Dekeyser, ``Sensitivity calculation for monte
  carlo particle simulations of neutrals in the plasma edge,'' {\em
  Contributions to Plasma Physics}, vol.~64, no.~7-8, p.~e202300138, 2024.

\bibitem{auroux}
D.~Auroux, P.~Ghendrih, L.~Lamerand, F.~Rapetti, and E.~Serre, ``Asymptotic
  behavior, non-local dynamics, and data assimilation tailoring of the reduced
  $\kappa$-$\epsilon$ model to address turbulent transport of fusion plasmas,''
  {\em Physics of Plasmas}, vol.~29, p.~102508, 10 2022.

\bibitem{lamerand}
L.~Lamérand, D.~Auroux, P.~Ghendrih, F.~Rapetti, and E.~Serre, ``Inverse
  problem for determining free parameters of a reduced turbulent transport
  model for tokamak plasma,'' {\em Advances in Computational Mathematics},
  vol.~50, no.~39, pp.~1572--9044, 2024.

\bibitem{nocedal}
J.~Nocedal and S.~Wright, {\em Numerical Optimization}.
\newblock Springer, 2006.

\bibitem{divshape}
W.~Dekeyser, D.~Reiter, and M.~Baelmans, ``Automated divertor target design by
  adjoint shape sensitivity analysis and a one-shot method,'' {\em Journal of
  Computational Physics}, vol.~278, pp.~117--132, 2014.

\bibitem{magnfield1}
M.~Blommaert, M.~Baelmans, W.~Dekeyser, N.~Gauger, and D.~Reiter, ``A novel
  approach to magnetic divertor configuration design,'' {\em Journal of Nuclear
  Materials}, vol.~463, pp.~1220--1224, 2015.

\bibitem{griewank}
A.~Griewank and A.~Walther, {\em Evaluating Derivatives}.
\newblock SIAM, 2008.

\bibitem{sander}
S.~Van~den Kerkhof, M.~Blommaert, J.~Coenen, and M.~Baelmans, ``Optimized
  design of a tungsten–copper functionally graded material monoblock for
  minimal von mises stress meeting the material operational temperature
  window,'' {\em Nuclear Fusion}, vol.~61, p.~046050, mar 2021.

\bibitem{petsc1}
S.~Balay, S.~Abhyankar, M.~Adams, S.~Benson, J.~Brown, P.~Brune, K.~Buschelman,
  E.~Constantinescu, L.~Dalcin, A.~Dener, V.~Eijkhout, W.~Gropp, V.~Hapla,
  T.~Isaac, P.~Jolivet, D.~Karpeev, D.~Kaushik, M.~Knepley, F.~Kong, S.~Kruger,
  D.~May, L.~McInnes, R.~Mills, L.~Mitchell, T.~Munson, J.~Roman, K.~Rupp,
  P.~Sanan, J.~Sarich, B.~Smith, S.~Zampini, H.~Zhang, H.~Zhang, and J.~Zhang,
  ``{PETS}c {W}eb page.'' \url{https://petsc.org/}, 2021.

\bibitem{petsc2}
S.~Balay, S.~Abhyankar, M.~Adams, S.~Benson, J.~Brown, P.~Brune, K.~Buschelman,
  E.~Constantinescu, L.~Dalcin, A.~Dener, V.~Eijkhout, W.~Gropp, V.~Hapla,
  T.~Isaac, P.~Jolivet, D.~Karpeev, D.~Kaushik, M.~Knepley, F.~Kong, S.~Kruger,
  D.~May, L.~McInnes, R.~Mills, L.~Mitchell, T.~Munson, J.~Roman, K.~Rupp,
  P.~Sanan, J.~Sarich, B.~Smith, S.~Zampini, H.~Zhang, H.~Zhang, and J.~Zhang,
  ``{PETSc/TAO} users manual,'' Tech. Rep. ANL-21/39 - Revision 3.16, Argonne
  National Laboratory, 2021.

\bibitem{petsc3}
S.~Balay, W.~Gropp, L.~McInnes, and B.~Smith, ``Efficient management of
  parallelism in object oriented numerical software libraries,'' in {\em Modern
  Software Tools in Scientific Computing} (E.~Arge, A.~M. Bruaset, and H.~P.
  Langtangen, eds.), pp.~163--202, Birkh{\"{a}}user Press, 1997.

\bibitem{kmodel1}
R.~Coosemans, W.~Dekeyser, and M.~Baelmans, ``A new mean-field plasma edge
  transport model based on turbulent kinetic energy and enstrophy,'' {\em
  Contributions to Plasma Physics}, vol.~60, no.~5-6, p.~e201900156, 2020.

\bibitem{kmodel2}
R.~Coosemans, W.~Dekeyser, and M.~Baelmans, ``Turbulent kinetic energy in 2d
  isothermal interchange-dominated scrape-off layer e × b drift turbulence:
  Governing equation and relation to particle transport,'' {\em Physics of
  Plasmas}, vol.~28, p.~012302, 01 2021.

\bibitem{kmodel3}
R.~Coosemans, W.~Dekeyser, and M.~Baelmans, ``Bayesian analysis of turbulent
  transport coefficients in 2d interchange dominated exb turbulence involving
  flow shear,'' {\em Journal of Physics: Conference Series}, vol.~1785,
  p.~012001, feb 2021.

\bibitem{kmodel5}
R.~Coosemans, W.~Dekeyser, and M.~Baelmans, ``Mean-field transport equations
  and energy theorem for plasma edge turbulent transport,'' {\em Journal of
  Plasma Physics}, vol.~90, no.~2, p.~905900202, 2024.

\bibitem{pope}
S.~B. Pope, {\em Turbulent Flows}.
\newblock Cambridge University Press, 2000.

\bibitem{solpsmanual}
X.~Bonnin, D.~Coster, O.~Wenisch, K.~Kukushkin, A.~Kukushkin, M.~Stanojevic,
  S.~Voskoboyinikov, E.~Kaveeva, I.~Senichenkov, S.~Makarov, S.~Carli, and
  W.~Van~Uytven, ``Solps-iter user manual,'' tech. rep., ITER Organization,
  2026.

\bibitem{dewolf}
R.~De~Wolf, R.~Coosemans, W.~Dekeyser, and M.~Baelmans, ``Bayesian approach to
  parameter estimation and model validation for nuclear fusion reactor
  mean-field edge turbulence modelling,'' {\em Nuclear Fusion}, vol.~61,
  p.~046048, mar 2021.

\bibitem{x21data}
D.~Sales~de Oliveira, T.~Body, D.~Galassi, and C.~Theiler, ``Tcv-x21: an open
  dataset for the validation of edge turbulence models,'' Mar. 2024.

\bibitem{wensing21}
M.~Wensing, H.~Reimerdes, O.~Février, C.~Colandrea, L.~Martinelli,
  K.~Verhaegh, F.~Bagnato, P.~Blanchard, B.~Vincent, A.~Perek, S.~Gorno,
  H.~de~Oliveira, C.~Theiler, B.~P. Duval, C.~K. Tsui, M.~Baquero-Ruiz,
  M.~Wischmeier, T.~Team, and M.~Team, ``Solps-iter validation with tcv l-mode
  discharges,'' {\em Physics of Plasmas}, vol.~28, p.~082508, 08 2021.

\bibitem{tonello2024}
E.~Tonello, F.~Mombelli, O.~Février, G.~Alberti, T.~Bolzonella,
  G.~Durr-Legoupil-Nicoud, S.~Gorno, H.~Reimerdes, C.~Theiler, N.~Vianello,
  M.~Passoni, the TCV~Team, and the WPTE~Team, ``Modelling of power exhaust in
  tcv positive and negative triangularity l-mode plasmas,'' {\em Plasma Physics
  and Controlled Fusion}, vol.~66, p.~065006, apr 2024.

\bibitem{colandrea}
C.~Colandrea, {\em Investigation of scrape-off layer and divertor transport
  using infrared thermography and SOLPS-ITER simulations}.
\newblock PhD thesis, EPFL, Lausanne, 2024.

\bibitem{carpita2025}
M.~Carpita, E.~Tonello, C.~Colandrea, O.~Février, H.~Reimerdes, G.~Sun, and
  F.~Mombelli, ``Verification of techniques for accelerated and stable
  solps-iter simulations including plasma drifts,'' {\em Nuclear Fusion},
  vol.~65, p.~116001, sep 2025.

\bibitem{kaveeva2018}
E.~Kaveeva, V.~Rozhansky, I.~Senichenkov, I.~Veselova, S.~Voskoboynikov,
  E.~Sytova, X.~Bonnin, and D.~Coster, ``Speed-up of solps-iter code for
  tokamak edge modeling,'' {\em Nuclear Fusion}, vol.~58, p.~126018, oct 2018.

\bibitem{bohm}
P.~C. Stangeby, ``The bohm–chodura plasma sheath criterion,'' {\em Physics of
  Plasmas}, vol.~2, no.~3, pp.~702--706, 1995.

\bibitem{repodata}
S.~Carli, ``{Replication Data for: Calibration of cross-field transport models
  in SOLPS-ITER on the TCV-X21 case},'' 2026.

\bibitem{matlab}
T.~M. Inc., ``Curve fitting toolbox version: 24.2 (r2024b),'' 2024.

\bibitem{eich}
T.~Eich, A.~Leonard, R.~Pitts, W.~Fundamenski, R.~Goldston, T.~Gray,
  A.~Herrmann, A.~Kirk, A.~Kallenbach, O.~Kardaun, A.~Kukushkin, B.~LaBombard,
  R.~Maingi, M.~Makowski, A.~Scarabosio, B.~Sieglin, J.~Terry, A.~Thornton,
  A.~U. Team, and J.~E. Contributors, ``Scaling of the tokamak near the
  scrape-off layer h-mode power width and implications for iter,'' {\em Nuclear
  Fusion}, vol.~53, p.~093031, aug 2013.

\bibitem{stangeby}
P.~Stangeby, {\em The Plasma Boundary of Magnetic Fusion Devices (1st ed.)}.
\newblock CRC Press, 2000.

\bibitem{unstrpet25}
W.~Dekeyser, S.~V. den Kerkhof, N.~Horsten, S.~Carli, W.~V. Uytven,
  N.~Vervloesem, G.~Samaey, M.~Blommaert, and M.~Baelmans, ``{Simulation of
  tokamak scrape-off layer plasmas up to the first wall with SOLPS-ITER}.''
  Presented at the 20th International Workshop on Plasma Edge Theory in Fusion
  Devices (PET-20), September 2025.

\bibitem{laplace}
L.~Tierney and B.~Kadane, ``Accurate approximations for posterior moments and
  marginal densities,'' {\em Journal of the American Statistical Association},
  vol.~81, pp.~82--86, 1986.

\bibitem{papadim}
D.~I. Papadimitriou and C.~Papadimitriou, ``Bayesian uncertainty quantification
  of turbulence models based on high-order adjoint,'' {\em Computers \&
  Fluids}, vol.~120, pp.~82--97, 2015.

\bibitem{reposcript}
S.~Carli, ``{Code for: Calibration of cross-field transport models in
  SOLPS-ITER on the TCV-X21 case},'' 2026.

\end{thebibliography}

\end{document}